\documentclass[12pt,a4paper]{article}
\usepackage[a4paper]{geometry}
\usepackage{amssymb}
\usepackage{amsmath}
\usepackage{graphicx}

\numberwithin{equation}{section}

\newcommand{\lyxaddress}[1]{
\par {\raggedright #1
\vspace{1.4em}
\noindent\par}
}

\begin{document}

\title{QUASI-CLASSICAL ALTERNATIVES IN QUANTUM\ CHEMISTRY}

\author{V. Gineityte}

\maketitle

\lyxaddress{Institute of Theoretical Physics and Astronomy, Vilnius University, Gostauto
12, LT-01108 Vilnius, Lithuania}

\begin{abstract}
The article contains an overview of authors achievements in developement of
alternative quantum-chemical approaches oriented towards revival of the
classical tradition of qualitative chemical thinking instead of obtaining
numerical results. The above-mentioned tradition is concluded to be based
mainly on principles (rules) of additivity, transferability and locality of
molecular properties. Accordingly, model Hamiltonian matrices are used in
the approaches under developement (called quasi-classical alternatives),
wherein algebraic parameters play the role of matrix elements and these are
assumed to be transferable for similar atoms and/or atomic orbitals in
addition. Further, passing to delocalized descriptions of electronic
structures (as usual) is expected to be the main origin of difficulties
seeking to formulate quasi-classical alternatives. In the framework of the
canonical method of molecular orbitals (MOs), delocalization is shown to be
partially avoidable by invoking a recently-suggested approach to secular
(eigenvalue) equations for model Hamiltonian matrices, wherein the usual
initial imposing of the zero-determinant condition is replaced by a certain
reformulation of the problem itself. The most efficient way of achieving the
same end, however, is shown to consist in passing to non-canonical
one-electron problems. The latter may be exemplified by the
block-diagonalization problem for the relevant Hamiltonian matrix following
from the Brillouin theorem and yielding non-canonical (localized) MOs and by
the commutation equation for the respective one-electron density matrix
(charge- bond order matrix). In this connection, most of attention is paid
in the article to perturbative solutions of the above-mentioned
non-canonical problems and to their implications, including common
quantum-chemical descriptions of entire classes of chemical compounds.
\end{abstract}

\section{Introduction. Quantum-mechanical and classical perspectives on
molecular structure and reactivity}

Chemistry experienced a period of a great invasion of quantum mechanics
during the last several decades. Consequently, numerical solutions of
various approximate versions of the Schr\"{o}dinger equation (such as the
Hartree-Fock (HF) equation (see e.g. [1-3])) started to play a central role
in the theoretical chemistry instead of traditional qualitative reasoning
based on simple models of interatomic interactions. The term 'computational
chemistry' is currently used to refer to this new branch of chemical science
[4-6].

Significant achievements of computations are beyond any doubt, especially so
far as quantitative aspects of molecular structures and properties are
concerned. Moreover, characteristics of both isolated molecules and chemical
reactions often are predictable nowadays as accurately as it is required and
thereby computations are able to precede or even to replace expensive
experiments. Unfortunately, these achievements go with essential losses.
Indeed, the most fundamental and fruitful classical generalities seem to be
left outside the contemporary theoretical chemistry, e.g. series and/or
classes of related compounds, functional groups (fragments), derivatives,
reaction center, various common effects including the inductive one, etc.
The same refers also to the main principles (rules) of the qualitative
chemical thinking, namely additivity, transferability and locality of
molecular properties, as well as to intuitition-based relations, e.g.
between local structures and local properties. This implies in summary that
the classical tradition of thinking about common trends in chemical
behaviour becomes broken off and we actually have to content ourselves with
studies of particular cases only. The aim of contributions overviewed in
this article consists in both revival of the above-discussed classical
tradition of qualitative thinking and its realization in terms of quantum
chemistry. These attempts are jointly called quasi-classical alternatives.

To reveal the reasons why the above-mentioned gap between the contemporary
quantum chemistry and the classical generalities arises, let us start with
perspectives on molecular structures, in general, and on their similarities,
in particular, underlying the quantum mechanics and the classical chemistry.

Molecules are considered as consisting of electrons and of nuclei in quantum
mechanics [2,3,7]. Accordingly, numbers of these particles are the principal
parameters both \ of the Schr\"{o}dinger equation and of its approximate
versions. Hence, specifying of these numbers is imperative to obtain a
solution, the latter thereby always referring to an individual molecule.
Although some similarity between two or several compounds may be revealed 'a
posteriori' (i.e. after comparison of the final results), no immanent
relation may be established between the relevant principal equations even if
the chemical structures of compounds concerned are very similar (e.g.
methane and ethane). Another important feature of these solutions (we mean
here of wave functions) consists in their generally delocalized nature
[1-3,8]. In other words, these functions usually embrace the entire system
under study and depend on its whole constitution in a rather intricate
manner (except for special cases when additional conditions of localization
are imposed [8]).

By contrast, chemical constitutions of molecules (see e.g. [9]) are defined
in terms of atoms involved in the given compound and the way these are bound
together (chemical bonds). Moreover, exact meanings of both atoms in
molecules and bonds are not essential in practice. Accordingly, compounds
consisting of uniform groups of atoms (elementary fragments) connected one
with another in a same manner are \textit{a priori} considered to be similar
whatever the actual number of these groups. As a result, a certain set of
similar molecules arises that is usually referred to as a class of related
compounds and regarded as a single object characterized by specific
properties (e.g. alkanes, acids [10,11], etc.). In other words, the
definition of the principal classical generality (i.e. of a class of
molecules) is based on common peculiarities of local constitutions, whereas
the overall structure of a particular molecule (the global structure) may
vary in a wide range. For example, the local structure of alkanes involves
both four-valency of carbon atoms and tetrahedral arrangement of respective
quartets of bonds, whilst the global structure embraces the total number of
atoms, the presence of cycles, branchings, etc. Accordingly, the relation
'structure/properties' is expected to reveal itself in two ways: i) as a
relation between local structures and local properties and ii) as that
between the global constitution and global properties. The first of these
aspects usually concerns common properties of the whole class under study
and is alternatively referred to as the principle (rule) of locality. Given
that certain details of structures and properties are ignored in addition,
the transferability rule for local characteristics follows [This rule is
usually applied to characteristics of atoms, bonds and functional groups of
related compounds [9]]. It also deserves adding here that the principles of
locality and transferability provide a basis for definition of the reaction
center (Section 9). Meanwhile, the relation between global structures and
global properties refers to individual representatives of the class and
gives birth to the well-known principle (rule) of additivity for properties
ascribed to the whole molecule [8].

The above comparative discussion demonstrates dramatic differences between
the two alternative perspectives on molecules and thereby their low
compatibility [12]. This fact evidently is among the main reasons of the
above-asserted gap between the computational and the classical branches of
chemistry. On the other hand, the same discussion gives us some hints
concerning the principal ways of developement of quasi-classical approaches:
First, the principle of locality should be realized as extensively as
possible in the approach concerned and preserved until obtaining the final
description of the whole system (if possible). Second, the transferability
rule should be invoked to unify quantum-mechanical problems for similar
molecules. Employment of models proves to be especially helpful in achieving
the latter end as discussed in Section 2 in a detail. So far as realization
of the locality principle is concerned, the overall state of things is
somewhat more complicated.

To clarify this point, let us turn again to the well-known canonical HF
equation [1-3] for molecules based on the concept of one-electron orbitals
(molecular orbitals (MOs)). These orbitals (MOs) are known to be
additionally expressed in the form of linear combinations of atomic orbitals
(AOs). The latter approach (usually called the LCAO approximation [1-3])
evidently is of classical origin and may be regarded as an important step
towards realization of the locality principle. Indeed, elements of the
representation matrix of the self-consistent Fockian are attached to
separate atoms and/or their pairs in the basis of AOs. Besides, employment
of hybrid AOs (HAOs) as basis functions \ [13,14] instead of usual AOs is
even more appropriate as the largest off-diagonal Fockian elements then
correspond to chemical bonds. These important achievements, however, become
lost almost entirely after passing to the basis of the standard (canonical)
MOs (CMOs), because just these MOs usually are delocalized over the whole
system under study and depend on its overall structure [1,2,8] as it is the
case with wave functions in general. Moreover, identical chemical bonds and
their groups are not, as a rule, accompanied by equivalent CMOs [8].
Finally, the link between CMO sets of related compounds is even more
complicated.

Two options seem to be possible in this situation. The first (conservative)
option consists in invoking additional approaches and/or modifications to
preserve the locality principle in the framework of the canonical method of
MOs. Some achievements of this type are discussed in Sections 3, 4 and 13.
The second (radical) option lies in turning to alternative one-electron
problems with respect to the canonical one, namely to the Brillouin theorem
(block-diagonalization problem) and the commutation equation for the
one-electron density matrix. The relevant achievements (the so-called PNCMO
theory) \ are discussed in Sections 5-12.

\section{The quasi-classical nature of the H\"{u}ckel model}

The very offspring of quantum chemistry was associated with developements of
simple models of electronic structures of polyatomic molecules that are
generally much closer to the classical chemistry as compared to the
contemporary numerical methods. The so-called 'curly arrow chemistry' based
on the well-known octet rule [15], the simplest version of the valence bond
(VB) method usually referred to as the resonance theory [16] and the H\"{u}%
ckel theory of molecular orbitals (HMO theory) [17,18] may be mentioned here
as the most outstanding examples. In this Section, we will dwell on the H%
\"{u}ckel model, both the standard HMO theory and studies overviewed in this
article are based on.

First of all, we should come to an agreement about the meaning of the term
'the H\"{u}ckel model'. Let us start with a notation that the original HMO
theory [19,20] was based on solution of secular (eigenvalue) equations for
respective model Hamiltonian matrices. As it turned out recently (see
Sections 5 and 6), however, applications of the H\"{u}ckel type Hamiltonian
matrices are not restricted to solutions of these most popular equations.
Moreover, just the employments of these matrices beyond the limits of the
canonical MO method offers us new quasi-classical alternatives. This implies
the term 'the H\"{u}ckel model' to mean much more than the standard HMO
theory.

Another important point here concerns the qualitative content of the H\"{u}%
ckel model. To discuss this aspect in a more detail, let us start with
recalling that the original version of the HMO theory [19,20] was intended
to be among the first semiempirical methods of calculation of electronic
structures of polyatomic molecules. The principal elements of the relevant
model Hamiltonian matrices (i.e. parameters $\alpha $\ and $\beta $) were
accordingly determined on the basis of the best coincidence of results of
calculation for a few reference compounds with the relevant experimental
data. It is no surprise, therefore, that understanding of the HMO theory as
an extremely approximate method of calculation of electronic structures is a
widespread viewpoint up to now. This popular attitude is additionally
supported by an obvious fact that the H\"{u}ckel type Hamiltonian matrix may
be considered as a rough approximation to the respective matrix of the
self-consistent Fockian. The qualitative content of the H\"{u}ckel model,
however, has little to do either with the numerical values of its parameters
or with computations. The main point here is that the structure of the model
Hamiltonian matrix reflects the spatial and/or chemical constitution of the
respective molecule including both local and global aspects. Indeed, certain
basis orbitals (AOs or HAOs) are assumed to correspond to diagonal elements (%
$\alpha $) of the H\"{u}ckel type Hamiltonian matrices ($\mathbf{H}$),
whilst the off-diagonal elements ($\beta $) represent the interactions
between these orbitals. The latter, in turn, are usually supposed to be
proportional to overlap integrals [21] being directly dependent on both the
respective internuclear distance and the spatial arrangement of basis
orbitals. As a result, the above-mentioned interrelation arises between the
structure of the matrix $\mathbf{H}$\ and that of the given molecule. For
conjugated and/or aromatic hydrocarbons, this relation is known to acquire
an extremely simple form, namely, the relevant model Hamiltonian matrix $%
\mathbf{H}$\ is proportional to the adjacency matrix of the graph describing
the structure of the C-skeleton of the given molecule [22]. In the case of
saturated hydrocarbons, an analogous relation also may be established
(Sections 3 and 4). Thus, we will concentrate ourselves on the
above-described qualitative aspects when using the term 'the H\"{u}ckel
model' and/or the 'H\"{u}ckel type model Hamiltonian matrix' throughout this
paper.

Furthermore, just the above-described relations between the H\"{u}ckel type
Hamiltonian matrices and chemical structures of molecules serve as the
principal argument for the quasi-classical nature of the H\"{u}ckel model
[In Ref.[12] the model has been considered as taking an intermediate place
in between quantum mechanics and the classical chemistry on the same basis].
Other arguments for this important conclusion follow from essential common
features between the model under discussion and the chemical perspective on
molecular structures:

First, neither the basis orbitals (AOs or HAOs) nor the one-electron
Hamiltonian operator underlying the H\"{u}ckel type Hamiltonian matrix $%
\mathbf{H}$ \ are defined explicitly in the model as it is the case with the
classical atoms in molecules and chemical bonds. Second, the role of the
number of electrons in the formation of electronic structures is extremely
reduced in the H\"{u}ckel model (Molecular orbitals are sought here without
regard to the number of electrons). Finally, the way the similarity of
related molecules is desribed in the model resembles that of the classical
chemistry. The last point deserves a separate discussion.

Thus, uniform $\alpha $\ and $\beta $\ parameters are usually accepted 
\textit{a priori} for conjugated and/or aromatic hydrocarbons [17,21] in
accordance with the classical transferability rule. This implies that less
important details of structures of individual systems are ignored for the
benefit of generality. Consequently, common model Hamiltonian matrices are
constructable for some simple series of related compounds (e.g. for
polyenes, polyacenes, etc.) and thereby common solutions of the relevant
secular equations are obtainable that contain the numbers of similar
fragments as the principal parameters (see Ref.[18] for an overview of these
solutions). The series of molecules concerned then becomes described as a
single object in the H\"{u}ckel model as it is the case with the classical
chemistry. Hydrocarbons and their heteroatom-containing derivatives is
another example of similar molecules. The respective H\"{u}ckel Hamiltonian
matrices usually differ one from another only in the value of a single
parameter, viz. of the Coulomb parameter ($\alpha _{X}$) referring to the
site of substitution [21,23]. As a result, the two similar problems may be
regarded as a single problem, wherein the parameter $\alpha _{X}$\ plays the
role of perturbation [23]. This way of investigation is evidently in line
with the classical concept of the derivative.

Therefore, the quasi-classical nature of the H\"{u}ckel model (in the
above-described qualitative sense) is beyond any doubt. That is why just
this model has been chosen to underly the attempts to construct the
analogues for the classical rules and generalities overviewed below. It is
evident that no numerical values of the Hamiltonian matrix elements are
required to achieve this end. Instead, algebraic methods are employed when
dealing with the relevant principal equations. Consequently, we actually
return to the 'world of deduction and modeling'[24].

\section{An alternative way of dealing with eigenvalue equations for
Hamiltonian matrices}

As already mentioned, the present Section is devoted to quasi-classical
alternatives in the framework of the canonical theory of MOs. In this
connection, we will dwell on secular (eigenvalue) equations for H\"{u}ckel
type model Hamiltonian matrices ($\mathbf{H}$) that yield the so-called
one-electron properties of molecules, viz. CMOs and respective one-electron
energies (coinciding with eigenfunctions and eigenvalues of the matrix $%
\mathbf{H}$, respectively). Moreover, eigenvalues referring to occupied CMOs
are additionally related to ionisation potentials of molecules measured
experimentally by the photoelectron spectroscopy [25] (cf. the Koopmans
theorem [2,26]). Accordingly, squares of the MO LCAO coefficients represent
the extents of participation of separate AOs in the ionisation of molecule
from the respective energy level.

Let us start with a notation that the principal rules of the qualitative
chemical thinking (Section 1) are usually applied to the so-called
'collective' properties of molecules [23], e.g. heats of formation and/or
atomisation, dipole moments, polarizabilities, etc. (the term 'collective'
is of quantum chemical origin here and indicates the property concerned to
be determined by all electrons and/or by all occupied one-electron states of
the system). Nevertheless, there are reasons to expect that the classical
principle of locality manifests itself in the one-electron characteristics
too. Indeed, existance of common peculiarities both of one-electron energy
spectra and of CMOs of related molecules follows from numerous theoretical
and experimental facts exemplified below and these peculiarities may be
traced back to similar local structures of compounds concerned. The most
outstanding theoretical fact worth mentioning is that CMOs of extended
compounds are expressible as linear combinations of those of elementary
fragments provided that the interfragmental interactions are weak and
thereby the perturbation theory is applicable (cf. the so-called PMO theory
[23]). This implies the local constitutions of CMOs of related molecules to
depend upon those of respective elementary fragments. [A question arises
here immediately whether these achievements may be generalized to the case
of comparable intra- and interfragmental interactions]. So far as
experimental arguments for common features of one-electron spectra of
similar molecules are concerned, the photoelectron spectra (PES) of alkanes
[27-30] are especially illustrative. Thus, two groups of ionization
potentials (two energy bands) reveal themselves in these spectra at 17-26
and 10-16 eV, respectively. The first of these bands (the so-called
high-energy band (HEB)) is of particular interest in our context. Indeed,
the HEB\ of an alkane C$_{N}$H$_{2N+2}$ or C$_{N}$H$_{2N}$ always consists
of $N$ peaks, the level distribution pattern of which closely resembles the
spectrum of the respective simple chain G$_{0}$ (Fig. 1). Given that the
band concerned is traced back to ionization of electrons mostly from $2s$\
AOs of carbon atoms (this is the simplest and commonly accepted
interpretation [27-30]), the above-mentioned level distribution pattern
seems to be due to similar local environments of these AOs over the chain,
i.e. to the uniform local constitution of alkanes. Among facts under present
interest, common spectral properties of alternant hydrocarbons [22,31] also
may be mentioned that may be found in almost all quantum chemistry textbooks
[A question here concerns the presence of common elementary fragments (see
below)].

It is evident that revealing of the influence of local structure upon
one-electron characteristics and thereby developement of the desired
quasi-classical alternative becomes feasible if we avoid an immediate
passing to the basis of delocalized MOs (CMOs). In this connection, an
alternative way of dealing with secular equations for Hamiltonian matrices ($%
\mathbf{H}$) of related chemical compounds has been suggested in Ref.[32],
the essence of which consists in an inverted order of the principal
operations vs. the standard one when solving the problem. Indeed, the first
step of the standard solution procedure coincides with imposing the
zero-determinant requirement and search for eigenvalues, whereas the second
one lies in obtaining the eigenfunctions. By contrast, we start with
reformulating the secular problem itself in order to reveal the structures
of eigenfunctions, whilst the eigenvalues follow from the second step. The
initial reformulating of the problem resolves itself into eliminating most
of variables (MO LCAO coefficients) by preserving only one of them for each
elementary fragment of the compound concerned. This variable-elimination
(reduction) procedure yields an effective $N\times N-$ dimensional secular
problem, wherein a single equation corresponds to each of $N$ elementary
fragments. Elements of the relevant reduced Hamiltonian matrix (to be
diagonalized during the second step)\ generally depend on the energy
variable $\varepsilon $ and the notation $\widetilde{\mathbf{H}}(\varepsilon
)$\ is used below in this connection. It is evident that diagonal and
off-diagonal elements of this new matrix represent elementary fragments and
their interactions, respectively. The off-diagonal elements often prove to
be additionally reducible to either 1 or 0 and thereby describe the
adjacencies of fragments. Given that this is the case, the local structure
of the given chain becomes represented implicitly via $\varepsilon $%
-dependent intrafragmental elements $\widetilde{H}_{ii}(\varepsilon )$,
whereas the respective global structure keeps to be reflected explicitly in
the overall constitution of the reduced matrix $\widetilde{\mathbf{H}}%
(\varepsilon )$. Moreover, any element $\widetilde{H}_{ii}(\varepsilon )$\
was shown to depend mostly on the structure of the Ith elementary fragment
and (to a certain extent) on its nearest environment [Elements $\widetilde{H}%
_{ii}(\varepsilon )$\ prove to be uniform or \ almost uniform for the same
elementary fragments in analogous environments]. Further, the above-desribed
variable-elimination procedure yields expressions for eigenfunctions of the
initial Hamiltonian matrix $\mathbf{H}$\ (i.e. for CMOs) in the form of
linear combinations of $N$ $\varepsilon -$dependent \ non-orthogonal basis
orbitals that are more or less localized on individual elementary fragments
and depend on constitution of the latter. In some cases, these basis
orbitals prove to be additionally related to MOs of elementary fragments.
Coefficients of the above-specified combinations, in turn, coincide with
those of eigenvectors of the reduced matrix $\widetilde{\mathbf{H}}%
(\varepsilon )$ and thereby are determined by the relevant global structure.
In summary, the roles of local and of global structures in the formation of
both one-electron spectra and CMOs are expected to follow from the first and
the second step of the new approach, respectively. Thus, a feasibility of
their separation arises, at least in principle. It also deserves emphasizing
that no assumptions about relatively weak interfragmental interactions are
invoked here and thereby a certain generalization of the above-mentioned
perturbation- theory- based approaches [23] is actually obtained.

Admitedly, acomplishing of the above-outlined alternative is not an easy
problem in practice. This especially refers to the ultimate diagonalization
of the $\varepsilon -$dependent reduced Hamiltonian matrix $\widetilde{%
\mathbf{H}}(\varepsilon )$. In this connection, three particular cases may
be distinguished that are characterized by relatively simple second steps.

a) The case, when an energy-variable- independent reduced Hamiltonian matrix
($\widetilde{\mathbf{H}}$) is obtainable [33], may be mentioned in the first
place. The well-known even alternant hydrocarbons (AHs) serve as an
excellent example here. To demonstrate this, let us introduce the $2N-$%
dimensional basis of $2p_{z}$\ AOs of carbon atoms $\{\chi \}$ and divide it
into two $N-$dimensional subsets $\{\chi ^{\ast }\}$\ \ and $\{\chi ^{\circ
}\}$ as usual [22, 31, 33, 34].\ The common H\"{u}ckel Hamiltonian matrix of
AHs may be then represented in terms of four submatrices (blocks) as follows%
\begin{equation}
\mathbf{H=}\left\vert 
\begin{array}{cc}
\mathbf{0} & \mathbf{B} \\ 
\mathbf{B}^{+} & \mathbf{0}%
\end{array}%
\right\vert ,  \tag{3.1}
\end{equation}%
where the non-zero blocks $\mathbf{B}$ and $\mathbf{B}^{+}$ contain
resonance parameters of chemical bonds (see Ref. [33] for details) and the
superscript + here and below designates the Hermitian-conjugate (transposed)
matrix. As originally shown in [35], the initial $2N\times 2N-$dimensional
secular problem for the matrix $\mathbf{H}$ of Eq.(3.1) resolves itself into
two $N\times N-$dimensional problems 
\begin{equation}
(\mathbf{BB}^{+})\mathbf{U=}\varepsilon ^{2}\mathbf{U,\qquad }(\mathbf{B}^{+}%
\mathbf{B)V=}\varepsilon ^{2}\mathbf{V,}  \tag{3.2}
\end{equation}%
where $\mathbf{U}$ and $\mathbf{V}$ are column- matrices of MO LCAO
coefficients referring to subsets subsets of AOs $\{\chi ^{\ast }\}$\ \ and $%
\{\chi ^{\circ }\},$\ respectively, and $\varepsilon $\ is the usual energy
variable. It is easily seen that any matrix problem of Eq.(3.2) may be
reformulated into a secular equation of the standard form for an $%
\varepsilon -$dependent effective Hamiltonian matrix, e.g.%
\begin{equation}
\widetilde{\mathbf{H}}(\varepsilon )=\mathbf{BB}^{+}+\varepsilon
(1-\varepsilon )\mathbf{I},  \tag{3.3}
\end{equation}%
where $\mathbf{I}$\ here and below stands for the unit matrix of the
relevant dimension (coinciding with $N$ in our case). Moreover, matrices $%
\mathbf{BB}^{+}$\ and $\widetilde{\mathbf{H}}(\varepsilon )$ commute one
with another and thereby possess a common set of eigenvectors contained
within the column-matrix $\mathbf{U}$. Consequently, the $\varepsilon -$%
dependent matrix $\widetilde{\mathbf{H}}(\varepsilon )$ of Eq.(3.3) may be
successfully replaced by the energy- variable- independent matrix $%
\widetilde{\mathbf{H}}=\mathbf{BB}^{+}$\ that was shown to represent [33]
adjacencies of elementary fragments of AHs specified below. Accordingly, a
generalized ($\varepsilon -$dependent) representation has been obtained for
eigenfunctions of the matrix $\mathbf{H}$ of Eq.(3.1) and thereby CMOs of
AHs, viz.%
\begin{equation}
\psi (\varepsilon )=\frac{1}{\sqrt{2}}\mathop{\displaystyle \sum }\limits%
_{i=1}^{(\ast )}\varphi _{i}(\varepsilon )U_{i},  \tag{3.4}
\end{equation}%
where the sum embraces AOs of the starred subset $\{\chi ^{\ast }\},$ $U_{i}$%
\ stand for elements of the column- matrix $\mathbf{U}$\ and $\varphi
_{i}(\varepsilon )$\ are $\varepsilon -$dependent basis orbitals of the
following constitution 
\begin{equation}
\varphi _{i}(\varepsilon )=\chi _{i}^{\ast }+\varepsilon ^{-1}%
\mathop{\displaystyle \sum }\limits_{k=1}^{(\circ )}\chi _{k}^{\circ
}B_{ki}^{+}  \tag{3.5}
\end{equation}%
containing a sum over AOs of the remaining (unstarred) subset $\{\chi
^{\circ }\}$ and referred to [33] as generalized basis orbitals (GBOs).
After substituting specific eigenvalues $\varepsilon _{m},$\ equations (3.4)
and (3.5) yield the usual expressions for respective CMOs of AHs. It is also
seen that each generalized basis orbital $\varphi _{i}(\varepsilon )$\ is
attached to a certain AO\ of the starred subset $\chi _{i}^{\ast }$ and
contains additional contributions of nearest-neighboring unstarred AOs $\chi
_{k}^{\circ }.$ This implies the constitution of the particular basis
orbital $\varphi _{i}(\varepsilon )$\ to be determined mostly by the nearest
environment of its principal AO $\chi _{i}^{\ast }$ and thereby by the local
structure [although a certain influence of the global structure also
manifests itself via the energy variable $\varepsilon $]. Consequently, an
interrelation has been concluded between local constitutions of AHs and
local shapes of their CMOs. A more detailed analysis of Eq.(3.5) showed that
three types of basis orbitals $\varphi _{i}(\varepsilon )$\ are actually
peculiar to AHs in accordance with three possible valencies of carbon atoms,
namely two-, three- and four-center orbitals. Moreover, the above-enumerated
orbitals correspondingly coincide with generalized ($\varepsilon -$%
dependent) MOs of ethene, allyle and trimethylenmethane. Thus, just these
rather simple systems were concluded to play the role of common elementary
fragments of AHs. Existance of these common fragments, in turn, formed the
basis for a conclusion that AHs may be considered as a separate class of
chemical compounds.

b) The case of weak actual dependences of elements of the reduced
Hamiltonian matrix upon energy variable $\varepsilon $\ may be mentioned as
a second one [36,37]. Elements $\widetilde{H}_{ii}(\varepsilon )$ may be
approximated by an appropriate constant (or a few constants) in this case.
The above-discussed HEB region of PES of alkanes [27-30] is an excellent
example here. A simple one-parameter H\"{u}ckel type Hamiltonian matrix
[28-30] designated by $\mathbf{H}_{1}$ proves to be an adequate model in
this case. The mean value of resonance integrals between bonding bond
orbitals (BBOs) of the nearest (geminal) bonds ($\beta $)\ playes the role
of the only parameter of this matrix. Moreover, the matrix $\mathbf{H}_{1}$\
is representable as follows%
\begin{equation}
\mathbf{H}_{1}=\alpha \mathbf{I+}\beta \mathbf{A(}G_{H}\mathbf{),}  \tag{3.6}
\end{equation}%
where $\mathbf{A(}G_{H}\mathbf{)}$\ stands for the adjacency matrix (AM) of
the graph $G_{H}$\ (further referred to as the Hamiltonian matrix graph) and 
$\alpha $\ is\ the relevant averaged Coulomb parameter (Note that the
equality $A_{ii}\mathbf{(}G_{H}\mathbf{)=}1$ for any $i$ \ has been accepted
for convenience). Graphs $G_{H}$\ are exhibited in Fig.1. It is seen that
full subgraphs (tetrahedrons) correspond to quartets of bonds attached to
the same carbon atom in these graphs. It is also noteworthy that subspectra
of matrices $\mathbf{A(}G_{H}\mathbf{)}$ corresponding to the HEB region
reflect both the above-discussed similarity to spectra of respective simple
chains $G_{0}$\ and the distinctive features of HEBs of alkanes.

\begin{figure}[htb]
  \begin{center}
    \includegraphics[width=0.9\textwidth]{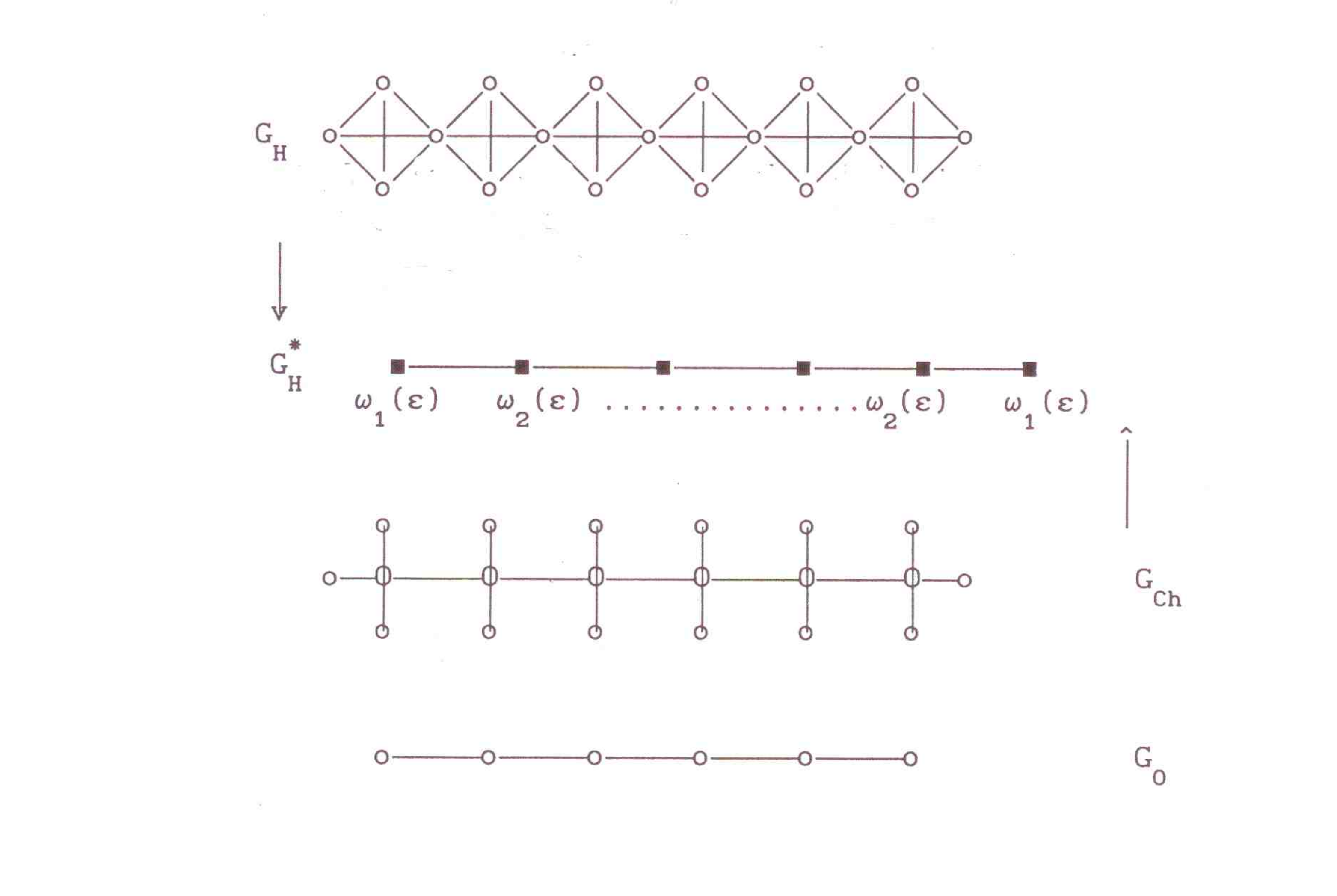}
  \end{center}
  \caption{The Hamiltonian matrix graph of the hexane molecule ($G_{H}$), the
  chemical graph of the same hydrocarbon in terms of atoms ($G_{Ch}$), and
  their common reduced form ($G_{H}^{\ast }\equiv G_{Ch}^{\ast }$). The
  relevant reduction procedures are indicated by arrows. The one-dimensional
  chain ($G_{0}$) also is shown.}
\end{figure}

Application of the above-described alternative approach to secular problems
for the AMs $\mathbf{A(}G_{H}\mathbf{)}$\ yields an $N-$dimensional reduced
problem for any alkane C$_{N}$H$_{2N+2}$ instead of the initial $3N+1-$%
dimensional problem, where $3N+1$ coincides with\ the respective total
number of bonding orbitals and/or chemical bonds. The most general way of
performing the appropriate reduction procedure [37] was based on
constructing a new variable $z=(C_{a}+C_{b}+C_{c}+C_{d})/\varepsilon $\ for
each full subgraph (tetrahedron) of the graph $G_{H}$\ containing four
vertices $a,b,c$ and $d$ and in reformulating the initial secular problem in
terms of $N$ variables $z_{k},k=1,2,..N$ ($C_{i},i=a,b..,$ stand here for
coefficients of eigenvectors of the AM $\mathbf{A(}G_{H}\mathbf{)}$). As a
result, an equation of the following form%
\begin{equation}
\mathop{\displaystyle \sum }\limits_{m}z_{m}+\omega _{v}(\varepsilon
)z_{p}=\varepsilon z_{p}  \tag{3.7}
\end{equation}%
has been derived for each (pth) tetrahedron, where $p=1,2...N$ and the sum
over $m$ embraces the variables $z_{m}$\ corresponding to neighboring
tetrahedrons with respect to the pth one. The functions $\omega
_{v}(\varepsilon )$\ of Eq.(3.7) depend on both the energy variable ($%
\varepsilon $) and the valency of the respective (i.e. pth) vertex of the
reduced graph $G_{H}^{\ast }$\ (Fig.1) according to the general expression 
\begin{equation}
\omega _{v}(\varepsilon )=3+\frac{4-v}{\varepsilon }.  \tag{3.8}
\end{equation}%
These functions evidently contain information about local structures of the
initial graph $G_{H}$\ including both the given tetrahedron and its nearest
environment. As is seen from Eq.(3.7) and Fig.1, the reduced graph $%
G_{H}^{\ast }$\ resembles the respective simple chain $G_{0}$\ except for a
more involved expression for diagonal elements of the respective new AM $%
\mathbf{A}(G_{H}^{\ast })$\ coinciding with $\omega _{v}(\varepsilon )$\ of
Eq.(3.8). Hence, the extent of similarity between graphs $G_{H}$\ \ and $%
G_{0}$\ (in respect of both structures and spectra) depends on the behaviour
of functions $\omega _{v}(\varepsilon )$\ within the $\varepsilon -$\ region
under interest.

The band limits of spectra of graphs $G_{H}^{\ast }$\ and thereby of $G_{H}$%
\ were shown to be conditioned by the equality $-2<[\varepsilon -\omega
_{2}(\varepsilon )]<2.$\ This requirement yields two intervals $\Delta
\varepsilon _{1}=(-0.37;-1)$\ \ and \ $\Delta \varepsilon _{2}=(2;5.37)$,
the latter corresponding to the HEB. A weak dependence of functions $\omega
_{v}(\varepsilon )$\ on $\varepsilon $\ and insignificant differences
between these functions for various valencies ($v$) have been established in
the above-specified region [36,37]. This principal result indicates the
local structures to be almost uniform over the chain in accordance with the
above expectation. Meanwhile, the global structures of particular alkanes
prove to be represented by respective simple chains $G_{0}$. \ Furthermore,
the same result allowed all diagonal elements of the reduced AM $\mathbf{A}%
(G_{H}^{\ast })$\ to be replaced by a single constant ($\varpi $), which
exceeds 3 as Eq.(3.8) indicates. Hence, the influence of the common local
structure of alkanes has been concluded to resolve itself mostly to a shift
of the whole level system by the value $\varpi -1$\ in the spectral region
concerned. Just this fact explains the observed similarity between HEBs of
alkanes and spectra of the simple chains $G_{0}$ [27-30]. Besides, no
assumption is required here about a significant energy gap between $2s$ and $%
2p$ AOs in contrast to ascribing the HEB to ionization of electrons from the 
$2s$ AOs.

\ For eigenfunctions ($\Psi _{i}$) of the initial AMs $\mathbf{A(}G_{H}%
\mathbf{),}$\ expressions like those of Eqs. (3.4) and (3.5) have been
obtained, where the variables $z_{k},k=1,2,..N$ play the role of
coefficients of linear combinations. These coefficients, in turn, may be
approximated by respective values for the relevant simple chain $G_{0}$. So
far as generalized ($\varepsilon -$dependent) basis orbitals of alkanes are
concerned, these also depend on the above-specified valency ($v$). As a
result, four distinct orbitals $\eta _{v}(\varepsilon ),v=1,2,3,4$\ have
been derived. Thus, the nearest neighborhood of the given tetrahedron also
playes some role in the formation of these orbitals. Nevertheless, a more
detailed analysis of these basis functions within the HEB region $\Delta
\varepsilon _{2}$ allowed us to conclude their dependence upon the valency $%
v $ to\ be actually insignificant. Moreover, the orbitals $\eta
_{v}(\varepsilon )$\ for any \ $v$ resemble the lowest completely symetric
MO of methane. Hence, the above-discussed interelation between local
structures of compounds concerned and those of respective CMOs proves to be
additionally supported.

c) The last example coincides with the so-called regular quasi-one-
dimensional systems characterized by translational symmetry [32,38]. In
other words, we turn now to chain-like molecules of cyclic global
constitution. These systems are known to serve as models of polymers and
solid state [39]. Accordingly, concepts and methods of the solid-state
theory [39-41] are most commonly applied to study the relevant one-electron
spectra. In this context, our approach proves to be a conseptual alternative
to the above-mentioned popular theory. This point deserves a somewhat more
detailed discussion.

Indeed, application of the solid state theory starts with taking into
account the translational symmetry of the whole chain and thereby its global
constitution. The usual way of doing this lies in passing to the basis of
delocalized Bloch functions being an analogue of CMOs. Moreover, a \ $\cos (%
\mathbf{ka})-$\ like\ dispersion curve usually corresponds to each Bloch
function and/or to each translationally- symmetric subchain of the whole
chain, where $\mathbf{k}$ and $\mathbf{a}$\ stand for the quasi-momentum
vector and the elementary cell's position vector, respectively. The second
step of the solid state theory coincides with diagonalization of a block of
the transformed Hamiltonian matrix corresponding to an elementary cell and
thereby with taking into account the local structure. This implies the final
dispersion curves to follow from the 'interaction' between the
above-specified $\cos (\mathbf{ka})-$\ like\ curves in this standard theory.
By contrast, our initial step consists in the regard for local constitution
of the same chain by reducing the initial system of secular equations into
an effective $N\times N-$\ dimensional problem as discussed above, where a
single equation corresponds to each of $N$ elementary cells and/or
fragments. Meanwhile, the translational symmetry may be taken into
consideration later when solving the reduced secular problem. It deserves
emphasizing here that symmetry properties of the initial system may be
easily preserved when performing the variable-elimination (reduction)
procedure. To this end, the retained variables (MO LCAO coefficients) should
be chosen to coincide with those referring to equivalent AOs. Finally, the
reduced secular problem proves to be representable by a certain new
effective chain, wherein extra bonds generally arise vs. the original ones,
e.g. between second neighboring pairs of AOs. As a result, the final
dispersion curves appear to be superpositions of $\cos (\mathbf{ka})-,\cos (2%
\mathbf{ka})-,$ etc. elementary curves.

For illustration, let us consider just the secular problem for the
above-described one-parameter Hamiltonian matrix ($\mathbf{H}_{1}$) of a
cyclic alkane C$_{N}$H$_{2N}$. The first step of our analysis consists in
reducing this problem into an $N-$dimensional one by means of passing to the
set of similar variables $z_{p},p=1,2,..N$. \ As already mentioned, the
relevant reduced chain $G_{H}^{\ast }$\ resembles the simple one ($G_{0}$)
except for diagonal parameters $\omega _{v}(\varepsilon )$\ defined by
Eq.(3.8). (It is evident that $\omega _{2}(\varepsilon )$ remains in the
present case). The translational symmetry of the reduced chain also may be
taken into account as usual. As a result, we obtain an implicit form of the
dispersion relation, viz.%
\begin{equation}
\varepsilon =\omega _{2}(\varepsilon )+2\cos [(\frac{2\pi }{N})j],  \tag{3.9}
\end{equation}%
where $j=1,2...N$\ and the function $\omega _{2}(\varepsilon )$\ is defined
by Eq.(3.8). Using the above-introduced vectors \ $\mathbf{k}$ and $\mathbf{a%
}$, the above relation may be reformulated as follows 
\begin{equation}
\varepsilon =\omega _{2}(\varepsilon )+2\cos (\mathbf{ka}).  \tag{3.10}
\end{equation}%
After substituting Eq.(3.8) and expressing $\varepsilon ,$\ the usual
explicit form of the dispersion relation may be easily obtained. The
above-exhibited forms of Eqs.(3.9) and (3.10), however, offer us a new
interpretation of dispersion curves. As already mentioned, the function $%
\omega _{2}(\varepsilon )$\ contains information about the local structure
of the chain. As is seen from the above formulae, the same function now
playes the role of an additive component of our dispersion relation and
describes the deviation of the actual dispersion curve from the standard $%
\cos (\mathbf{ka})-$\ like shape. Thus, this deviation may be unambiguously
ascribed to the effect of the local (internal) constitution of the chain.
Analogous implicit forms of the level density function also have been
derived.\ Finally, an eigenfunction ($\Psi _{i}$)\ of the initial AM $%
\mathbf{A(}G_{H}\mathbf{)}$\ takes the form of a Bloch sum of $N$ energy-
variable- dependent local-structure- determined basis orbitals $\eta
_{2}(\varepsilon )$.

Efficiency of the approach under present discussion is especially evident if
we turn to regular chains of more involved constitutions. In particular, the
secular problem for an extended model Hamiltonian matrix of the cyclic
polyethylene chain has been successfully studied [38], wherein resonance
parameters between BBOs of vicinal C-C(C-H) bonds have been taken into
consideration along with those of the above-described matrix $\mathbf{H}_{1}$%
. [This new model Hamiltonian matrix describes the whole spectrum of this
polymer adequately]. The implicit form of the final dispersion relation for
polyethylene then contains three additive components, viz. \ 
\begin{equation}
\varepsilon =\delta (\varepsilon )+2\tau (\varepsilon )\cos (\mathbf{ka}%
)+2\sigma (\varepsilon )\cos (2\mathbf{ka}),  \tag{3.11}
\end{equation}%
where 
\begin{align}
\delta (\varepsilon ) =&-\varepsilon ^{2}+4\varepsilon +2+2s(1+2s), 
\nonumber \\
\tau (\varepsilon ) =&\varepsilon -s(\varepsilon -5+2s),\quad \sigma
(\varepsilon )=s(9-2\varepsilon )  \tag{3.12}
\end{align}%
and $s$ stands for the resonance parameter between bonding orbitals of two
vicinal gauche-arranged bonds [for trans-arranged bonds, the same parameter
equals to $-2s$]. The energy- variable- dependent function $\delta
(\varepsilon )$ describes effective interactions inside a separate CH$_{2}$%
-group embedded into the chain and proves to be a generalization of the
function $\omega _{2}(\varepsilon )$\ of Eqs. (3.9) and (3.10). Meanwhile,
the remaining functions ($\tau (\varepsilon )$ and $\sigma (\varepsilon )$)
represent analogous interactions between pairs of first- and
second-neighboring CH$_{2}$-groups, respectively. Accordingly, a
superposition of $\cos (\mathbf{ka})-$ like and $\cos (2\mathbf{ka})-$ like
increments arises in the dispersion relation of polyethylene. The actual
shapes of dispersion curves prove to be determined by relative mean values
of functions $\tau (\varepsilon )$ and $\sigma (\varepsilon )$\ within the $%
\varepsilon $\ region under interest. On this basis, a new accounting has
been suggested for emergence of an unusual minimum within the low-energy
branch of the dispersion curves of polyethylene situated at a low symmetry
point of the first Brillouin zone ($k\approx 0.6\pi /a$), namely this
minimum has been established to appear owing to considerable values of
effective interactions between the second-neighboring CH$_{2}$-groups within
the respective energy interval. Similarity and differences between these
results and those of the standard solid state theory are discussed in Ref.
[32] in a detail.

In summary, the results of this Section support manifestation of the
classical locality principle in the formation of both one-electron spectra
and CMOs of molecules. Moreover, the above-described alternative approach
makes it possible to study the roles of local and of global structures
separately.

\section{Invoking of the concept of the Line graph}

The alternative approach of Section 3 is applicable to H\"{u}ckel type
Hamiltonian matrices of molecules, the latter being based on description of
these systems in terms of orbitals and their interactions (Section 2).
Meanwhile, atoms and interatomic (chemical) bonds play the role of the
principal terms in the classical chemistry (Section 1). Hence, an important
question is whether one-electron characteristics of molecules (including
energy spectra) may be related to (local and/or global) classical chemical
structures in terms of atoms and bonds, i.e. to (local and/or global)
peculiarities of the relevant chemical formulae (graphs). The present
Section addresses studies in this direction.

Let us start with emphasizing that, generally, there is no one-to-one
correspondence between basis orbitals (e.g. atomic orbitals (AOs)) and
atoms. For example, one and four AOs (HAOs) represent any hydrogen and any
carbon atom in alkanes, respectively. As a result, an alkane C$_{N}$H$%
_{2N+2} $ containing $3N+2$ atoms is described by $6N+2$ basis orbitals.
Although the latter number may be reduced substantially by passing to the
basis of bond orbitals (BOs) [28] and confining oneself to the subset of
bonding BOs (BBOs) (Section 3), the above-specified one-to-one
correspondence is still not ensured (we obtain $3N+1$ bonding BO for an
alkane C$_{N}$H$_{2N+2}$ ). Just this fact makes difficult any
straightforward interpretation of one-electron properties and/or spectra of
molecules in terms of peculiarities of the relevant chemical formulae and/or
graphs. Nevertheless, the above-specified end may be achieved by invoking
the concept of the so-called Line graph [42].

To demonstrate this, let us dwell again on one-electron spectra of alkanes
(Section 3). These were shown to result from secular equations for AMs $%
\mathbf{A(}G_{H}\mathbf{)}$\ of the relevant Hamiltonian matrix graphs $G_{H}
$\ \ referring to the basis of BBOs (Fig.1). From the one-to-one
correspondence between BBOs and chemical bonds it follows immediately that
the same graph represents the interactions (adjacencies) of chemical bonds.
Thus, a chemical graph in terms of bonds has been formally introduced for
any alkane. Let this new graph to be denoted by $G_{ch}^{b}$\ and note that $%
G_{H}\equiv G_{ch}^{b}.$\ 

According to the standard definition [42], the vertices of the unique Line
graph $L(G)$ with respect to the given graph $G$ correspond to the edges of
the latter, whereas the edges of the new graph $L(G)$\ represent the
adjacencies of the edges of the initial graph $G$ (Note that the two edges
are called adjacent if they possess a common vertex). Moreover, secular
polynomials (SPs) of any pair of graphs $G$ and $L(G)$ are interrelated as
follows%
\begin{equation}
P_{L(G)}(\lambda -2)=\lambda ^{p-q}P[\lambda \mathbf{I-A(}G\mathbf{)-D(}G%
\mathbf{)],}  \tag{4.1}
\end{equation}%
where $\lambda $\ stands for the SP variable. The right-hand side of
Eq.(4.1) contains characteristics of the graph $G$, viz. the number of edges
($p$) and that of vertices ($q$), as well as the respective adjacency matrix 
$\mathbf{A(}G\mathbf{)}$\ and the diagonal matrix of valencies $\mathbf{D}(G)
$. The notation $P[...]$ stands for the SP of the total matrix within the
braces. The left-hand side of Eq.(4.1) contains the SP of the graph $L(G)$,
where $\lambda -2$\ plays the role of the variable.

Let us return again to the case of alkanes. It is evident that the
above-discussed graph in terms of chemical bonds ($G_{ch}^{b}$) is the Line
graph with respect to the usual chemical graph of an alkane in terms of
atoms ($G_{ch}^{a}$), i.e. $G_{ch}^{b}=L(G_{ch}^{a})$. Hence, graphs $%
G_{ch}^{b}$\ and $G_{ch}^{a}$\ describe the chemical structure of an alkane
in terms of bonds and atoms, respectively. This implies the same information
to be involved within both $G_{ch}^{b}$\ and $G_{ch}^{a}$. Then, the
unambiguity requirement for the relation structure- spectrum served as a
basis for imposing the isospectrality condition on the AMs of the
above-mentioned graphs. To this end, just the relation of Eq.(4.1) has been
applied. To guarrantee a direct link between the SPs of graphs $G_{ch}^{b}$\
and $G_{ch}^{a}$\ necessary for ensuring their isospectrality, a modified AM
\ $\mathbf{B}(G_{ch}^{a})$ of the graph $G_{ch}^{a}$\ has been defined, viz.%
\begin{equation}
\mathbf{B}(G_{ch}^{a})=\mathbf{A}(G_{ch}^{a})+\mathbf{D}(G_{ch}^{a}). 
\tag{4.2}
\end{equation}%
Having in mind the choice $A_{ii}(G_{ch}^{b})=1$\ (Section 3), we obtain the
values of $B_{ii}(G_{ch}^{a})$\ equal to 3 and 0 for C and H atoms,
respectively. Furthermore, equalities $p-q=-1$\ and $p-q=0$\ refer to
non-cyclic and cyclic alkanes, correspondingly. Thus, the index $s(s=1,0)$
has been ascribed to the difference \ $q-p$. Finally, a new variable $%
\varepsilon =\lambda -1$\ may be introduced. From Eq.(4.1) we then obtain 
\begin{equation}
(\varepsilon +1)^{s}P[\mathbf{A}(G_{ch}^{b})](\varepsilon )=P[\mathbf{B}%
(G_{ch}^{a})](\varepsilon ),  \tag{4.3}
\end{equation}%
i.e. a coincidence between the SPs of graphs $G_{ch}^{b}$\ and $G_{ch}^{a}$\
accurate to a constant factor. As a result, the graphs $G_{ch}^{b}$\ and $%
G_{ch}^{a}$\ are isospectral apart from the additional root $\varepsilon
^{\prime }=-1$\ appearing for non-cyclic alkanes.

The so-called high-energy bands (HEBs) of alkanes have been traced back to
definite subspectra of graphs $G_{H}\equiv G_{ch}^{b}$ , namely to the
energy region $\Delta \varepsilon _{2}=(2;5.37)$\ (Section 3). The
above-concluded isospectrality of graphs $G_{ch}^{b}$\ and $G_{ch}^{a}$\
then implies the same HEB to be related also to the analogous subspectra of
the usual chemical graphs of alkanes in terms of atoms ($G_{ch}^{a}$). Just
the latter achievement allowed an interpretation of these energy bands in
terms of peculiarities of the relevant chemical formulae. This possibility
is even more surprising if we have in mind that AMs of graphs $G_{ch}^{a}$\
are not related to Hamiltonian matrices of systems concerned. It is also
evident that both principal graphs of alkanes (i.e. $G_{ch}^{b}$\ and $%
G_{ch}^{a}$) are reducible to the same graph $G_{ch}^{b\ast }\equiv $ $%
G_{ch}^{a\ast }\equiv G_{H}^{\ast }$. For the graph $G_{ch}^{b}\equiv G_{H},$%
\ the relevant procedure has been described in Section 3. To reduce the
graph $G_{ch}^{a}$\ into $G_{ch}^{a\ast },$\ no more is required as to
eliminate the coefficients at the monovalent vertices (H atoms) ($C_{h}$)
from the secular problem for the matrix $\mathbf{B}(G_{ch}^{a})$. As a
result, an $N-$dimensional problem for the reduced matrix $\mathbf{B}%
(G_{ch}^{a\ast })$\ coinciding with $\mathbf{A}(G_{ch}^{b\ast })\equiv 
\mathbf{A}(G_{H}^{\ast })$\ easily follows. On this basis, a new
interpretation of individual members of the expression for $\omega
_{v}(\varepsilon )$ of Eq.(3.8) may be formulated in terms of local chemical
structure, viz. the term 3 may be considered as the contribution of the
carbon atom, whereas the $\varepsilon -$dependent term ($(4-v)/\varepsilon $%
) proportional to the number of the adjacent hydrogen atoms ($4-v$) reflects
the contribution of the latter. Accordingly, the observed similarity between
HEBs of alkanes and the spectra \ of the simple chains $G_{0}$\ (Section 3)
may be accounted for by relatively weak effect of C-H bonds on spectra of
respective C-skeletons. Moreover, the effect is almost uniform over the
whole chain (see also Section 13).

Let us turn now to eigenfunctions of our adjacency matrices [37].\ It is
evident that an eigenfunction ($\Phi _{i}$) of the matrix $\mathbf{B}%
(G_{Ch}^{a})$\ may be ascribed to any energy level ($\varepsilon _{i}$) of
the spectral region $\Delta \varepsilon _{2}$\ along with the former
function $\Psi _{i}$\ (Section 3) referring to AMs \ $\mathbf{A}%
(G_{H})\equiv \mathbf{A}(G_{ch}^{b})$. As opposed to $\Psi _{i}$, however,
the eigenfunction $\Phi _{i}$\ has nothing to do with the relevant
Hamiltonian matrix and thereby with canonical MOs of alkanes. Further, the
above-mentioned reducibility of both graphs $G_{ch}^{b}$\ and $G_{Ch}^{a}$\
\ into the same graph ($G_{ch}^{b\ast }\equiv $ $G_{ch}^{a\ast }$) yields an
expression for $\Phi _{i}$\ of the following form 
\begin{equation}
\Phi _{i}=\mathop{\displaystyle \sum }\limits_{k=1}^{N}\mu _{k}(\varepsilon
_{i})z_{ki}\ ,  \tag{4.4}
\end{equation}%
where the coefficients $z_{ki}$\ follow from eigenvectors of the reduced AMs
of graphs $G_{ch}^{b\ast }\equiv $ $G_{ch}^{a\ast }$\ as previously (Section
3) and $\mu _{k}(\varepsilon )$\ are the relevant $\varepsilon -$dependent
basis functions pertinent to individual CH$_{r}-$ like subgraphs of the
graph $G_{Ch}^{a}.$ These\ generalized orbitals, in turn, are expressible in
terms of certain unspecified basis functions of carbon ($\kappa _{(c)k}$)
and of hydrogen atoms ($\kappa _{(h)kp}$), viz. 
\begin{equation}
\mu _{k}(\varepsilon )=\kappa _{(c)k}+\frac{1}{\varepsilon }%
\mathop{\displaystyle \sum }\limits_{m=1}^{r}\kappa _{(h)km},  \tag{4.5}
\end{equation}%
where the sum over $m$ embraces all the hydrogen atoms attached to the kth
carbon atom. The coefficient $z_{ki}$\ now represents the contribution of
the kth CH$_{r}-$ group to the eigenfunction $\Phi _{i}$. Moreover, the same
coefficient describes the increment of the kth carbon atom as substituting
of Eq.(4.5) into Eq.(4.4) shows. In this connection, squares of coefficients 
$z_{ki}$\ have been interpreted as the extents of participation of separate
carbon atoms in the ionization of molecule from the respective energy level $%
\varepsilon _{i}$\ [37]. These characteristics evidently are determined by
the global structure of the chain. To derive the extent of participation of
the attached hydrogen atom(s), the square $\mid z_{ki}\mid ^{2}$ should be
multiplied by an additional factor $1/\varepsilon ^{2}$ as Eq.(4.5)
indicates. Thus, participations of hydrogen atoms are related to those of
the neighboring carbon atom and vice versa and this relation is independent
of the global structure of the chain. It also deserves mentioning here that
no need actually arises for specifying the basis functions $\kappa _{(c)k}$
and $\kappa _{(h)kp}.$ This fact is in line with the chemical perspective on
molecular world (Section 2).

A more detailed analysis of the relation of Eq.(4.3) has been carried out in
Ref.[43]. Coincidence has been established between local terms of SPs $P[%
\mathbf{A}(G_{ch}^{b})]$ and $P[\mathbf{B}(G_{ch}^{a})]$\ \ corresponding to
definite subgraphs of graphs $G_{ch}^{b}$ and $G_{ch}^{a}.$ On this basis,
non-canonical expressions have been derived for these SPs in terms of the
same second-class Chebyshev polynomials $U_{p}(Q)$\ of the variable $Q$
proportional to the SP $q_{2}(\varepsilon )$\ of an isolated CH$_{r}$-like
subgraph. Common spectral properties of graphs $G_{ch}^{b}$ and $G_{ch}^{a}$
have been then related to those of $U_{p}(Q)$.

Therefore, invoking of the concept of the Line graph yields non-trivial
relations between one-electron characteristics of molecules and
peculiarities of their chemical formulae. Moreover, we actually obtain an
analogue of the classical principle of locality regarding one-electron
spectra.

\section{Choice of the block-diagonalization problem instead of the
secular equation}

As discussed already, the delocalized nature of the usual canonical MOs
(CMOs) is the main origin of difficulties in revealing quantum-chemical
analogues for classical chemical concepts. Again, the well-known success of
various additive models (schemes) in evaluating collective properties of
extended compounds (e.g. dipole moments) [8,9] indicates the feasibility of
a localized quantum chemical approach to electronic structures. The concept
of electron pairs pertaining to separate bonds also gives us a similar hint.

In the framework of the MO method, the above-anticipated approach may be
realized by invoking alternative sets of one-electron orbitals (MOs), in
particular the so-called localized MOs (LMOs) [1,2,8,21]. The latter may be
obtained either indirectly, i.e. by transforming the set of occupied CMOs
into that of LMOs using various localization criteria [8] or directly by
means of the Brillouin theorem [8, 44-47]. Passing to the basis of
delocalized CMOs may be entirely avoided just in the latter case. This fact
is among the principal reasons of our choice of the Brillouin theorem as the
method to be employed. Another reason consists in no need for specific
localization criteria when applying this theorem.

Among particular forms of the Brillouin theorem there is a zero value
requirement for an off-diagonal element of the Fockian operator referring to
an occupied and a vacant MO [48]. In its matrix form, this requirement
resolves itself into the zero matrix condition for the occupied-vacant
off-diagonal block (submatrix) of the total Fockian matrix in the basis of
non-canonical MOs (NCMOs) being sought [44-46]. As a result, the
block-diagonalization problem for the Fockian matrix actually arises (see
below), which becomes a real alternative to the usual diagonalization
problem and/or to the secular equation. As with the latter, the H\"{u}ckel
model may be invoked in the new problem so that the Fockian matrix becomes
replaced by the relevant H\"{u}ckel type Hamiltonian matrix containing 
\textit{a priori} uniform (transferable) parameters for analogous atoms and
bonds in related chemical compounds.

As opposed to the usual secular (eigenvalue) equation, the
block-diagonalization problem does not yield a unique set of LMOs [Note that
no unique NCMOs may be defined in contrast to CMOs [48]]. To reduce the
extent of this natural ambiguity, the block-diagonalization problem has been
supplemented with an orthogonality requirement for NCMOs (LMOs) being sought
in our studies [49-54] overviewed also in [55]. Meanwhile, orthogonality of
basis orbitals is not imperative [52].

Let the system under study to be represented by a certain $N$-dimensional
basis set $\{\Phi \},$ wherein the relevant H\"{u}ckel type Hamiltonian
matrix $\mathbf{H}$\ and the overlap matrix of basis orbitals $\mathbf{S}$\
are defined. Moreover, the system is assumed to contain an even number of
electrons (say $2n$). The most general form of the overall matrix problem
then resolves itself into two requirements [52], viz. 
\begin{equation}
\mathbf{C}^{+}\mathbf{HC=}\left\vert 
\begin{array}{cc}
\mathbf{E}_{1}^{(n\times n)} & \mathbf{0}^{(n\times s)} \\ 
\mathbf{0}^{(s\times n)} & \mathbf{E}_{2}^{(s\times s)}%
\end{array}%
\right\vert \equiv \mathbf{E},\quad \mathbf{C}^{+}\mathbf{SC=I,}  \tag{5.1}
\end{equation}%
where $\mathbf{E}_{1}^{(n\times n)}$\ and $\mathbf{E}_{2}^{(s\times s)}$\
are the so-called eigenblocks of the initial Hamiltonian matrix $\mathbf{H}$%
\ referring to subspaces of double-occupied and vacant one-electron orbitals
(NCMOs), respectively [Note that $s=N-n$], $\mathbf{C}$\ stands for the
relevant representation matrix of NCMOs (LMOs) being sought and $\mathbf{E}$
is the block-diagonal matrix of the right-hand side of the first relation.
Given that the basis orbitals $\{\Phi \}$ are orthonormalized in addition,
the second relation of Eq.(5.1) takes the form of an unitarity condition for
the transformation matrix $\mathbf{C}$\ , viz.%
\begin{equation}
\mathbf{C}^{+}\mathbf{C=I.}  \tag{5.2}
\end{equation}%
The above-described matrix problems, in turn, may be alternatively
represented as follows%
\begin{equation}
\mathbf{HC=SCE,\qquad HC=CE.}  \tag{5.3}
\end{equation}%
The second relation of Eq.(5.3) has been conveniently called the eigenblock
equation for the matrix $\mathbf{H.}$\ The first relation is then a
generalization of the latter to the case of a non-orthogonal basis set. [For
saturated systems, the basis set orthogonality assumption has been
substantiated in [56]].

Although the overall forms of equations shown in Eq.(5.3) closely resemble
those of the usual secular problems for matrices, their solution is much
more complicated. Again, solutions of Eqs.(5.1)-(5.3) discussed below prove
to be of a more general nature vs. those of secular equations, namely these
are expressed in terms of entire submatrices (blocks) of the initial
Hamiltonian matrix and embrace entire classes of molecules.

In this Section, we will dwell on the perturbative solution of the
eigenblock equation. In this connection, certain additional requirements
have been imposed on the initial Hamiltonian matrix $\mathbf{H.}$\
[Non-perturbative solutions also are possible for some particular cases and
these are discussed in Section 12]. The perturbative solution of the first
matrix problem of Eq.(5.3) has been obtained and analyzed in Ref.[52].
Because of its rather cumbersome nature, however, this result acquired no
further applications. Thus, let us confine ourselves to the more simple
second problem.

Let the basis set $\{\Phi \}$ to consist of $n$ double-occupied orbitals and
of $s$ vacant ones. Besides, the bonding and antibonding orbitals of
separate bonds (Sections 3 and 4) may be referred to here as examples.
Further, let the above-specified orbitals to be collected into two subsets $%
\{\Phi _{1}\}$\ and $\{\Phi _{2}\}$\ [ Notations $\{\Phi _{(+)}\}$\ and \ $%
\{\Phi _{(-)}\}$\ used previously also will be preserved]. To be able to
look for the matrix $\mathbf{C}$\ and for the eigenblocks $\mathbf{E}%
_{1}^{(n\times n)}$\ and $\mathbf{E}_{2}^{(s\times s)}$\ in the form of
power series (i.e. as sums of contributions of various orders ($k$)), the
subsets $\{\Phi _{1}\}$\ and $\{\Phi _{2}\}$\ will be assumed to be
separated by a substantial energy gap vs. the intersubset interactions. In
this connection, let the matrix $\mathbf{H}$\ to contain a sum of zero and
first order matrices ($\mathbf{H}_{(0)}$ and $\mathbf{H}_{(1)},$\
respectively), the former taking a block-diagonal form [51, 53-55], viz.%
\begin{equation}
\mathbf{H=H}_{(0)}+\mathbf{H}_{(1)}=\left\vert 
\begin{array}{cc}
\mathbf{E}_{(+)} & \mathbf{0} \\ 
\mathbf{0} & -\mathbf{E}_{(-)}%
\end{array}%
\right\vert +\left\vert 
\begin{array}{cc}
\mathbf{T} & \mathbf{R} \\ 
\mathbf{R}^{+} & \mathbf{Q}%
\end{array}%
\right\vert ,  \tag{5.4}
\end{equation}%
where the minus sign in front of $\mathbf{E}_{(-)}$\ is introduced for
further convenience. The blocks $\mathbf{E}_{(+)},$ $\mathbf{E}_{(-)},%
\mathbf{T,R}$ and $\mathbf{Q}$ correspond here to subsets $\{\Phi _{(+)}\}$\
and \ $\{\Phi _{(-)}\}$\ \ and to their interaction, and the supersript +
represents the Hermitian- conjugate (transposed) counterpart of the matrix $%
\mathbf{R.}$ It should be emphasized here that no specifying either of
internal constitutions of the above-enumerated blocks or their dimensions is
required [In this connection, the superscripts $(n\times n),(n\times s),etc.$
are omitted in Eq.(5.4) and below for simplicity]. Finally, members ($%
\mathbf{C}_{(k)}$) of the power series for the matrix $\mathbf{C}$\ also may
be represented in terms of four blocks of respective dimensions, viz.%
\begin{equation}
\mathbf{C}_{(k)}\mathbf{=}\left\vert 
\begin{array}{cc}
\mathbf{C}_{11}^{(k)} & \mathbf{C}_{12}^{(k)} \\ 
\mathbf{C}_{21}^{(k)} & \mathbf{C}_{22}^{(k)}%
\end{array}%
\right\vert ,  \tag{5.5}
\end{equation}%
where the order parameter ($k$) takes the upper position for convenience. It
also deserves mentioning here that the above-formulated problem proves to be
a matrix generalization of a two-level problem, wherein non-commutative
quantities (submatrices) stand instead of usual (one-dimensional) Coulomb
and resonance parameters [50]. In this connection, the perturbation theory
(PT) used for its solution (see below) has been referred to as the
non-commutative Rayleigh- Schr\"{o}dinger PT (NCRSPT) [53,55] (see also
Section 13).

Let us turn now to an overview of solution of the eigenblock equation of
Eq.(5.3). Let us note first that a certain ambiguity in determining NCMOs
still remains even after imposing an orthogonality condition on the latter.
In this connection, let the zero order member $\mathbf{C}_{(0)}$\ to
coincide with the unit matrix ($\mathbf{I}$). It is noteworthy that choice $%
\mathbf{C}_{(0)}=\mathbf{I}$\ complies with all requirements concerned. This
implies that we may actually confine ourselves to NCMOs (LMOs) of the
basis-orbital- and-tail constitution. Further, the two off-diagonal blocks ($%
\mathbf{C}_{12}^{(k)}$ and $\mathbf{C}_{21}^{(k)}$) of the kth order matrix $%
\mathbf{C}_{(k)}$\ $(k=1,2,3...)$ of Eq.(5.5) were shown to be expressible
via the same matrix $\widetilde{\mathbf{G}}_{(k)}$, i.e. 
\begin{equation}
\mathbf{C}_{12}^{(k)}=\widetilde{\mathbf{G}}_{(k)},\qquad \mathbf{C}%
_{21}^{(k)}=-\widetilde{\mathbf{G}}_{(k)}^{+}.  \tag{5.6}
\end{equation}%
Matrices $\widetilde{\mathbf{G}}_{(k)},(k=1,2,3...),$ in turn, proved to be
conditioned by equations of the common form, viz.\ 
\begin{equation}
\mathbf{E}_{(+)}\widetilde{\mathbf{G}}_{(k)}+\widetilde{\mathbf{G}}_{(k)}%
\mathbf{E}_{(-)}+\widetilde{\mathbf{V}}_{(k)}\mathbf{=0}  \tag{5.7}
\end{equation}%
containing a series of matrices $\widetilde{\mathbf{V}}_{(k)},(k=1,2,3...)$
defined as follows%
\begin{align}
\mathbf{V}_{(1)} &=\mathbf{R,\qquad V}_{(2)}=\mathbf{TG}_{(1)}-\mathbf{G}%
_{(1)}\mathbf{Q,}  \nonumber \\
\widetilde{\mathbf{V}}_{(3)} &=\mathbf{TG}_{(2)}-\mathbf{G}_{(2)}\mathbf{Q-}%
\frac{1}{2}\mathbf{(RG}_{(1)}^{+}\mathbf{G}_{(1)}+\mathbf{G}_{(1)}\mathbf{G}%
_{(1)}^{+}\mathbf{R+\mathbf{G}_{(1)}R\mathbf{G}_{(1)}^{+}),}etc\mathbf{.} 
\tag{5.8}
\end{align}%
[The symbol \symbol{126} serves to distinguish between the principal
matrices of the present Section resulting from the block-diagonalization
problem and those following from the commutation equation of Section 6. The
distiction concerned, however, actually manifests itself starting from $k=3$
only and thereby the symbol \symbol{126} is omitted for $k=1$ and $k=2$ in
Eq.(5.8)].

Again, the blocks $\mathbf{C}_{11}^{(k)}$ and $\mathbf{C}_{22}^{(k)}$\
taking the diagonal positions within the same matrices $\mathbf{C}_{(k)}$\
of Eq.(5.5) are expressible via matrices $\widetilde{\mathbf{G}}_{(k)}$\ of
lower orders, e.g.%
\begin{align}
\mathbf{C}_{11}^{(2)} =&-\frac{1}{2}\mathbf{G}_{(1)}\mathbf{G}%
_{(1)}^{+},\quad \mathbf{C}_{22}^{(2)}=-\frac{1}{2}\mathbf{G}_{(1)}^{+}%
\mathbf{G}_{(1)},  \nonumber \\
\mathbf{C}_{11}^{(3)} =&-\frac{1}{2}(\mathbf{G}_{(1)}\mathbf{G}_{(2)}^{+}+%
\mathbf{G}_{(2)}\mathbf{G}_{(1)}^{+}),\quad etc.  \tag{5.9}
\end{align}%
[Note that $\mathbf{C}_{11}^{(1)}$ $=\mathbf{C}_{22}^{(1)}=\mathbf{0}$].
Finally, similar expressions follow for members of power series for
eigenblocks $\mathbf{E}_{1}$\ and $\mathbf{E}_{2}$\ (Section 13).

Let the occupied NCMOs ($\psi _{1,i}$) to be collected into the ket-vector $%
\mid \Psi _{1}>.$ The latter is then expressible as follows%
\begin{equation}
\mid \Psi _{1}>=\mid \Phi _{1}>\mathbf{C}_{11}+\mid \Phi _{2}>\mathbf{C}_{21}%
\mathbf{,}  \tag{5.10}
\end{equation}%
where $\mid \Phi _{1}>$\textbf{\ }and\textbf{\ }$\mid \Phi _{2}>$\ are
ket-vectors correspondingly embracing the subsets of basis functions $\{\Phi
_{1}\}$ and $\{\Phi _{2}\}.$ Substituting the power series expansions for
submatrices $\mathbf{C}_{11}$ and $\mathbf{C}_{21}$ into Eq.(5.10) then
yields the following result%
\begin{equation}
\mid \Psi _{1}>=\mid \Phi _{1}>\mathbf{(I-}\frac{1}{2}\mathbf{\mathbf{G}%
_{(1)}\mathbf{G}_{(1)}^{+})}-\mid \Phi _{2}>(\mathbf{G}_{(1)}^{+}+\mathbf{G}%
_{(2)}^{+})\mathbf{,}  \tag{5.11}
\end{equation}%
where terms to within the second order inclusive are shown. It is seen that
each occupied LMO is attached to an individual basis orbital in accordance
with the equality $\mathbf{C}_{(0)}=\mathbf{I}$ and contains a certain tail
extending over the remaining occupied and vacant basis functions in
addition. Moreover, tails of the former type are determined by elements of
the matrix \ $\mathbf{\mathbf{G}_{(1)}\mathbf{G}_{(1)}^{+}}$, whilst those
embracing vacant basis orbitals coincide with elements of the sum of the
principal matrices $\widetilde{\mathbf{\mathbf{G}}}\mathbf{_{(k)}}$. This
implies that explicit expressions for LMOs are actually obtainable if matrix
equations of Eq.(5.7) may be solved algebraically.

General solutions of matrix problems like that of Eq.(5.7) are known to take
the form of an integral [57]. This integral, however, yields no explicit
relations between elements of matrices $\widetilde{\mathbf{G}}_{(k)}$\ and $%
\widetilde{\mathbf{V}}_{(k)}$\ and\ thereby does not permit us to study the
dependence of LMOs upon the structure of the given system. In this
connection, some particular cases deserve distinguishing that allow the
above-specified relations to be obtained.

First, let us assume the zero order submatrices $\mathbf{E}_{(+)}$ and $%
\mathbf{E}_{(-)}$\ of our Hamiltonian matrix $\mathbf{H}$\ \ to take
diagonal forms, i.e. 
\begin{equation}
E_{(+)ij}=\varepsilon _{(+)i}\delta _{ij},\quad \ E_{(-)lm}=\varepsilon
_{(-)l}\delta _{lm},  \tag{5.12}
\end{equation}%
where $\varepsilon _{(+)i}$ and $\varepsilon _{(-)l}$\ coincide with
one-electron energies of separate basis orbitals $\varphi _{(+)i}$ and $%
\varphi _{(-)l}$. Relations of Eq.(5.12) imply all interorbital interactions
to be first order terms vs. energy gaps between occupied and vacant basis
orbitals.

Let us now define the so-called fragmentary molecules as those consisting of
certain weakly-interacting elementary fragments, e.g. chemical bonds, phenyl
rings, etc. In other words, intrafragmental resonance parameters will be
assumed here to exceed the interfragmental ones considerably. Accordingly,
let our basis functions $\{\Phi \}$ to coincide with eigenfunctions of
intrafragmental blocks of the relevant Hamiltonian matrix. These orbitals
will be then localized on separate elementary fragments and referred to as
fragmental orbitals (FOs). It is evident that the above-defined fragmentary
systems meet the requirement of Eq.(5.12). It deserves adding here that this
definition embraces numerous classes of chemical compounds in the chemical
sense.

Before passing to specific properties of LMOs of fragmentary molecules, let
us note that matrix equations of Eq.(5.7) may be solved algebraically [53]
under the assumption of Eq.(5.12). For the first and second order elements,
we obtain \ 
\begin{equation}
G_{(1)il}^{(f)}=-\frac{R_{il}}{\varepsilon _{(+)i}+\varepsilon _{(-)l}}, 
\tag{5.13}
\end{equation}%
\begin{equation}
G_{(2)il}^{(f)}=\frac{1}{\varepsilon _{(+)i}+\varepsilon _{(-)l}}\left\{ %
\mathop{\displaystyle \sum }\limits_{(+)j}\frac{T_{ij}R_{jl}}{\varepsilon
_{(+)j}+\varepsilon _{(-)l}}-\mathop{\displaystyle \sum }\limits_{(-)r}\frac{%
R_{ir}Q_{rl}}{\varepsilon _{(+)i}+\varepsilon _{(-)r}}\right\} ,  \tag{5.14}
\end{equation}%
whilst elements of higher orders take somewhat more cumbersome forms (see
e.g.[58]). Sums over $(+)j$\ and $(-)r$ correspondingly embrace here all
occupied and all vacant FOs of the system. The superscript $(f)$ indicates
the relevant formulae to refer just to fragmentary systems. Elements of
Eqs.(5.13) and (5.14) have been interpreted as the direct (through-space)
interaction between orbitals $\varphi _{(+)i}$ and $\varphi _{(-)l}$ and as
the relevant indirect interaction by means of a single mediator,
respectively. Both occupied ($\varphi _{(+)j}$) and vacant FOs ($\varphi
_{(-)r}$) are able to play the role of mediators in the second interaction
as Eq.(5.14) shows. To be an efficient mediator, however, the orbitals
concerned should interact directly with both $\varphi _{(+)i}$ and $\varphi
_{(-)l}.$ Accordingly, elements of higher orders (i.e. $\widetilde{G}%
_{(3)il},\widetilde{G}_{(4)il},etc.$) are interpretable as indirect
interactions of the same orbitals by means of a chain-like set of $k-1$
mediators.

Additivity of indirect interactions with respect to contributions of
individual mediators is among the principal conclusions following from
analysis of expressions for separate elements $\widetilde{G}_{(k)il}^{(f)}$\
[In particular, this property is easily seen from Eq.(5.14)]. Furthermore,
the well-known extinction of resonance parameters when the distance between
orbitals concerned grows [21] allows us to expect an analogous behaviour of
elements $\widetilde{G}_{(k)il}^{(f)}.$ Just this circumstance forms the
basis for locality and transferability of interorbital interactions
[provided that the resonance parameters involved are transferable]. Finally,
elements of matrix products $\mathbf{\mathbf{G}_{(1)}\mathbf{G}_{(1)}^{+}}$\
are interpretable as indirect interaction between occupied basis orbitals
via vacant ones and also may be easily shown to be additive with respect to
increments of the latter. After turning to Eq.(5.11) we may then conclude
that additivity, locality and transferability are the principal distinctive
features of LMOs of fragmentary molecules.

Let us introduce now some more particular cases. First, let us define the
so-called simple fragmentary systems as those consisting of chemical bonds
and lone electron pairs only. It is evident that FOs then coincide with
bonding and antibonding bond orbitals (BOs) [Note that orbitals of lone
electron pairs may be considered as a particular case of BOs [51]). Further,
homogeneous fragmentary molecules may be distinguished, wherein uniform
elementary fragments are assumed to be contained. Given that the latter
coincide with chemical bonds in addition, the term 'simple homogeneous
compounds' seems to be adequate. Let us dwell now just on the latter case.

One electron energies of all bonding BOs (BBOs) ($\varepsilon _{(+)i}$) and
of all antibonding BOs (ABOs) ($\varepsilon _{(-)l}$) take uniform values
for these systems and thereby may be replaced by constants $\varepsilon
_{(+)}$ and $\varepsilon _{(-)},$\ respectively. Further, let the energy
reference point to be chosen in the middle of the energy gap between the
subsets of basis orbitals, whilst the energy unit will coincide with
parameters \ $\varepsilon _{(+)}$ and/or $\varepsilon _{(-)}$. Consequently,
the equality $\mathbf{E}_{(+)}=$ $\mathbf{E}_{(-)}=\mathbf{I}$ follows that
allows entire matrices $\widetilde{\mathbf{G}}_{(k)}^{(f)}$\ of the
above-defined simple homogeneous systems to be expressed algebraically in
terms of entire submatrices of the initial Hamiltonian matrix [49-51] as
follows%
\begin{equation}
\mathbf{G}_{(1)}^{(f)}=-\frac{1}{2}\mathbf{R,}\qquad \mathbf{G}_{(2)}^{(f)}=%
\frac{1}{4}(\mathbf{TR}-\mathbf{RQ}),etc.  \tag{5.15}
\end{equation}%
For the ket-vector of occupied LMOs, we accordingly obtain 
\begin{equation}
\mid \Psi _{1}>=\mid \Phi _{1}>\mathbf{(I-}\frac{1}{8}\mathbf{\mathbf{R}%
R^{+})}+\frac{1}{2}\mid \Phi _{2}>[\mathbf{R}^{+}+\frac{1}{2}(\mathbf{R}^{+}%
\mathbf{T-QR}^{+})]\mathbf{.}  \tag{5.16}
\end{equation}%
On the whole, expressibility of electronic structure characteristics in
terms of entire Hamiltonian matrix blocks seems to be among the most
important distinctive features of simple homogeneous systems.

To illustrate the above-overviewed results, let us consider the class of
alkanes as an example [49,50,56]. The C-C and C-H bonds may be regarded as
uniform in this case because of negligible differences in the relevant
Coulomb and resonance parameters. As a result, alkanes meet the definition
of simple homogeneous systems. As already mentioned, bond orbitals (BBOs and
ABOs) play the role of FOs in this case. Consequently, LMOs of the
bonding-bond- orbital- and-tail constitution prove to be peculiar to alkanes
that possess first order tails embracing ABOs of the remaining bonds. These
tails are determined by elements of the matrix $\frac{1}{2}\mathbf{R}^{+}$\
and thereby by direct interactions between BBOs and ABOs. [Because of the
definition of BOs as eigenfunctions of the relevant $2\times 2-$dimensional
Hamiltonian matrix blocks, the equality $R_{ii}=0$ is valid for any $i$.
That is why the ABO of the same bond does not contribute to the first order
tail of the respective LMO]. The above-discussed extinction of resonance
parameters when the distance between bonds concerned grows [21] then implies
the most significant first order tails to embrace ABOs of the
nearest-neighboring bonds with respect to the given one. Numbers of such
neighbours, in turn, are constant for all C-C and for all C-H bonds in
alkanes and equal to six and three, respectively. Thus, similar structures
are expected for LMOs belonging to all C-C and to all C-H bonds provided
that the relevant elements $R_{ji}$\ are transferable [such an assumption is
supported by estimations [28, 59]]. Finally, LMOs referring to all C-H bonds
are predicted to resemble an equivalent MO of methane. Analysis of the
respective second order tails may be found in Ref.[49].

Therefore, choice of the non-canonical (i.e. block-diagonalization) problem
instead of the standard eigenvalue equation allows general solutions to be
obtained for extended classes of chemical compounds. These solutions, in
turn, yield common expressions for the relevant LMOs. Finally, LMOs of
fragmentary molecules are shown to obey the classical rules of locality,
transferability and additivity (in contrast to the canonical MOs).

\section{The direct way of obtaining the one-electron density matrix}

One-electron density matrix (DM) is among the most fundamental
quantum-mechanical characteristics of molecule describing the respective
charge distribution and thereby related to numerous observed properties.
Moreover, the DM is a unique characteristic [1,2] of the given system in
contrast to NCMOs and/or LMOs.

The most popular way of obtaining the DM of a certain molecule consists in
constructing a projector to the relevant set of occupied one-electron
orbitals [1,2,48], the latter usually coinciding with the canonical MOs
(CMOs) [Although NCMOs (or LMOs) also are able to play this role as
discussed in the next Section]. Molecular orbitals, in turn, usually are
expressed as linear combinations of certain basis functions, e.g. of AOs.
Consequently, elements of the relevant representation matrix of the DM
(commonly referred to as the charge- bond order (CBO) matrix) are generally
determined by sums over all occupied MOs of products of respective
coefficients of the above-specified linear combinations (MO LCAO
coefficients), i.e. by all electrons of the given system. This implies both
the CBO matrix $\mathbf{P}$\ and the resulting charge distribution to belong
to collective properties of molecules (Section 3). That is why applicability
of classical principles of locality, additivity and transferability both to
separate elements of the CBO matrix and to the overall charge distribution
is among natural expectations.

It is evident that relations are required between elements of the CBO
matrix, on the one hand, and those of the initial Hamiltonian matrix, on the
other hand, to prove the above anticipations in practice. These relations,
however, are extremely intricate and difficult to analyze when using the
above-outlined indirect way of derivation of the DM, especially if
delocalized CMOs play the role of one-electron orbitals. This implies an
involved nature of the overall link between charge distributions and the
relevant chemical structures to say nothing about interrelations between CBO
matrices and/or charge distributions of similar chemical compounds. Just
these circumstances stimulated our turning to an alternative (direct) way of
obtaining the same matrix by means of solution of the so-called commutation
equation. It deserves an immediate emphasizing that no passing to the basis
of either CMOs or NCMOs is required when applying this less known
alternative approach. Instead, the following system of matrix equations
should be solved [60] 
\begin{equation}
\lbrack \mathbf{H,P}]_{-}=\mathbf{0,\quad P}^{2}=2\mathbf{P};\quad Spur%
\mathbf{P}=2n,  \tag{6.1}
\end{equation}%
where $\mathbf{H}$\ stands for the initial Hamiltonian matrix and $2n$
coincides with the total number of electrons as previously, and the notation 
$[\mathbf{..,..}]_{-}$ indicates a commutator of matrices. The first
relation of Eq.(6.1) (the commutation condition) is the main physical
requirement determining the matrix $\mathbf{P}$ and resulting from the Dirac
equation for the time-independent Hamiltonian. The remaining relations are
additional system-structure- independent restrictions following from the
idempotence requirement ($\mathbf{\Pi }^{2}\mathbf{=\Pi }$) for the
projector $\mathbf{\Pi }=\frac{1}{2}\mathbf{P}$ and the charge conservation
condition, respectively.

For the specific model Hamiltonian matrix defined by Eq.(5.4), the matrix
problem of Eq.(6.1) may be solved perturbatively [51] analogously to the
block-diagonali- zation problem of Section 5. To this end, any member $%
\mathbf{P}_{(k)}$\ of the power series for the CBO matrix $\mathbf{P}$ also
has been represented in terms of four blocks as shown in Eq.(5.5). Moreover,
comparative analysis of matrix problems of Eqs. (5.1) and (6.1) for the case 
$\mathbf{S=I}$ (Section 5) showed them to yield similar intermediate
equations [51]. On this basis, a deep interrelation has been concluded
between these problems (i.e. between the commutation equation for the DM and
the Brillouin theorem) for the matrix $\mathbf{H}$\ of Eq.(5.4)\ and thereby
between the relevant solutions (i.e. the CBO matrix $\mathbf{P}$ and the LMO
representation matrix $\mathbf{C}$). Indeed, the two matrices prove to be
largely similar and representable in terms of the same entire blocks of the
initial Hamiltonian matrix without specifying either the structures or
dimensions of the latter. In particular, the off-diagonal blocks of the
correction $\mathbf{P}_{(k)}$\ take the form 
\begin{equation}
\mathbf{P}_{12}^{(k)}=-2\mathbf{G}_{(k)},\qquad \mathbf{P}_{21}^{(k)}=-2%
\mathbf{G}_{(k)}^{+},  \tag{6.2}
\end{equation}%
where $\mathbf{G}_{(k)}$\ are conditioned by matrix equations like that of
Eq.(5.7). Elements of these blocks represent bond orders between FOs of
opposite initial occupation. It is evident that the relations of Eq.(6.2)
closely resemble those shown in Eq.(5.6). Expressions for separate terms of
power series for submatrices $\mathbf{P}_{11}$ and $\mathbf{P}_{22}$
determining both the occupation numbers of FOs and bond orders between FOs
of the same initial occupation and taking the diagonal positions within the
CBO matrix $\mathbf{P}$, in turn, are of the following form%
\begin{align}
\mathbf{P}_{11}^{(0)} =&2\mathbf{I,\quad P}_{22}^{(0)}=\mathbf{P}%
_{11}^{(1)}=\mathbf{P}_{22}^{(1)}=\mathbf{0,}  \nonumber \\
\mathbf{P}_{11}^{(2)} =&-2\mathbf{G}_{(1)}\mathbf{G}_{(1)}^{+},\quad 
\mathbf{P}_{22}^{(2)}=2\mathbf{G}_{(1)}^{+}\mathbf{G}_{(1)},  \nonumber \\
\mathbf{P}_{11}^{(3)} =&-2(\mathbf{G}_{(1)}\mathbf{G}_{(2)}^{+}+\mathbf{G}%
_{(2)}\mathbf{G}_{(1)}^{+}),\quad etc.  \tag{6.3}
\end{align}%
that closely resemble the respective formulae of Eq.(5.9). The actual extent
of similarity between corrections $\mathbf{P}_{(k)}$\ and $\mathbf{C}_{(k)}$%
, however, depends on that between matrices \ $\widetilde{\mathbf{V}}_{(k)}$
and $\mathbf{V}_{(k)}$\ (see Eq.(5.7)) and thereby on the value of the order
parameter $k$:

The starting members of the series of matrices $\widetilde{\mathbf{V}}_{(k)}$
and $\mathbf{V}_{(k)}$ are uniform, i.e. $\widetilde{\mathbf{V}}_{(1)}$ $=%
\mathbf{V}_{(1)}$ and $\widetilde{\mathbf{V}}_{(2)}$ $=\mathbf{V}_{(2)}.$
This implies coincidences between respective principal matrices $\widetilde{%
\mathbf{G}}_{(k)}$ and $\mathbf{G}_{(k)}$ referring to $k=1$ and $k=2$.
Equalities $\widetilde{\mathbf{G}}_{(1)}$ $=\mathbf{G}_{(1)}$ and $%
\widetilde{\mathbf{G}}_{(2)}$ $=\mathbf{G}_{(2)}$ along with the common form
of matrix relations of Eqs.(5.9) and (6.3) then indicate a large extent of
similarity between corrections $\mathbf{P}_{(1)}$\ and $\mathbf{C}_{(1)}$,
as well as between $\mathbf{P}_{(2)}$\ and $\mathbf{C}_{(2)}.$ Moreover,
these corrections prove to be actually expressible via the same submatrices $%
\mathbf{G}_{(1)},$ $\mathbf{G}_{(2)},$ $\mathbf{G}_{(1)}\mathbf{G}_{(1)}^{+}$
and $\mathbf{G}_{(1)}^{+}\mathbf{G}_{(1)}$ with distinct numerical
coefficients [51]. So far as corrections of the same series of higher orders
($k=3,4...$) are concerned, the relations between $\mathbf{P}_{(k)}$\ and $%
\mathbf{C}_{(k)}$ become of a somewhat more involved nature [54].
Nevertheless, expressibility of corrections $\mathbf{P}_{(k)}$\ and $\mathbf{%
C}_{(k)}$ via the same principal matrices $\mathbf{G}_{(k)}$\ \ is still
preserved. Let us consider this point in more detail.

Let us note first that distinct expressions are obtained for matrices $%
\widetilde{\mathbf{V}}_{(k)}$ and $\mathbf{V}_{(k)}$ of Eq.(5.7) for $%
k=3,4...$ As for instance, the third order matrix $\mathbf{V}_{(3)}$ takes
the form%
\begin{equation}
\mathbf{V}_{(3)}=\mathbf{TG}_{(2)}-\mathbf{G}_{(2)}\mathbf{Q-(RG}_{(1)}^{+}%
\mathbf{G}_{(1)}+\mathbf{G}_{(1)}\mathbf{G}_{(1)}^{+}\mathbf{R)}  \tag{6.4}
\end{equation}%
that differs from that of $\widetilde{\mathbf{V}}_{(3)}$ as comparison of
Eqs.(5.8) and (6.4) shows. As a result, different matrices $\widetilde{%
\mathbf{G}}_{(k)}$ and $\mathbf{G}_{(k)}$\ follow for $k=3,4...$ These
matrices, however, are actually conditioned by similar matrix equations
shown in Eq.(5.7). It is no surprise in this connection that certain matrix
relations connecting $\widetilde{\mathbf{G}}_{(k)}$ and $\mathbf{G}_{(k)}$\
may be established [54], e.g.%
\begin{equation}
\mathbf{G}_{(3)}=\widetilde{\mathbf{G}}_{(3)}-\frac{1}{2}\mathbf{G}_{(1)}%
\mathbf{G}_{(1)}^{+}\mathbf{G}_{(1)}.  \tag{6.5}
\end{equation}%
[Relations of this type follow from coincidence between expressions for the
CBO matrix derived directly as described above and indirectly by
constructing a projector to the set of occupied LMOs of Section 5]. It is
evident that Eq.(6.5) allows us to eliminate the matrix $\widetilde{\mathbf{G%
}}_{(3)}$\ from the expression for the correction $\mathbf{C}_{(3)}$\ and
thereby to represent the latter in terms of matrices $\mathbf{G}_{(1)},$ $%
\mathbf{G}_{(2)}$ and $\mathbf{G}_{(3)}$ and their products as it is the
case with the correction $\mathbf{P}_{(3)}$.

In summary, the CBO matrix $\mathbf{P}$\ and the LMO representation matrix $%
\mathbf{C}$ are expressible via the same submatrices and thereby depend on
the constitution of the given system in the same manner. Thus, the matrix $%
\mathbf{P}$\ also may be concluded to belong to the localized way of
representing electronic structures along with LMOs. Accordingly, the rules
of locality, additivity and transferability established in the previous
Section may be easily extended to elements of the CBO matrix.

To demonstrate this, let us dwell on diagonal elements of the CBO matrix $%
\mathbf{P}$\ (i.e. $P_{11,ii}$ and $P_{22,ll}$) following from expressions
like that of Eq.(6.3) and representing occupation numbers (populations) of
basis orbitals ($\varphi _{1,i}$ and $\varphi _{2,l},$ respectively).
Notations $X_{(+)i}$ and $X_{(-)l}$ will correspondingly stand below just
for these populations. Members of the respective power series will be
accordingly denoted by $X_{(+)i}^{(k)}$ and $X_{(-)l}^{(k)}$. Additivity of
matrix products ( $\mathbf{G}_{(1)}\mathbf{G}_{(1)}^{+},\mathbf{G}_{(1)}%
\mathbf{G}_{(2)}^{+},etc.$) with respect to mediators (Section 5) implies
expressibility of occupation numbers $X_{(+)i}$ and $X_{(-)l}$ in the
following forms [61] 
\begin{equation}
X_{(+)i}=2-\mathop{\displaystyle \sum }\limits_{(-)m}x_{(+)i,(-)m},\quad
X_{(-)l}=\mathop{\displaystyle \sum }\limits_{(+)j}x_{(-)l,(+)j}.  \tag{6.6}
\end{equation}%
Sums over $(+)j$ and over $(-)m$ embrace here the occupied and vacant basis
orbitals, respectively, and $x_{(+)i,(-)m}$ stand for partial populations
transferred between orbitals of opposite initial occupation that meet the
charge conservation condition , i.e.%
\begin{equation}
x_{(+)i,(-)m}=x_{(-)m,(+)i}.  \tag{6.7}
\end{equation}%
From Eq.(6.3), expressions for members of the power series for $x_{(+)i,(-)m}
$\ easily follows, viz. 
\begin{align}
x_{(+)i,(-)m}^{(2)} =&2(G_{(1)im})^{2},\quad
x_{(+)i,(-)m}^{(3)}=4G_{(1)im}G_{(2)im}  \nonumber \\
x_{(+)i,(-)m}^{(4)} =&4G_{(1)im}G_{(3)im}+2G_{(1)im}(\mathbf{G}_{(1)}%
\mathbf{G}_{(1)}^{+},\mathbf{G}_{(1)})_{im}+2(G_{(2)im})^{2},etc. 
\tag{6.8}
\end{align}%
Thus, additivity of actual occupation numbers of initially-occupied basis
functions with respect to contributions of various electron-accepting
orbitals proves to be supported.

Let us dwell now on particular cases defined in Section 5. \ For fragmentary
compounds, elements $G_{(k)im}$\ represent various types of interactions
between occupied and vacant FOs as it was the case with $\widetilde{G}%
_{(3)im}$\ (Section 5). Accordingly, $x_{(+)i,(-)m}^{(2)}$\ of Eq.(6.8)
depends on the absolute value of the direct interaction between FOs $\varphi
_{1,i}$ and $\varphi _{2,m}$. As a result, extinction of the second order
partial transferred population $x_{(+)i,(-)m}^{(2)}$\ follows
straightforwardly from the analogous properties of elements $G_{(1)im}$ when
the distance between the above-specified orbitals grows. Further, the
relevant third order increment $x_{(+)i,(-)m}^{(3)}$\ does not vanish if the
orbitals $\varphi _{1,i}$ and $\varphi _{2,m}$\ interact one with another
both directly and indirectly by means of a single mediator (say $\varphi
_{s} $). This condition evidently cannot be met for remote FOs $\varphi
_{1,i}$ and $\varphi _{2,m}$. An analogous conclusion may be drawn also for $%
k=4$ and thereby the local nature of occupation numbers in general becomes
supported. Transferability of these numbers, in turn, follows from the
above-mentioned locality provided that the relevant elements of the initial
Hamiltonian matrix are transferable.

In the case of simple homogeneous systems consisting of uniform bonds
(Section 5), expressions for matrices $\mathbf{G}_{(k)}$\ in terms of entire
blocks of the initial Hamiltonian matrix (see Eq.(5.15)) allow analogous
formulae to be obtained for members \ ($\mathbf{P}_{(k)}$) of the power
series for the CBO matrix $\mathbf{P}$. As a result, the first and second
order terms of the series concerned are as follows [49,50]%
\begin{equation}
\mathbf{P}_{(1)}=\left\vert 
\begin{array}{cc}
\mathbf{0} & \mathbf{R} \\ 
\mathbf{R}^{+} & \mathbf{0}%
\end{array}%
\right\vert ,\quad \mathbf{P}_{(2)}=\frac{1}{2}\left\vert 
\begin{array}{cc}
-\mathbf{RR}^{+} & \mathbf{RQ-TR} \\ 
\mathbf{QR}^{+}-\mathbf{R}^{+}\mathbf{T} & \mathbf{R}^{+}\mathbf{R}%
\end{array}%
\right\vert .  \tag{6.9}
\end{equation}%
(Note that the zero order member $\mathbf{P}_{(0)}$\ follows
straightforwardly from Eq.(6.1) and contains matrices $2\mathbf{I}$\ and $%
\mathbf{0}$ in its diagonal positions). It is seen that matrices $\mathbf{%
R,RR}^{+},\mathbf{R}^{+}\mathbf{R}$ and $\mathbf{RQ-TR}$ play the role of
building blocks here as it was the case with LMOs (see Eq.(5.16)). This
implies the LMOs of simple homogeneous systems and the respective CBO matrix
to be extremely similar localized representations of electronic structures.

Finally, let us dwell on the case of alkanes for illustration and recall
that BBOs and ABOs of separate chemical bonds play the role of basis
functions in this case. From expressibility of both the LMO representation
matrix $\mathbf{C}$\ and the CBO matrix $\mathbf{P}$\ via the same
submatrices then follows immediately that bond orders between BOs of alkanes
are proportional to the respective tails of LMOs [49]. As a result, these
two characteristics prove to be equivalent measures of the electron
delocalization. Another straightforward conclusion here is that most
significant bond orders embrace the neighboring (geminal) pairs of bonds.
These bond orders are determined by the direct (through-space) interactions
between BOs of these bonds of the opposite initial occupation. Accordingly,
zero bond orders easily follow for BOs of the same bond.

On the whole, the direct way of obtaining the DM yields general expressions
for CBO matrices of extended classes of chemical compounds as it was the
case with the Brillouin theorem for LMOs (Section 5). Moreover, both the
common CBO matrix of fragmentary molecules $\mathbf{P}$\ and the relevant
LMO representation matrix $\mathbf{C}$\ largely resemble one another,
especially in respect of their dependences upon particular Hamiltonian
matrix blocks and their products. Additivity, transferability and locality
of the CBO matrix elements then are among natural consequences of the
above-mentioned similarity. Since CBO matrix elements are known to determine
local charge distributions, a relation may be concluded between the latter
and peculiarities of the relevant local structures.

\section{Alternative expressions for occupation numbers of basis orbitals
in terms of delocalization coefficients of LMOs}

The fundamental concept of electron pairs pertinent to individual chemical
bonds suggested by Lewis [62] may be regarded as taking an intermediate
place in between the classical and modern theoretical chemistry. This
concept forms the basis for popular interpretations of various structural
changes in organic compounds (including chemical reactions) in terms of
shifts of separate pairs of electrons (cf. the so-called 'curly arrow
chemistry' [15]). It is also essential in our context that these
interpretations are in line with the classical principle of locality: Only a
few pairs are usually assumed to undergo essential shifts, namely those
referring to the center of the given effect or process, whilst the remaining
pairs are considerd as inactive.

As discussed already, the occupation number (population) of a basis function
is defined in quantum chemistry as a sum of increments of all occupied
one-electron states (cf. the collective nature of charge distribution). This
definition evidently yields no relation between an alteration in the
occupation number of a certain basis orbital and reshaping of a single pair
of electrons. Again, just the above-expected relations are required to
reflect the above-outlined Lewis perspective on the CBO matrix. In this
connection, we will look for possibilities of getting rid of sums over all
occupied one-electron orbitals (MOs) in the expressions for occupation
numbers of basis orbitals in the present Section.

The above-mentioned standard definition of an occupation number follows from
the projector to the relevant set of occupied one-electron orbitals. In the
present section, we will assume the localized MOs of Section 5 to play the
role of orbitals underlying the projector instead of CMOs as usual. \ Good
prospects of this alternative originate from the feasibility of extending
the above-discussed one-to-one correspondence between basis orbitals and
LMOs to embrace alterations in the relevant characteristics due to
interaction. Before implementing this scheme, however, some additional
definitions are required.

Let us start with expressions for an individual occupied LMO $\psi _{(+)i}$
\ and a vacant one $\psi _{(-)m}$\ following from formulae like that of
Eq.(5.10), viz. 
\begin{align}
\psi _{(+)i} =&\mathop{\displaystyle \sum }\limits_{(+)j}\varphi
_{(+)j}C_{11,ji}+\mathop{\displaystyle \sum }\limits_{(-)l}\varphi
_{(-)l}C_{21,li},  \nonumber \\
\psi _{(-)m} =&\mathop{\displaystyle \sum }\limits_{(+)j}\varphi
_{(+)j}C_{12,jm}+\mathop{\displaystyle \sum }\limits_{(-)l}\varphi
_{(-)l}C_{22,lm},  \tag{7.1}
\end{align}%
where sums over over $(+)j$ and $(-)l$ here and below embrace the occupied
and vacant basis functions, respectively. It is seen that elements of
submatrices $\mathbf{C}_{21}$ and $\mathbf{C}_{12}$\ reflect tails of LMOs
of the intersubset type (i.e. the intersubset delocalization), whereas those
of the remaining blocks ($\mathbf{C}_{11}$ and $\mathbf{C}_{22}$) represent
tails of the intrasubset nature (intrasubset delocalization). Moreover,
diagonal elements of submatrices $\mathbf{C}_{11}$ and $\mathbf{C}_{22}$\
describe renormalization of basis orbitals $\varphi _{(+)i}$ \ and $\varphi
_{(-)m},$\ respectively, when building up the LMOs. In this connection,
matrices $\mathbf{C}_{11}$ and $\mathbf{C}_{22}$\ will be called the
renormalization matrices for convenience.

Let us now define the so-called partial intersubset delocalization
coefficients of LMOs $\psi _{(+)i}$\ and $\psi _{(-)m}$\ over particular
basis functions, e.g. over $\varphi _{(-)l}$ and $\varphi _{(+)j},$
respectively [50,51,54]. Let these coefficients to be correspondingly
denoted by $d_{(+)i,(-)l}$ and $d_{(-)m,(+)j}.$\ The defitions concerned are
as follows%
\begin{align}
d_{(+)i,(-)l} =&\mid C_{21,li}\mid ^{2}=\mid G_{(1)li}^{+}+G_{(2)li}^{+}+%
\widetilde{G}_{(3)li}^{+}+\widetilde{G}_{(4)li}^{+}+...\mid ^{2},  \nonumber
\\
d_{(-)m,(+)j} =&\mid C_{12,jm}\mid ^{2}=\mid G_{(1)jm}+G_{(2)jm}+\widetilde{%
G}_{(3)jm}+\widetilde{G}_{(4)jm}+...\mid ^{2}.  \tag{7.2}
\end{align}%
It is easily seen that the partial delocalization coefficients concerned
also are expressible in the form of power series, i.e. as sums over
parameter $k$ of contributions $d_{(+)i,(-)l}^{(k)}$ and $d_{(-)m,(+)j}^{(k)}
$. Further, let total delocalization coefficients of the same LMOs over all
vacant and over all occupied basis functions to be defined as follows%
\begin{equation}
D_{(+)i}=\mathop{\displaystyle \sum }\limits_{(-)l}d_{(+)i,(-)l},\quad
D_{(-)m}=\mathop{\displaystyle \sum }\limits_{(+)j}d_{(-)m,(+)j}.  \tag{7.3}
\end{equation}%
From Eqs.(7.2) and (7.3) it follows that $D_{(+)i}$\ and $D_{(-)m}$\
actually coincide with diagonal elements of the following matrices 
\begin{equation}
\mathbf{D}_{(+)}=\mathbf{C}_{21}^{+}\mathbf{C}_{21},\quad \mathbf{D}_{(-)}=%
\mathbf{C}_{12}^{+}\mathbf{C}_{12}  \tag{7.4}
\end{equation}%
that may be referred to as the intersubset delocalization matrices.
Substituting the power series for submatrices $\mathbf{C}_{21}$ and $\mathbf{%
C}_{12}$ (Section 5) into Eq.(7.4) yields analogous series for matrices $%
\mathbf{D}_{(+)}$ and $\mathbf{D}_{(-)},$ the starting members of which are
proportional to renormalization matrices of the same order, e.g.%
\begin{equation}
\mathbf{D}_{(+)}^{(2)}=-2\mathbf{C}_{11}^{(2)},\quad \mathbf{D}%
_{(+)}^{(3)}=-2\mathbf{C}_{11}^{(3)}.  \tag{7.5}
\end{equation}%
Meanwhile, expressions for next terms of the series are of a somewhat more
involved nature [54], e.g. 
\begin{equation}
\mathbf{D}_{(+)}^{(4)}=-2\mathbf{C}_{11}^{(4)}-(\mathbf{C}_{11}^{(2)})^{2}. 
\tag{7.6}
\end{equation}

Let us turn now to the definition of the one-electron DM $P(\mathbf{r}\mid 
\mathbf{r}^{\prime })$\ as a projector [1,2] to the subset of occupied LMOs $%
\{\Psi _{1}\},$\ viz. 
\begin{equation}
P(\mathbf{r}\mid \mathbf{r}^{\prime })=2\left\vert \Psi _{1}(\mathbf{r)}%
\right\rangle \left\langle \Psi _{1}(\mathbf{r}^{\prime })\right\vert , 
\tag{7.7}
\end{equation}%
where $\mathbf{r}$\ represents the position of an electron in the real
space, and $\left\vert \Psi _{1}(\mathbf{r)}\right\rangle $\ and $%
\left\langle \Psi _{1}(\mathbf{r}^{\prime })\right\vert $\ correspondingly
stand for the ket-vector of occupied LMOs (see Eq. (5.10)) and the relevant
bra-vector. After substituting expressions like that of Eq.(5.10) into
Eq.(7.7), we obtain%
\begin{equation}
P(\mathbf{r}\mid \mathbf{r}^{\prime })=\mathop{\displaystyle \sum }\limits%
_{I,J=1}^{2}\left\vert \Phi _{I}(\mathbf{r)}\right\rangle \mathbf{P}%
_{IJ}\left\langle \Phi _{J}(\mathbf{r}^{\prime })\right\vert ,  \tag{7.8}
\end{equation}%
where $\mathbf{P}_{I,J},I=1,2;J=1,2$\ are multidimensional elements (blocks)
of the representation of the DM in terms of two subsets of basis functions \
\ $\{\Phi _{1}\}$ and $\{\Phi _{2}\}$ introduced in Section 5. The
expression of Eq.(7.8) serves as a matrix generalization of the well-known
bilinear form of the DM in terms of individual basis functions [1]. For
blocks $\mathbf{P}_{IJ},$\ we accordingly obtain%
\begin{equation}
\mathbf{P}_{11}=2\mathbf{C}_{11}\mathbf{C}_{11}^{+};\quad \mathbf{P}_{22}=2%
\mathbf{C}_{21}\mathbf{C}_{21}^{+};\quad \mathbf{P}_{12}=2\mathbf{C}_{11}%
\mathbf{C}_{21}^{+}.  \tag{7.9}
\end{equation}%
Let the submatrices $\mathbf{P}_{11}$\ and $\mathbf{P}_{22}$\ to be
alternatively denoted by $\mathbf{X}_{(+)}$\ and $\mathbf{X}_{(-)}$\ (in
accordance with designations $X_{(+)i}$\ and $X_{(-)m}$\ of Section 6
standing for the relevant diagonal elements) and called the intrasubset
population matrices. Substituting the power series for $\mathbf{C}_{11}$ and 
$\mathbf{C}_{21}$ (Section 5) into the expression of Eq.(7.9) followed by
employment of interdependences like that shown in Eqs.(7.5) and (7.6) yields
the following principal results 
\begin{equation}
\mathbf{X}_{(+)}=2(\mathbf{I}-\mathbf{D}_{(+)}),\qquad \mathbf{X}_{(-)}=2%
\mathbf{D}_{(-)}.  \tag{7.10}
\end{equation}%
Thus, intrasubset population matrices prove to be proportional to respective
intersubset delocalization matrices [54]. After passing to diagonal elements
within Eq.(7.10), we accordingly obtain 
\begin{equation}
X_{(+)i}=2(I-D_{(+)i}),\qquad X_{(-)m}=2D_{(-)m}.  \tag{7.11}
\end{equation}%
Hence, the actual population of the basis function $\varphi _{(+)i}$\ (or $%
\varphi _{(-)m}$) is determined only by the shape of the respective 'own'
LMO $\psi _{(+)i}(\psi _{(-)m}).$ Moreover, the population of the basis
function $\varphi _{(+)i}$\ lost due to the interorbital interaction is
proportional to the total intersubset delocalization coefficient of the
above-specified exclusive LMO. In other words, the one-to-one correspondence
between basis orbitals and LMOs (seen from the equality $\mathbf{C}_{(0)}=%
\mathbf{I}$) is now replenished by the following rule: The more delocalized
the orbital $\varphi _{(+)i}$\ becomes when building up the respective LMO $%
\psi _{(+)i},$ the more charge it loses and vice versa. This result may be
also interpreted as a kind of simultaneous separability of both charge
redistribution and delocalization into increments of separate pairs of
electrons. Further, expressing both sides of Eq.(7.11) in the form of power
series yields analogous relations between $X_{(+)i}^{(k)}$ and $%
D_{(+)i}^{(k)}$, and between $X_{(-)m}^{(k)}$ and $D_{(-)m}^{(k)}.$\
Finally, invoking expressions like those of Eqs.(6.6) and (7.3) yields
proportionalities between partial increments, i.e. 
\begin{equation}
x_{(+)i,(-)l}^{(k)}=2d_{(+)i,(-)l}^{(k)},\quad
x_{(-)m,(+)j}^{(k)}=2d_{(-)m,(+)j}^{(k)}.  \tag{7.12}
\end{equation}%
Thus, the more population is transferred between orbitals $\varphi _{(+)i}$\
and $\varphi _{(-)l},$\ the larger is the partial delocalization coefficient
of the LMO \ \ $\psi _{(+)i}$ over the same vacant basis function and vice
versa. If we recall the definition of this coefficient as square of the
relevant tail of the LMO $\psi _{(+)i}$ (see Eq.(7.2)), we may also conclude
a more significant tail of this LMO over the vacant orbital $\varphi _{(-)l}$%
\ to correspond to a more efficient charge transfer between basis functions $%
\varphi _{(+)i}$\ and $\varphi _{(-)l}$\ and vice versa. Irrelevance of the
intrasubset delocalization in the formation of charge redistributions also
is among the conclusions. This result evidently causes no surprise.
Immediate reasons why the respective terms vanish in the expressions for
occupation numbers are clarified in Ref.[54]. It deserves mentioning finally
that relations of Eqs.(7.10) and (7.11) indicate both populations of basis
orbitals and delocalization coefficients of LMOs to be characterized by
common properties. It is evident that locality, additivity and
transferability (established in Sections 5 and 6) are expected to be among
these properties.

On the whole, the results of this Section may be summarized as parallelism
between charge (re)distribution and delocalization. Moreover,\ we obtain a
certain quantum-chemical analogue of the Lewis perspective on charge
(re)distribution and thereby of the 'curly arrow chemistry'. Two differences
between this analogue and its classical version deserve mentioning: First,
LMOs are not localized completely in contrast to the relevant classical
model. Second, the approach suggested allows comparisons of relative extents
of shifts (reshapings) of separate pairs of electrons for related compounds
and/or for alternative routes of the same process. Just the second point may
be regarded as an important advantage of the new approach over the classical
one. For illustration, we will confine ourselves to comparisons of related
compounds in this Section (alternative routes of reactions are discussed in
Section 9). To this end, let us turn to studies of alkanes and their
heteroatom-containing derivatives [63] (cf. the so-called inductive effect).
Rules governing the additional interbond charge transfer due to introduction
of a heteroatom (Z) along with the related extra delocalization of LMOs and
thereby of separate pairs of electrons have been formulated there. As a
result of additivity of both characteristics with respect to contributions
of individual bonds, an increase was observed both in the total population
of the Z-C$_{\alpha }$ bond and in the extent of delocalization of the
respective single pair of electrons when the size of the hydrocarbon
fragment grows (e.g. when passing from substituted methanes to ethanes).
Furthermore, the changing nature of the nearest-neighboring bond when
passing from 1-mono- to 1,1$^{\prime }$-disubstituted compounds was shown to
give rise to suppression of delocalization of LMOs associated with the C$%
_{\alpha }$-Z$_{1}$ bond and thereby to reductions in the relevant
occupation numbers. These conclusions proved to be in aggreement with the
relevant experimental trends. Finally, both electron density redistributions
and reshapings of LMOs due to introduction of the heteroatom were shown to
be proportional to the extents of delocalization of LMOs in the parent
hydrocarbon. Accordingly, the well-known short-range nature of the inductive
effect has been accounted for by a weak interbond delocalization in alkanes.
This interpretation evidently is in line with the classical perspective.

Finally, the principal result of this Section may be alternatively
formulated as a kind of one-orbital representation of populations lost
(acquired) by individual basis orbitals. This state of things closely
resembles the one-orbital representations of ionization potentials in the
canonical MO method known as the Koopmans' theorem [26]. If we recall that
representation of this type are not achievable for ionization potentials and
for charge (re)distributions in the NCMO and the CMO methods, respectively,
the present results support the complementary nature of the above-mentioned
principal approaches of quantum chemistry [34].

\section{Alternative definitions of total energies and their implications}

Total energies of molecules also rank among the most popular quantum-
chemical characteristics. Moreover, these are comparable to experimental
data even more directly, i.e. to heats of formation and/or atomization. The
latter, in turn, are known to be expressible as sums of transferable bond
increments [8], at least for saturated molecules, e.g. alkanes and their
derivatives. Thus, total energies belong to properties the classical
principles are expected to be applicable to.

In the framework of the H\"{u}ckel model, the standard definition of the
total energy of any molecular system ($\mathcal{E}$) contains a sum of
one-electron energies referring to occupied MOs multiplied by their
occupation number 2. Inasmuch as one-electron energies of delocalized
(canonical) MOs usually depend on the structure of the system in an
intricate manner, the above-mentioned standard definition is not the most
suitable one for revealing the analogues of the classical rules within total
energies. Hence, alternative definitions of the same characteristic should
be invoked.

In the non-canonical MO method based on the Brillouin theorem (see
Eq.(5.1)), the relevant total energy may be defined as follows%
\begin{equation}
\mathcal{E}=2Spur\mathbf{E}_{1}^{(n\times n)}=2Spur\{\mathbf{C}^{+}\mathbf{HC%
}\}_{11},  \tag{8.1}
\end{equation}%
where $\mathbf{E}_{1}^{(n\times n)}$\ is the eigenblock of the Hamiltonian
matrix referring to the subset of occupied LMOs and the symbol $\{...\}_{11}$%
\ \ stands here for the submatrix of the matrix product $\mathbf{C}^{+}%
\mathbf{HC}$\ taking the first diagonal position. Thus, the total energy $%
\mathcal{E}$ \ may be obtained without passing to the basis of CMOs as it
was the case with the transformation matrix $\mathbf{C}$\ (see Sections 5
and 13). In analogy with obtaining the DM on the basis of LMOs (Section 7),
employment of Eq.(8.1) may be referred to as the indirect way of derivation
of the total energy. Meanwhile, the direct way consists in the use of the
relation between the total energy and the one-electron DM (CBO matrix) $%
\mathbf{P}$ [60], viz.%
\begin{equation}
\mathcal{E}=Spur(\mathbf{PH}).  \tag{8.2}
\end{equation}

Let us dwell now on the case of a nearly block- diagonal Hamiltonian matrix $%
\mathbf{H}$ of Eq.(5.4) containing a block-diagonal zero order contribution $%
\mathbf{H}_{(0)}$ and a certain first order term $\mathbf{H}_{(1)}.$\ Both
the CBO matrix $\mathbf{P}$\ and the LMO representation matrix $\mathbf{C}$\
are then expressible in the form of power series (Sections 5 and 6). The
same accordingly refers to total energies defined by either Eq.(8.1) or
Eq.(8.2), both of these formulae evidently yielding the same result. The
first five terms of the series concerned are as follows [61,64,65]%
\begin{align}
\mathcal{E}_{(0)} =&2Spur\mathbf{E}_{(+)},\qquad \mathcal{E}_{(1)}=2Spur%
\mathbf{T,}  \nonumber \\
\mathcal{E}_{(2)} =&-2Spur(\mathbf{G}_{(1)}\mathbf{R}^{+}),\qquad \mathcal{E%
}_{(3)}=-2Spur(\mathbf{G}_{(2)}\mathbf{R}^{+}),  \nonumber \\
\mathcal{E}_{(4)} =&-2Spur[(\mathbf{G}_{(3)}+\mathbf{G}_{(1)}\mathbf{G}%
_{(1)}^{+}\mathbf{G}_{(1)})\mathbf{R}^{+}],  \tag{8.3} \\
\mathcal{E}_{(5)} =&-2Spur[(\mathbf{G}_{(4)}+\mathbf{G}_{(2)}\mathbf{G}%
_{(1)}^{+}\mathbf{G}_{(1)}+\mathbf{G}_{(1)}\mathbf{G}_{(2)}^{+}\mathbf{G}%
_{(1)}+\mathbf{G}_{(1)}\mathbf{G}_{(1)}^{+}\mathbf{G}_{(2)})\mathbf{R}%
^{+}],etc.  \nonumber
\end{align}%
The sum of zero and first order members of Eq.(8.3) coincides with the total
energy of the set of isolated occupied basis functions in accordance with
the expectation, whereas that of the remaining corrections describes
stabilization (or destabilization) of the given system vs. the
above-specified set. In particular, the second order correction $\mathcal{E}%
_{(2)}$\ proved to be a generalization [66] to the case of a block- diagonal
zero order Hamiltonian matrix $\mathbf{H}_{(0)}$\ of the well-known Dewar
formula [23], the latter corresponding to a diagonal form of $\mathbf{H}%
_{(0)}$\ and following from the standard Rayleigh- Schr\"{o}dinger
perturbation theory. Accordingly, the subsequent corrections of Eq.(8.3)
starting with $k=3$ may be regarded as an extension of the usual second
order approximation. Finally, additivity of the total energy with respect to
pairs of basis functions of opposite initial occupation ($\varphi _{(+)i}$\
and $\varphi _{(-)l}$) easily follows from Eq.(8.3), although separate
increments generally are non-local in their nature.

Expressibility of \ the initial Hamiltonian matrix in the form of a sum of
zero and first order members as shown in Eq.(5.4) along with application of
Eq.(8.2) allows us also to define two components within any correction $%
\mathcal{E}_{(k)}$, viz. 
\begin{equation}
\mathcal{E}_{(k)}=\mathcal{E}_{(k)}^{(\alpha )}+\mathcal{E}_{(k)}^{(\beta )},
\tag{8.4}
\end{equation}%
where 
\begin{equation}
\mathcal{E}_{(k)}^{(\alpha )}=Spur\mathbf{(P}_{(k)}\mathbf{H}_{(0)}\mathbf{)}%
,\quad \mathcal{E}_{(k)}^{(\beta )}=Spur\mathbf{(P}_{(k-1)}\mathbf{H}_{(1)}%
\mathbf{)}.  \tag{8.5}
\end{equation}%
(Note that Eq.(8.1) also may be successfully used for the same purpose).
Substituting the formulae for $\mathbf{P}_{(k)}$\ (Section 6) along with a
definite algebraic procedure based on employment of Eq.(5.7) yields the
following general relation [61]%
\begin{equation}
(k-1)\mathcal{E}_{(k)}^{(\beta )}=-k\mathcal{E}_{(k)}^{(\alpha )}  \tag{8.6}
\end{equation}%
for any $k$ that indicates the components $\mathcal{E}_{(k)}^{(\alpha )}$
and $\mathcal{E}_{(k)}^{(\beta )}$ to be of opposite signs. Moreover, the
absolute value of $\mathcal{E}_{(k)}^{(\beta )}$ always exceeds that of $%
\mathcal{E}_{(k)}^{(\alpha )},$ i.e. $\mid \mathcal{E}_{(k)}^{(\beta )}\mid
>\mid \mathcal{E}_{(k)}^{(\alpha )}\mid .$ Consequently, it is the sign\ of $%
\mathcal{E}_{(k)}^{(\beta )}$\ that determines the actual sign of the total
kth order energy $\mathcal{E}_{(k)}.$ Finally, the relation of Eq.(8.6)
implies the total correction $\mathcal{E}_{(k)}$\ to be alternatively
representable as follows%
\begin{equation}
\mathcal{E}_{(k)}=-\frac{1}{k-1}\mathcal{E}_{(k)}^{(\alpha )},\qquad 
\mathcal{E}_{(k)}=\frac{1}{k}\mathcal{E}_{(k)}^{(\beta )}.  \tag{8.7}
\end{equation}%
Let us dwell now on the case of fragmentary molecules (Section 5), the zero
order Hamiltonian matrix of which ($\mathbf{H}_{(0)}$) takes a diagonal form
containing one-electron energies of FOs. Accordingly, the component\ $%
\mathcal{E}_{(k)}^{(\alpha )}$\ may be easily shown to involve populations
of FOs only. After an additional expressing the latter in terms of partial
increments as shown in Eq.(6.6), the component $\mathcal{E}_{(k)}^{(\alpha
)} $\ takes the form [61]

\begin{equation}
\mathcal{E}_{(k)}^{(\alpha )}=-\mathop{\displaystyle \sum }\limits_{(+)i}%
\mathop{\displaystyle \sum }\limits_{(-)l}x_{(+)i,(-)l}^{(k)}(\varepsilon
_{(+)i}+\varepsilon _{(-)l}),  \tag{8.8}
\end{equation}%
and depends upon populations transferred between FOs of opposite initial
occupation. In this connection, this component has been interpreted as the
charge transfer energy. Meanwhile, the remaining component ($\mathcal{E}%
_{(k)}^{(\beta )}$) describes the effect of formation of new bond orders
between FOs upon the kth order energy due to interfragmental interaction
[Note that $\mathcal{E}_{(k)}^{(\beta )}$\ does not contain populations of
basis orbitals because one-electron energies of FOs always may be entirely
included into the zero order matrix $\mathbf{H}_{(0)}$\ and thereby diagonal
elements of the first order member $\mathbf{H}_{(1)}$\ vanish]. It is also
noteworthy that bond orders determining the component $\mathcal{E}%
_{(k)}^{(\beta )}$\ originate from members of the (k-1)th order of the power
series for the DM $\mathbf{P}$. If we recall now that populations of FOs of
the kth order (i.e. $X_{(+)i}^{(k)}$\ and $X_{(-)l}^{(k)}$) are expressed in
terms of products of the principal matrices of lower orders ($\mathbf{G}%
_{(k-1)},\mathbf{G}_{(k-2)},etc.$) that, in turn, represent intersubset bond
orders being formed within previous terms of the same series (see Eq.(6.3)),
the above-established interdependence between components $\mathcal{E}%
_{(k)}^{(\alpha )}$ and $\mathcal{E}_{(k)}^{(\beta )}$ causes no surprise.
Moreover, charge redistribution of the kth order may be then considered as a
consequence (or counter-effect) of formation of intersubset bond orders
within terms of lower orders of the same series for the DM $\mathbf{P}$.
Thus, we have to do here with a certain gradual reorganization of bonding,
the energetic increments of which are interrelated and governed by Eq.(8.6).
Furthermore, the above-drawn conclusions concerning absolute values of $%
\mathcal{E}_{(k)}^{(\alpha )}$ and $\mathcal{E}_{(k)}^{(\beta )}$\ allow us
to expect that stabilization of our fragmentary system vs. the set of
isolated occupied FOs (if any) is entirely due to formation of new bond
orders, and the subsequent charge redistribution actually reduces this
stabilizing effect. Nevertheless, the absolute value of the relevant
stabilization energy is proportional to that of the respective charge
transfer energy $\mathcal{E}_{(k)}^{(\alpha )}$ as the first relation of
Eq.(8.7) shows. Indeed, the minus signs of Eqs.(8.7) and (8.8) cancel out
one another and we obtain [61]%
\begin{equation}
\mathcal{E}_{(k)}=\frac{1}{k-1}\mathop{\displaystyle \sum }\limits_{(+)i}%
\mathop{\displaystyle \sum }\limits_{(-)l}x_{(+)i,(-)l}^{(k)}(\varepsilon
_{(+)i}+\varepsilon _{(-)l}).  \tag{8.9}
\end{equation}%
On this basis, an interpretation of the total kth order energy as the charge
transfer energy also becomes acceptable (in spite of its somewhat
oversimplified nature). This interpretation along with Eq.(8.9) provides us
with a quantum- chemical analogue of the intuition-based relation between
stabilization and charge redistribution. If we recall finally the
above-discussed proportionality between $x_{(+)i,(-)l}^{(k)}$\ and $%
d_{(+)i,(-)l}^{(k)}$\ exhibited in Eq.(7.12), we obtain that [54] 
\begin{equation}
\mathcal{E}_{(k)}=\frac{2}{k-1}\mathop{\displaystyle \sum }\limits_{(+)i}%
\mathop{\displaystyle \sum }\limits_{(-)l}d_{(+)i,(-)l}^{(k)}(\varepsilon
_{(+)i}+\varepsilon _{(-)l}).  \tag{8.10}
\end{equation}%
As a result, the correction $\mathcal{E}_{(k)}$\ is accordingly
interpretable as the energy of the intersubset delocalization of
initially-localized pairs of electrons. Hence, the well-known intuition-
based relation between delocalization and stabilization [10,67] also
acquires a quantum- chemical support.

Let us turn now to the case of simple fragmentary systems consisting of
individual chemical bonds and lone electron pairs (if any) (Section 5). Each
of these two-center fragments evidently is represented by two basis
functions. [Note that a faked ABO may be introduced without affecting the
final results [51] in the case of a lone electron pair]. Let the Ith
fragment to be accordingly described by the occupied orbital (BBO) $\varphi
_{(+)i}$\ and its vacant counterpart (ABO) $\varphi _{(-)l}$. As a result,
the component $\mathcal{E}_{(k)}^{(\alpha )}$ and thereby the total kth
order energy $\mathcal{E}_{(k)}$ is representable as a sum of contributions $%
\mathcal{E}_{(k)I}$ of individual fragments (bonds), where \ 
\begin{equation}
\mathcal{E}_{(k)I}=\frac{1}{k-1}\mathop{\displaystyle \sum }%
\limits_{(-)l}x_{(+)i,(-)l}^{(k)}(\varepsilon _{(+)i}+\varepsilon _{(-)l})=%
\frac{2}{k-1}\mathop{\displaystyle \sum }\limits_{(-)l}d_{(+)i,(-)l}^{(k)}(%
\varepsilon _{(+)i}+\varepsilon _{(-)l}).  \tag{8.11}
\end{equation}%
It is seen that each individual contribution to the energy correction $%
\mathcal{E}_{(k)}$ is still additive with respect to increments of other
fragments (ABOs). Thus, a two-fold additivity of the kth order energy may be
concluded. If we now confine ourselves to simple homogeneous systems and
choose a negative energy unit coinciding with $\varepsilon _{(+)i}$, the
equality $\varepsilon _{(+)i}=\varepsilon _{(-)i}=1$ for any $i$ becomes
acceptable (Section 5). Employment of the latter within Eqs. (8.9) and
(8.10) yields \ 
\begin{equation}
\mathcal{E}_{(k)}=\frac{2}{k-1}\mathop{\displaystyle \sum }\limits_{(+)i}%
\mathop{\displaystyle \sum }\limits_{(-)l}x_{(+)i,(-)l}^{(k)}=\frac{4}{k-1}%
\mathop{\displaystyle \sum }\limits_{(+)i}\mathop{\displaystyle \sum }%
\limits_{(-)l}d_{(+)i,(-)l}^{(k)}.  \tag{8.12}
\end{equation}%
After invoking Eqs.\ (7.3) and (7.4), the above expression takes the form \ 
\begin{equation}
\mathcal{E}_{(k)}=\frac{4}{k-1}D_{(+)}^{(k)},  \tag{8.13}
\end{equation}%
where $D_{(+)}^{(k)}$\ is defined as follows%
\begin{equation}
D_{(+)}^{(k)}=\mathop{\displaystyle \sum }\limits_{(+)i}D_{(+)i}^{(k)}=Spur%
\mathbf{D}_{(+)}^{(k)}  \tag{8.14}
\end{equation}%
and may be referred to as the kth order complete delocalization coefficient
[50] of occupied LMOs. Thus, proportionality between delocalization and
stabilization takes an especially simple form in this case.

As already mentioned (Section 5), the principal matrices $\mathbf{G}%
_{(k)}^{{}}$\ are expressible via entire submatrices of the initial
Hamiltonian matrix in the case of similar homogeneous compounds as shown in
Eq.(5.15). The relation between first order matrices $\mathbf{G}_{(1)}$\ and 
$\mathbf{R}$ is especially noteworthy here. The point is that just this
relation allows us to substitute $-2\mathbf{G}_{(1)}$ for $\mathbf{R}$
within any energy correction of Eq.(8.3) and thereby to express the latter
in terms of matrices $\mathbf{G}_{(k)}$\ only [64,65], e.g.%
\begin{align}
\mathcal{E}_{(2)} =&4Spur(\mathbf{G}_{(1)}\mathbf{G}_{(1)}^{+}),\quad 
\mathcal{E}_{(3)}=4Spur(\mathbf{G}_{(2)}\mathbf{G}_{(1)}^{+}),  \nonumber \\
\mathcal{E}_{(4)} =&4Spur[(\mathbf{G}_{(3)}+\mathbf{G}_{(1)}\mathbf{G}%
_{(1)}^{+}\mathbf{G}_{(1)})\mathbf{G}_{(1)}^{+}],  \tag{8.15} \\
\mathcal{E}_{(5)} =&4Spur[(\mathbf{G}_{(4)}+3\mathbf{G}_{(2)}\mathbf{G}%
_{(1)}^{+}\mathbf{G}_{(1)})\mathbf{G}_{(1)}^{+}],etc.  \nonumber
\end{align}%
Moreover, alternative expressions for corrections $\mathcal{E}_{(4)}$ and $%
\mathcal{E}_{(5)}$\ have been derived that contained the principal matrices
up to $\mathbf{G}_{(2)}$\ and $\mathbf{G}_{(3)},$\ respectively, viz.%
\begin{align}
\mathcal{E}_{(4)} =&4Spur(\mathbf{G}_{(2)}\mathbf{G}_{(2)}^{+})-4Spur(%
\mathbf{G}_{(1)}\mathbf{G}_{(1)}^{+}\mathbf{G}_{(1)}\mathbf{G}_{(1)}^{+}), 
\tag{8.16} \\
\mathcal{E}_{(5)} =&4Spur[(\mathbf{G}_{(3)}-\mathbf{G}_{(1)}\mathbf{G}%
_{(1)}^{+}\mathbf{G}_{(1)})\mathbf{G}_{(2)}^{+}].  \nonumber
\end{align}%
Formulae of the above-exemplified type underly the original approach of
Refs. [64,65] to evaluating and rationalizing relative stabilities of simple
homogeneous compounds of similar constitutions. This approach proved to be
especially successfull in the case of conjugated hydrocarbons modelled as
sets of weakly-interacting initially-double (C=C) bonds, including isomers
of dienes, individual Kekul\'{e} valence structures of benzenoids, etc.
Local nature of stabilization of these hydrocarbons vs. the relevant sets of
isolated C=C bonds and thereby the dependence of the stabilization energy
upon local peculiarities of constitution was among principal conclusions
here [64]. Moreover, a large extent of additivity of this extra energy with
respect to transferable increments of definite substructures (fragments) has
been established. On the whole, these results indicate stabilization
energies of conjugated hydrocarbons to be of quasi-classical nature.
Finally, the analogy between the perturbative approach under discussion and
models based on the concepts of conjugated circuits [68] and paths [69]
deserves mentioning here.

In summary, the results overviewed in Sections 5-8 form the basis of the
perturbative non-canonical theory of molecular orbitals further referred to
as the PNCMO theory [55]. The subsequent Sections 9-12 are devoted to
applications of this theory.\ 

\section{Analysis of heterolytic organic reactions in terms of direct and
indirect participation of separate fragments}

In accordance with the classical principle of locality (Section 1), a
definite functional group is regarded as taking part in the given chemical
process directly and it is usually referred to as the reaction center [70].
Again, the remaining parts of molecules (e.g. substituents) are supposed to
participate in the same process indirectly by exerting certain
electron-donating or accepting effects upon this center. Moreover, the
extents of these effects are usually considered to be quite different at
various stages of the reaction [10,71]. Extinction of the indirect influence
when the distance between the given fragment and the reaction center grows
also is among the expectations. Besides, the above-outlined concepts
originated from studies of heterolytic (i.e. electrophilic and nucleophilic)
organic reactions, the latter being often accordingly referred to as the
classical ones. Specific features of homolytic (pericyclic) reactions are
discussed in Section 10.

Let us note immediately that the standard quantum- chemical approaches to
early stages of organic reactions [11,23, 67,70, 72-74] allow no revealing
of roles of separate fragments. The point is that these approaches are based
on an initial passing into the basis of delocalized (canonical) MOs of
isolated compounds and thereby on the concept of a non-local intermolecular
interaction even if local fragments of extended compounds actually come into
contact.

To formulate quantum-chemical analogues of the above-discussed classical
concepts and to be able to discuss chemical reactions in terms of local
structures and/or interactions, the PNCMO theory of Sections 5-8 has been
applied both to the general case of two interacting fragmentary molecules A
and B [54, 75, 76] and to some specific reactions [54, 58, 77-83]. This
alternative methodology has been jointly called the semilocalized approach
to chemical reactivity. Electron density redistributions among basis
orbitals (FOs) of both participants of the given process along with the
related delocalization coefficients of LMOs [54] are the principal
characteristics here instead of the total intermolecular interaction energy
or constitutions of canonical MOs as usual [although the total energy also
may be derived either on the basis of Eq.(8.2) or by means of Eqs. (8.9) and
(8.10) as discussed in Section 10].

Let us start with the general additivity property of the total DM (CBO
matrix) $\mathbf{P}$ of two interacting molecules A and B [76]. Indeed, this
matrix was shown to be representable as follows%
\begin{equation}
\mathbf{P}=\mathbf{P}^{(A)}\oplus \mathbf{P}^{(B)}+\delta \mathbf{P,} 
\tag{9.1}
\end{equation}%
where the first term coincides with the direct sum of DMs of isolated
molecules A and B, and the second one ($\delta \mathbf{P}$) is a correction
originating from the intermolecular interaction and vanishing if this
interaction turns to zero. An analogous relation may be easily proven for
the LMO representation matrix $\mathbf{C}$\ too. Such an additivity property
of the principal matrices $\mathbf{P}$\ and $\mathbf{C}$\ implies that new
contributions to elements of the latter arising as a result of the
intermolecular contact may be studied separately and independently from the
intramolecular ones. As a result, a general formula has been derived and
analyzed for an alteration ($\delta X_{(+)i}^{(A)}$) in the population of a
certain initially-occupied FO ($\varphi _{(+)i}^{(A)}$) of the molecule A
due to its contact with the molecule B. As with the DM $\mathbf{P}$\ itself
(Section 6), this alteration also takes the form of the power series with
respect to intermolecular resonance parameters. Separate terms of this
series, in turn, are expressible as sums of partial increments ($\delta
x_{(+)i,(-)l}^{(A,k))}$) referring to various initially-vacant FOs (k stands
here for the order parameter as previously). Finally, the relevant
alterations in delocalization coefficients of LMOs (i.e. $\delta
D_{(+)i}^{(A)}$ and $\delta D_{(-)m}^{(A)}$) also take analogous forms
because of interrelations%
\begin{equation}
\delta X_{(+)i}^{(A)}=2\delta D_{(+)i}^{(A)},\qquad \delta
X_{(-)m}^{(A)}=2\delta D_{(-)m}^{(A)}  \tag{9.2}
\end{equation}%
that may be easily proven on the basis of the results of Section 7.

Before turning to an overview of both general and specific results of the
semilocalized approach, let us recall some definitions of Ref.[76]. Thus,
the directly interacting (contacting) fragments of molecules A and B have
been called the reaction centers and denoted by RC(A) and RC(B). (The direct
contact implies that the intermolecular resonance parameters take non-zero
values for pairs of FOs of only these fragments). Further, the fragments of
molecules A and B, the FOs of which interact directly only with those of
reaction centers of their own molecules (but not with orbitals of opposite
molecule) have been referred to as the nearest-neighboring fragments and
denoted by NN(A) and NN(B), respectively. Analogously, the next-nearest-
neighboring fragments NNN(A) and NNN(B) have been defined and so forth.
Given that all direct interorbital interactions vanish in our RC (i.e. $%
G_{(1)il}=0$\ for all $i$ and $l$ belonging to the RC), the term 'a simple
RC' has been used. It is evident that reaction centers containing only a
single fragment (the FOs of which are defined as eigenfunctions of the
respective Hamiltonian matrix block) comply with this definition. Otherwise,
we have to do with an extended RC. An assumption that both RC(A) and RC(B)
are simple reaction centers usually implies that only two elementary
fragments of our molecules actually come into contact. This case has been
referred to as that of a local intermolecular contact. Otherwise, we have to
do with the case of a non-local contact.

Let us return again to the population alteration $\delta X_{(+)i}^{(A)}.$\
It is evident that the underlying orbital $\varphi _{(+)i}^{(A)}$\ may
belong to any of the above-defined fragments (i.e. RC(A), NN(A), etc.) and,
consequently, distinct population alterations $\delta X_{(+)i}^{(A)}$\
actually arise. Analysis of separate members of the relevant power series
showed that the higher is the order parameter of the given increment ($k$),
the more distant fragments are embraced by the relevant charge
redistribution [76].

To illustrate the above-drawn general conclusion, let us start with second
order terms ($k=2$). These terms contain an intermolecular component only
that depends on squares of the relevant first order elements $G_{(1)il}$\
(see Eq.(6.8)) and represents the charge transfer between orbitals of
directly contacting fragments RC(A) and RC(B). The actual nature of these
RCs (i.e. whether they are simple or extended) plays no essential role in
the formation of these terms. By contrast, non-zero third order increments
arise only if at least one of reaction centers is an extended RC. Although
these contributions also embrace the reaction centers only, both intra- and
intermolecular components are possible here, and these describe charge
redistributions inside the extended RC and between the RC(A) and RC(B),
respectively.

On the whole, the above-discussed local charge redistributions demonstrate
the primary role of reaction centers in chemical processes and may be
regarded as quantum chemical analogues of the supposed direct participation
of the RC(A) and RC(B) fragments in the given reaction.

The fourth order terms of the same series represent additive components of
an indirect influence of a certain nearest-neighboring fragment NN(A) (e.g.
of a substituent) upon the reactivity of the whole system. Three principal
components may be mentioned here: i) The NN(A)\ fragment exerts an
additional electron-donating or accepting effect upon the reaction center of
its own molecule (RC(A)) under influence of the approaching reaction center
of the opposite molecule (RC(B)), because the latter offers its orbitals as
mediators for specific indirect intramolecular interaction; ii) The NN(A)
fragment exerts an analogous effect upon the reaction center of the opposite
molecule (RC(B)) owing to an ability of orbitals of its own reaction center
(RC(A)) to mediate specific indirect intermolecular interactions; iii) The
NN(A) fragment offers its orbitals as mediators for indirect interorbital
interactions that gives rise to additional electron density redistributions
both inside the reaction center of its own molecule and between RC(A) and
RC(B). The whole of the above-enumerated effects may be accordingly regarded
as a quantum chemical analogue of an indirect participation of the NN(A)
fragment in the given reaction.

In the case of a still more remote fragment (e.g. NNN(A)), terms of even
higher orders are required to describe the relevant effects. Thus,
extinction of an indirect influence is predicted when the distance between
the given fragment and the reaction center(s) grows. Moreover, the relative
importance of higher order terms may be expected to increase when passing
from the early stages of reaction to later ones. Finally, the
above-enumerated results may be easily reformulated in terms of reshapings
of LMOs of reacting molecules due to an intermolecular contact [54]. Thus,
the classical concepts (mental images) overviewed at the beginning of this
section acquire a full support.

This outcome of our analysis, however, is not the only one. Indeed, the
above-outlined general results form the basis for constructing abstract
semi-local models for specific reactions [54, 58, 77-83]. As opposed to
popular numerical studies referring to individual compounds and/or
processes, these models are intended for embracing all reactions of a
certain type in accordance with their chemical classification. These models
usually contain orbitals of the supposed reaction centers and of some
neighboring fragments of particular interest. Meanwhile, details of specific
systems may be easily ignored in these models for the benefit of both
simplicity and generality. As a result, the overall approach allows direct
qualitative comparisons of similar processes including alternative routes of
the same reaction and processes embracing reactants or reagents of related
chemical constitution (e.g. a hydrocarbon and its heteroatom-containing
derivative). Let us now turn to an overview of these important points and
the relevant achievements.

It is evident that a single initially-occupied (vacant) orbital $\varphi
_{(+)N}$ ($\varphi _{(-)E}$) may be successfully used to represent a
nucleophilic (electrophilic) reagent B when comparing alternative directions
of its attack upon the same reactant A [such a one-orbital model implies
that the reagent B consists only of a simple reaction center RC(B)].
Furthermore, absolute values of direct intermolecular interactions often may
be assumed to be uniform in the above-discussed comparisons. [The latter
assumption is based on the fact that alternative routes of a certain
reaction are most commonly characterized by different spatial arrangements
of non-neighboring fragments. For example, the $\alpha $- \ and $\beta $%
-attacks of electrophile upon a substituted ethene (Z-C$_{\alpha }$H=C$%
_{\beta }$H$_{2}$) are described by different positions of the reagent
relatively to the substituent Z [78]. This implies the absolute value of
direct intermolecular interactions and thereby the second order partial
transferred populations to play no important role in the formation of
predominant routes of reactions]. Finally, heterolytic organic reactions may
be classified on the basis of the order parameter ($k$) of the decisive
terms of power series [83].

For illustration, let us consider the so-called extended model [77] of the
bimolecular nucleophilic substitution (S$_{N}$2) process between a
substituted alkane Z-C$_{\alpha }$H$_{2}$-C$_{\beta }$H$_{2}$-\dots and
nucleophile ($Nu$), where Z stands for a heteroatom (nucleofuge). As opposed
to the usual local models of the same process containing the orbital of the
reagent ($\varphi _{(+)N}$) and the electron-accepting (antibonding) orbital
of the C$_{\alpha }$-Z bond ($\varphi _{(-)a}$) [23,70], orbitals of the C$%
_{\alpha }$-C$_{\beta }$ and/or C$_{\alpha }$-H bonds also have been
included into our model and resonance parameters between these extra
orbitals and the orbital $\varphi _{(+)N}$\ have been taken into account
explicitly. This implies an assumption about an extended reaction center
RC(A) containing four bonds (viz. the C$_{\alpha }$-Z bond and its three
geminal neighbours). As a result, non-zero third order charge
redistributions both inside the above-specified reaction center and between
orbitals $\varphi _{(+)N}$ and $\varphi _{(-)a}$\ were shown to be peculiar
to this reaction. Moreover, the partial population ($\delta
x_{(+)N,(-)a}^{(3)}$) transferred between orbitals $\varphi _{(+)N}$ and $%
\varphi _{(-)a}$\ indirectly and defined as follows%
\begin{equation}
\delta x_{(+)N,(-)a}^{(3)}=4G_{(1)Na}G_{(2)Na}\   \tag{9.3}
\end{equation}%
(see Eq. (6.8)) proved to be of the largest absolute value among the third
order increments concerned, where $G_{(1)Na}$ stands for the direct
interaction between orbitals $\varphi _{(+)N}$ and $\varphi _{(-)a}$\ and $%
G_{(2)Na}$\ coincides with the respective indirect interaction by means of
orbitals of the C$_{\alpha }$-C$_{\beta }$ and/or C$_{\alpha }$-H bonds. The
most important aspect here, however, consists in opposite signs of the
above-mentioned principal third order partial transferred populations for
alternative directions of the attack, viz. positive (negative) signs of the
latter correspond to the back (frontal) position of nucleophile (see also
[83]). When added (subtracted) to (from) the uniform positive second order
increments, these corrections ensure a larger (smaller) value of the total
population acquired by the Z-C$_{\alpha }$ bond for the back (frontal)
attack of nucleophile. It is natural to assume that the more population the
nucleofuge acquires, the easier it leaves. Hence, a greater efficiency of
the back attack unambiguosly follows from our analysis in aggreement with
the well-known experimental facts [10,23,70, 74, 84, 85]. Analogous
conclusions may also be easily drawn using partial delocalization
coefficients of LMOs as the principal terms [54]. In particular, a more
significant partial delocalization of the lone pair of electrons of
nucleophile over the Z-C$_{\alpha }$ bond is predicted to refer to the
predominant back attack vs. the frontal one.

Our next example serves to illustrate comparisons of relative reactivities
of related chemical compounds. We mean here the S$_{N}$2 reactions of
derivatives of substituted alkanes containing additional unsaturated groups
at the C$_{\alpha }$ atom. These derivatives are known to be generally
characterized by largely higher relative reactivities as compared to the
relevant alkyl analogues [11]. Moreover, a further increase of reactivity is
observed when the C=C bond is replaced by a C=O bond, e.g. in
halogen-substituted ketones (Z-C$_{\alpha }$H$_{2}$-RC$_{\beta }$=O)\ \ vs.
allyl halogenides (see [11] and references therein). In this connection,
increments have been comparatively analyzed [79] that represent the indirect
participation of orbitals of C$_{\beta }$=C$_{\gamma }$ and C$_{\beta }$=O
bonds in the decisive third order charge transfer between the
above-specified basis functions $\varphi _{(+)N}$ and $\varphi _{(-)a}.$\ It
turned out that participation of the bonding $\pi -$orbital of the C$_{\beta
}$=C$_{\gamma }$ bond in the indirect interaction $G_{(2)Na}$\ of Eq.(9.3)
contributes to lowering of the positive partial transferred population $%
\delta x_{(+)N,(-)a}^{(3)back}$\ and thereby of the relative reaction rate,
whereas that of the antibonding $\pi -$orbital gives rise to an opposite
effect. For highly electron-donating (soft) nucleophiles, the second
contribution was shown to predominate over the first one in addition. Just
this fact served to account for the higher reactivity of allyl halogenides
vs. their alkyl analogues. Finally, passing from the C$_{\beta }$=C$_{\gamma
}$ to the C$_{\beta }$=O\ bond proved to be accompanied by such changes in
shapes and one-electron energies of bond orbitals of the $\pi -$type that
ensure a significant reduction of the absolute value of the negative
increment of the bonding orbital and a simultaneous increase of the positive
contribution of the antibonding orbital to the population acquired by the Z-C%
$_{\alpha }$\ bond. This result formed the basis for a new interpretation of
the largely increased reactivity of $\alpha -$halocarbonyl compounds vs.
their hydrocarbon analogues.

Similarities between reactions of seemingly different nature also are among
possible outcomes of application of our approach. In this respect,
comparative analysis of early stages of the above-discussed S$_{N}$2
reactions, on the one hand, and of the Ad$_{N}$2 reactions of carbonyl
compounds [82], on the other hand, serves as an excellent example. Indeed,
electron-accepting substituents attached to the carbon atom of the carbonyl
group were shown to exert a rate-accelerating effect, the overall mechanism
of which closely resembles the above-overviewed mechanism of the indirect
influence of the C=O group in the S$_{N}$2 reactions of $\alpha $%
-halocarbonyl compounds. It also deserves adding here that the double C=O
bond has been represented in Ref.[82] by two equivalent bent bonds, only one
of them being under attack of nucleophile.

The aromatic electrophilic substitution (S$_{E}$2) reaction of pyridine is
another outstanding example of third order processes. The usual (canonical)
MOs of benzene played the role of basis functions in the relevant model [81]
along with a single vacant orbital of electrophile ($\varphi _{(-)E}$).
After taking into account the perturbation of the Coulomb parameter ($\alpha 
$) at the site of replacing the carbon atom by a nitrogen one, non-zero
direct intramolecular interactions arise in the pyridine molecule so that
the latter becomes an extended RC. As a result, third order partial
populations ($\delta x_{(+)2,(-)E}^{(3)}$) transferred between the HOMO of
the reactant ($\varphi _{(+)2}$) and the only orbital of the reagent ($%
\varphi _{(-)E}$) become responsible for alterations in relative
reactivities of carbon atoms when passing from benzene to pyridine.
Moreover, the corrections $\delta x_{(+)2,(-)E}^{(3)}$\ proved to be
negative quantities for all possible directions of the attack (i.e. ortho,
meta and para) in aggreement with well-known lowered reactivities of all
sites of pyridine vs. that of benzene. Finally, the correction $\delta
x_{(+)2,(-)E}^{(3)meta}$\ was shown to take the smallest absolute value vs.
the remaining ones in accordance with the highest reactivity of meta
position of pyridine.

To exemplify the fourth order processes, let us consider electrophilic
addition (Ad$_{E}$2) reaction of substituted ethenes (Z-C$_{\alpha }$H=C$%
_{\beta }$H$_{2}$). In the case of an electron-donating substituent (Z=D),
the relevant semilocal model [78] contains the occupied orbital of the
substituent D ($\varphi _{(+)d}$), the electron-accepting (vacant) orbital
of electrophile ($\varphi _{(-)E}$), as well as two bond orbitals of the
ethene fragment, i.e. the BBO $\varphi _{(+)e}$ and the ABO $\varphi _{(-)e}$%
. Direct interactions between orbitals $\varphi _{(+)d}$ and $\varphi
_{(-)E} $ have been ignored in this model because of their small values.
Consequently, our reactant (A) contains a simple RC consisting of the ethene
group and an NN(A) fragment coinciding with the substituent D. That is why
fourth order partial transferred populations play the decisive role in the
formation of distinct reactivities of the C$_{\alpha }$ and C$_{\beta }$
atoms under influence of the substituent D. Moreover, specific fourth order
contributions and the effects underlying the latter (enumerated below) are
in line with the above-discussed three principal components of an indirect
influence of an NN fragment: First, the substituent D exerts an additional
intramolecular electron-donating effect upon the ethene fragment under
influence of the approaching electrophile. This effect proves to be governed
by the following fourth order partial population 
\begin{equation}
\delta x_{(+)d,(-)e}^{(4)}=4G_{(1)de}G_{(3)de},  \tag{9.4}
\end{equation}%
where $G_{(1)de}$\ stands for the direct interaction between orbitals $%
\varphi _{(+)d}$ and $\varphi _{(-)e}$ and $G_{(3)de}$\ represents the
relevant indirect interaction mediated by orbitals $\varphi _{(+)e}$ and $%
\varphi _{(-)E}.$ Second, the substituent D exerts influence upon the
intermolecular charge transfer between the ethene fragment and electrophile
described by the following correction 
\begin{equation}
\delta x_{(+)e,(-)E}^{(4)}=4G_{(1)eE}G_{(3)eE}  \tag{9.5}
\end{equation}%
[The orbital $\varphi _{(+)d}$\ participates here as a mediator of the
indirect interaction $G_{(3)eE}$]. The indirect electron-donating effect of
the substituent D upon the electrophile may be mentioned as the third
contribution. This effect is represented by the fourth order partial
transferred population between orbitals $\varphi _{(+)d}$\ and $\varphi
_{(-)E}$, viz. 
\begin{equation}
\delta x_{(+)d,(-)E}^{(4)}=2(G_{(2)dE})^{2}.  \tag{9.6}
\end{equation}%
Furthermore, positive (negative) values of increments of Eqs.(9.4) and (9.5)
have been established to refer to attacks of electrophile upon the C$_{\beta
}$(C$_{\alpha }$) atoms. This implies that the approaching reagent
contributes to an increase of electron- donating effects both from the
substituent D to the ethene fragment and from the latter to the electrophile
itself provided the the C$_{\beta }$ atom is under attack. Otherwise, both
effects are predicted to be suppressed. So far as the increment of Eq.(9.6)
is concerned, a large (small) absolute value of the indirect interaction $%
G_{(2)dE}$\ and thereby of the partial transferred population $\delta
x_{(+)d,(-)E}^{(4)}$\ itself has been found for an attack of electrophile
upon the C$_{\beta }$(C$_{\alpha }$) atoms. This implies the indirect
electron-donating effect of the substituent D upon the reagent to be more
significant just for the $\beta $-attack.

Using interrelations between partial populations transferred between FOs and
the relevant delocalization coefficients (Section 7), we may reformulate the
above results in terms of LMOs [54] as follows: The lone pair of electrons
of the substituent D becomes delocalized more significantly both over the
antibonding orbital of the ethene fragment (C=C bond) and over the FO of
electrophile, if the reagent approaches the C$_{\beta }$\ atom vs. the C$%
_{\alpha }$\ atom. Similarly, the initially-localized pair of electrons of
the C=C bond proves to be delocalized more substantially over the FO of
electrophile in the case of the $\beta -$attack. In summary, the \
above-overviewed results indicate the C$_{\beta }$\ atom to be of greater
relative reactivity as compared to the C$_{\alpha }$\ atom in aggreement
with the well-known Markovnikov rule [10,11, 70,85].

The case of an electron-accepting substituent (Z=A) also has been considered
similarly [78]. As opposed to the above-discussed donor-containing system,
no indirect charge transfer is now possible between the substituent and the
reagent. As a result, two fourth order increments remain, namely partial
populations transferred from the initially-occupied orbital of ethene ($%
\varphi _{(+)e}$) to that of the substituent ($\varphi _{(-)a}$) and to the
orbital of electrophile ($\varphi _{(-)E}$). The relevant analysis showed
that these corrections are of positive (negative) signs for C$_{\alpha }$(C$%
_{\beta }$) attacks. This implies the electron-accepting effects both of the
substituent A and of the reagent (E) upon the C=C bond to become
strengthened for the $\alpha $-attack. Otherwise, the same effects are
predicted to be suppressed. Analogously, the pair of electrons of the C=C
bond becomes delocalized more substantially over the FO of electrophile just
for the $\alpha -$attack [54]. These results also are in line with
experimental facts [10, 11, 85].

It deserves adding finally that the above-overviewed results concerning the
Ad$_{E}$2\ reaction of substituted ethenes allows some conclusions to be
drawn the scope of applicability of which is rather wide. This primarily
refers to the above-demonstrated interdependence between the extent of an
electron-donating (accepting) effect of an external subsystem and the
structure of the remaining part of the whole system. In other words, we
actually have to do here with the variable nature of electron-donating and
accepting effects. Equivalence between an electrophile and an
electron-accepting substituent (A) also is noteworthy [both are represented
by an initially-vacant orbital]. It is no surprise in this connection that
similar results have been obtained also when studying intersubstituent
interactions [86].

Furthermore, the above-concluded variable nature of electron- donating
(accepting) effects concerns not only the relative extents of the latter,
but also their directions, namely a substituent proves to be able to turn
from donor to acceptor (or vice versa) if the constitution of the remaining
system is altered substantially. To illustrate this general statement, let
us refer to an analogous study of addition reactions of butadiene [58]. In
the relevant model, one of the two H$_{2}$C=CH-fragments has been supposed
to be under a direct attack of the reagent (either electrophile or
nucleophile) and thereby to coincide with a simple reaction center RC(A) of
the reactant A. Meanwhile, the remaining H$_{2}$C=CH-group has been assumed
to play the role of an NN(A) fragment (substituent), the latter being now
represented by both an initially-occupied (electron-donating) and an
initially-vacant (accepting) orbital. It has been demonstrated that the
effects of these orbitals upon the remaining fragments of the whole system
(i.e. upon the RC(A) and the reagent) may be considered independently
whatever the nature of the reaction. However, a strong interdependence has
been established between the actual relative extents of the above-specified
two components of the total effect of the H$_{2}$C=CH-group and the
electron-donating (accepting) properties of the reagent. Moreover, the H$_{2}
$C=CH-group was shown to manifest itself as an electron-donating (accepting)
substituent under influence of an electrophilic (nucleophilic) attack.

The stereoselective bimolecular $\beta $-elimination (E2) reactions of
substituted alkanes may be mentioned finally as an example of the fifth
order heterolytic processes [80]. As opposed to the competing S$_{N}$2
reaction of the same substrates discussed above, it is the H-C$_{\beta }$
bond that is assumed to be under a direct attack of an external base (B$^{..}
$) in this case. As a result, the above-mentioned bond and the base may be
considered as simple reaction centers RC(A) and RC(B), respectively.
Accordingly, the C$_{\alpha }$-C$_{\beta }$ and C$_{\alpha }$-Z bonds have
been correspondingly regarded as the NN(A) and the NNN(A) fragments. It is
no surprise in this connection that decisive additional indirect
electron-accepting effects of the C$_{\alpha }$-Z bond upon both C$_{\beta }$%
-H bond and the external base (B$^{..}$) proved to be represented by
specific fifth order partial transferred populations. Moreover, positive
(negative) signs of these populations for trans (cis) elimination processes
formed the basis for a conclusion about the former being the predominant
route of the given reaction.

In summary, alternative routes of heterolytic reactions were shown to be
accompanied by decisive partial transferred populations of opposite signs.

\section{Applications of the semilocalized approach to pericyclic
processes. The common selection rule for organic reactions}

Numerous new organic reactions have been discovered within the last several
decades including the so-called pericyclic processes [10, 11, 15].
Peculiarities of these reactions and rules governing them seem to differ
significantly from those of heterolytic processes discussed in Section 9.
Indeed, "a concerted reorganization of bonding occurs throughout a cyclic
array of continuosly bounded atoms "[11] in this case and thereby no local
reaction center is likely to be present. Another (an even more important)
distinctive feature of pericyclic processes consists in their high
stereospecificity that has been successfully accounted for by applying just
the CMO method (cf. the famous Woodward-Hoffmann rule [87-89]). The concept
of the H\"{u}ckel and M\"{o}bius aromaticity (cf. the Dewar-Zimmerman rule
[90-93]) also has been invoked to treat the same problem (Note that this
concept is traditionally applied to delocalized $\pi $-electron systems of
benzenoids). The above-enumerated circumstances usually create an impression
that delocalization is an immanent feature of pericyclic reactions in
contrast to heterolytic ones. In this context, a question of particular
interest is about applicability to pericyclic reactions of the PNCMO theory
of Sections 5-8 and thereby of the semilocalized perspective to chemical
processes in general. It is evident that an affirmative answer to this
principal question would imply feasibility of a unified theory for both
types of reactions in terms of direct and indirect interactions of localized
basis functions (FOs).

The thermal electrocyclic closure of polyenes containing N C=C bonds (C$%
_{2N} $H$_{2N+2}$) is among the most well-known examples of pericyclic
processes that often serves as a common model for all reactions of this
type. Thus, applicability of the PNCMO theory to initial open polyene chains
becomes a necessary condition here. This point has been explored in
Ref.[34]. The relevant zero and first order Hamiltonian matrices contained
resonance parameters between pairs of $2p_{z}$\ AOs referring to
initially-double (C=C) and initially-single (C-C) bonds, respectively.
Convergence of the power series for the CBO matrix and for the total energy
of $\pi $-electrons of open polyenes has been demonstrated for chains of a
small and medium size even if all carbon-carbon bonds are formally
represented by resonance parameters of uniform values. Thereupon, the same
approach has been employed to investigate the closure process [94]. To this
end, an additional resonance parameter ($\lambda _{1,2N}$) has been assumed
to emerge between the $2p_{z}$\ AOs $\chi _{1}$ and $\chi _{2N}$\ of the
terminal carbon atoms C$_{1}$\ and C$_{2N}$. This new parameter has been
considered as a small perturbation (vs. the intrachain ones) when studying
very early stages of reactions. Accordingly, alterations in total energies
of polyene chains due to perturbation $\lambda _{1,2N}$\ are expressible as
follows%
\begin{equation}
\Delta \mathcal{E}=\mathcal{E}_{(1)}=2P_{1,2N}\lambda _{1,2N},  \tag{10.1}
\end{equation}%
where $P_{1,2N}$\ stands for bond orders between the terminal AOs $\chi _{1}$
and $\chi _{2N}$\ in the initial (open) chains. Formulae for $P_{1,2N}$, in
turn, have been derived by invoking the retransformation procedure for the
DM described below (Section 11). It turned out that the signs of bond orders 
$P_{1,2N}$\ alternate with growing N values. Moreover, positive (negative)
signs of $P_{1,2N}$\ have been established for odd (even) total numbers of
C=C bonds. Consequently, positive (negative) $\lambda _{1,2N}$\ values are
required to ensure positive energy corrections $\Delta \mathcal{E}$\ of
Eq.(10.1) and thereby stabilization of the whole system (Note that negative
energy units were used). Positive (negative) resonance parameters $\lambda
_{1,2N}$\ and /or overlap integrals, in turn, were shown to correspond to
disrotatory (conrotatory) ways of closure. Thus, a conrotatory closure of
the chain followed for even $N$ values (e.g. butadiene) and a disrotatory
one was expected for odd $N$ values (e.g. hexadiene). Inasmuch as $4n$ and $%
4n+2$ electrons, respectively, correspond to these cases, the
above-established result coincides with the Woodward-Hoffmann rule [87-89].
An important point here is that no passing to delocalized (canonical) MOs
was required when obtaining this result.

Another essential aspect of the above-outlined derivation is that the
decisive bond orders $P_{1,2N}$\ depend on definite matrix elements ($%
G_{(k)il}$) describing the interactions between BOs of terminal bonds of an
open polyene chain either directly (e.g. in butadiene) or indirectly by
means of the remaining BOs of the chain. Accordingly, alternation of signs
of $P_{1,2N}$\ is determined by an analogous behaviour of the
above-specified interactions with growing number of intervening bonds.
Inasmuch as $\lambda _{1,2N}$\ refers to the direct overlap of the terminal
BOs owing the closure process, the energy correction $\Delta \mathcal{E}$\
of Eq.(10.1) seems to depend upon a certain roundabout interaction over the
whole cycle. The latter anticipation has been proven rigorously in Ref.[95],
where a somewhat different model of pericyclic processes has been suggested
that was intended for representing a later stage of the whole process. Both
the newly-emerging resonance parameter ($\lambda _{1,2N}$) and those
referring to initially-single (C-C) bonds were assumed to take similar and
small values relatively to parameters referring to the initially-double
(C=C) bonds in this alternative model. Application of the power series for
the DM then accordingly resulted into an analogue of the Woodward- Hoffmann
rule for pericyclic reactions in terms of the so-called roundabout
interaction of the newly-formed cycle ($\Omega _{(N)}$) also alternatively
referred to as topological factor. This characteristic has been defined
explicitly as a product of resonance parameters (or overlap integrals)
between BOs of neighboring C=C bonds over the whole cycle and contained a
definite $N$-dependent parity factor in addition. Besides, the overall
pattern of the reorganization of bonding during the reaction (including
alterations in orders of intrachain bonds) also was shown to be governed by
the newly-formulated analogue of the Woodward- Hoffmann rule. Finally,
analysis of the same reactions in terms of delocalization coefficients of
LMOs has been carried out [54]. The extents of delocalization of
initially-localized pairs of electrons of C=C bonds were shown here to obey
an analogue of the Woodward- Hoffmann rule too.

On the whole, the above-overviewed results demonstrate adequacy of the PNCMO
theory for investigation of pericyclic reactions and thereby
interpretability of these reactions in terms of direct and indirect
interactions of orbitals of initially-double (C=C) bonds.

In this connection, attempts have been made to construct a unified approach
to both types of reactions. Let us turn now to discussion of the most
outstanding achievements in this direction.

Analysis of early stages of the most popular electrophilic and nucleophilic
reactions showed certain cycles of basis orbitals [96] to be formed in this
case in some analogy with pericyclic processes. For example, the
initially-occupied orbital of nucleophile ($\varphi _{(+)N}$) interacts with
orbitals of both C$_{\alpha }$-Z and C$_{\alpha }$-C$_{\beta }$(C$_{\alpha }$%
-H) bonds, the latter also interacting one with another. This implies
closure of at least three-membered cycle to underly the S$_{N}$2 processes
of substituted alkanes. To represent the overlap topologies of these
newly-revealed cycles [96], a topological factor has been defined like that
discussed above ($\Omega _{(N)}$). Moreover, the popular concept of the
usual (H\"{u}ckel) cycle and of its M\"{o}bius analogue has been generalized
by defining them as cycles described by positive and negative topological
factors, respectively. Accordingly, the classical $4n+2/4n$\ rule
representing the H\"{u}ckel [19, 20] and M\"{o}bius aromaticities [97] has
been extended. The predominant routes of the most well-known heterolytic
reactions are then shown to be governed by the extended $4n+2/4n$\ rule, the
latter being an analogue of the Dewar-Zimmerman rule for pericyclic
reactions. Thus, a common description of both types of reactions has been
achieved in this case by invoking the concept H\"{u}ckel [19, 20] and M\"{o}%
bius aromaticities [97].

Another unified description of both homolytic and heterolytic processes has
been suggested in Ref.[83] and called the common selection rule for organic
reactions.

To discuss this rule, let us return to the power series for the total energy 
$\mathcal{E}$\ of fragmentary molecules, the separate members of which ($%
\mathcal{E}_{(k)}$) are represented either via partial populations ($%
x_{(+)i,(-)l}^{(k)}$)\ transferred between basis orbitals (FOs) of opposite
initial occupation as shown in Eq.(8.9) or in terms of the related partial
delocalization coefficients of LMOs ($d_{(+)i,(-)l}^{(k)}$) of Eq.(8.10).
Let us dwell on the first representation for convenience. As is seen from
Eq.(8.9), the correction $\mathcal{E}_{(k)}$\ contributes to stabilization
(destabilization) of the whole system if the partial populations $%
x_{(+)i,(-)l}^{(k)}$\ contained within the relevant definition are of
positive (negative) signs [Negative energy units are assumed to be chosen
here as previously so that $\varepsilon _{(+)i}+\varepsilon _{(-)l}>0$]. For
the second order member $\mathcal{E}_{(2)},$ positive signs of $%
x_{(+)i,(-)l}^{(2)}$ unambiguosly follow from Eq.(6.8.8) whatever the
particular FOs $\varphi _{(+)i}$ and $\varphi _{(-)l}.$\ So far as
increments of higher orders ($k=3,4...$) are concerned, positive (negative)
signs of the relevant partial transferred populations $x_{(+)i,(-)l}^{(k)}$\
are ensured if the interorbital interactions contained within the relevant
definitions of Eq.(6.8) (i.e. $G_{(1)il},G_{(2)il},G_{(3)il},etc$)\ are of
coinciding (opposite) signs for separate pairs of FOs. Thus, it is the signs
of direct and indirect interorbital interactions that may be expected to
determine the predominant ways of chemical reactions. In this context,
correlations between signs referring to different pairs of FOs $\varphi
_{(+)i}$ and $\varphi _{(-)l}$ also become essential. Three cases may be
distinguished here: The first one embraces reactions described by all (or
almost all) positive increments $x_{(+)i,(-)j}^{(k)}$ to the decisive k-th
order energy correction $\mathcal{E}_{(k)}$ so that the positive sign and
thereby the stabilizing nature of the latter is unambiguosly ensured. These
processes have been referred to as allowed k-th order reactions. The second
case embraces processes represented by all (or almost all) negative
increments $x_{(+)i,(-)l}^{(k)}$ and thereby by a negative k-th order
correction $\mathcal{E}_{(k)}.$ The term \textquotedblleft the forbidden
k-th order reactions\textquotedblright\ has been employed in this case. An
intermediate case also is possible here when the principal pairs of FOs of
opposite initial occupation yield contributions of different signs to the
above-specified corrections. Due to the strong correlation of signs of
interorbital interactions for different pairs of FOs peculiar to specific
reactions, the actual processes prove to belong to either allowed or
forbidden ones.

This general rule has been successfully \ applied to numerous specific
reactions including both heterolytic and pericyclic. These achievements are
overviewed in Ref. [83] in a detail. In this connection, we will confine
ourselves here to third order processes and discuss their heterolytic and
pericyclic representatives.

For the S$_{N}$2 reaction of substituted alkanes, the third order partial
population transferred between the electron-donating orbital of nucleophile (%
$\varphi _{(+)N}$) and the electron-accepting (antibonding) orbital of the C$%
_{\alpha }$-Z bond ($\varphi _{(-)a}$) playes the decisive role in the
choice of the predominant route as discussed in Section 9. The relevant
definition is exhibited in Eq.(9.3) and contains the direct interaction of
these orbitals ($G_{(1)Na}$) along with their indirect interaction ($%
G_{(2)Na}$) by means of orbitals of C$_{\alpha }$ -C$_{\beta }$ (C$_{\alpha
} $-H) bonds. Analysis of these matrix elements for frontal and back attacks
of nucleophile yields the following result 
\begin{equation}
G_{(1)Na}^{(b)}<0,\qquad G_{(1)Na}^{(f)}>0,\qquad G_{(2)Na}<0.  \tag{10.2}
\end{equation}%
The last relation of Eq.(10.2) indicates the sign of the indirect
interaction $G_{(2)Na}$\ to be independent of the position of nucleophile.
It is also seen that the back attack is represented by the principal first
and second order interorbital interactions of coinciding signs and thereby
may be considered as an allowed process. Meanwhile, the frontal attack is
characterized by the same interactions of opposite signs and proves to be
forbidden.

In the cace of the S$_{E}$2 reaction of pyridine, the decisive third order
partial transferred population ($\delta x_{(+)2,(-)E}^{(3)}$) refers to the
HOMO of the parent system (benzene) ($\varphi _{(+)2}$) and the orbital of
electrophile ($\varphi _{(-)E}$) and also is accordingly determined by
interactions $G_{(1)2E}$\ and $G_{(2)2E}.$ Moreover, the second order
(indirect) interaction $G_{(2)2E}$\ and thereby the correction $\delta
x_{(+)2,(-)E}^{(3)}$\ contain two components representing the mediating
effects of the LUMO of benzene $\varphi _{(-)5}$\ and of the HOMO itself
(the so-called self-mediating effect). These components are denoted below by
' and " , respectively. Analysis of the relevant expressions showed that $%
G_{(2)2E}^{\prime \prime }$\ and thereby $\delta x_{(+)2,(-)E}^{(3)\prime
\prime }$\ are negative quantities whatever the direction of the attack.
Hence, the predominant route of the given reaction proves to be determined
by the increment $\delta x_{(+)2,(-)E}^{(3)\prime }$\ containing the LUMO $%
\varphi _{(-)5}$\ as a mediator. \ The signs of the relevant indirect
interactions are actually conditioned by the structure of the LUMO $\varphi
_{(-)5}$\ itself. We then obtain%
\begin{equation}
G_{(2)2E}^{\prime (o)}>0,\qquad G_{(2)2E}^{\prime (m)}>0,\qquad
G_{(2)2E}^{\prime (p)}<0  \tag{10.3}
\end{equation}%
for the ortho-, meta- and para- positions of electrophile, respectively.
Meanwhile, the relevant direct interactions $G_{(1)2E}$\ are of the
following signs%
\begin{equation}
G_{(1)2E}^{(o)}<0,\qquad G_{(1)2E}^{(m)}>0,\qquad G_{(1)2E}^{(p)}>0 
\tag{10.4}
\end{equation}%
that are determined by the constitution of the HOMO of benzene $\varphi
_{(+)2}.$\ Comparison of Eqs.(10.3) and (10.4) allows us to conclude that
just the predominant meta attack of electrophile is characterized by the
principal interorbital interactions of coinciding signs and thereby may be
considered as an allowed process.

The electrocyclic closure of hexatriene (C$_{6}$H$_{8}$) and the Diels-Alder
reaction between butadiene and ethene are the most illustrative
representatives of third order pericyclic processes. These systems contain
three initially-double (C=C) bonds that form six-membered cycles during the
closure process. Thus, a single model may be actually applied to these
reactions and the relevant conclusions also closely resemble one another.
Let us consider the closure process of hexatriene as an example. Let the AOs 
$\chi _{1},\chi _{2},...\chi _{6}$ of this molecule to form six BOs that
acquire numbers 1,2 and 3, viz. $\varphi _{(+)1}(\varphi _{(-)1}),$\ $%
\varphi _{(+)2}(\varphi _{(-)2})$ and $\varphi _{(+)3}(\varphi _{(-)3}).$\
Resonance parameters between pairs of AOs inside the C=C bonds will be
assumed to coincide with 1, whereas those of C-C bonds will be denoted by $%
\kappa ,$\ where $0<\kappa <1.$\ Accordingly, $\lambda $\ will stand for the
newly-formed resonance parameter between the terminal AOs $\chi _{1}$\ and $%
\chi _{6}$. Orbitals of the terminal bonds C$_{1}$=C$_{2}$ and C$_{5}$=C$%
_{6} $, viz. $\varphi _{(+)1}$ and $\varphi _{(-)3},$ interact both directly
(owing to the parameter $\lambda $) and indirectly through orbitals of the
intervening C$_{3}$=C$_{4}$ bond. The latter interactions (i.e $G_{(2)13}$)
coincide with $\kappa ^{2}/8$ and are positive quantities whatever the
actual way of closure. Hence, a positive sign of the direct interaction $%
G_{(1)13}$\ also is required to ensure positive third order transferred
population $\delta x_{(+)1,(-)3}^{(3)}.$ Inasmuch as\ $G_{(1)13}$ equals to $%
\lambda /4$, the above condition proves to be met for positive $\lambda $\
values, the latter corresponding to a disrotatory way of closure.
Interactions of other pairs of BOs may be considered analogously. For
example, matrix elements describing the interactions between BOs $\varphi
_{(+)1}$ and $\varphi _{(-)2}$ are as follows%
\begin{equation}
G_{(1)12}=-\frac{\kappa }{4},\qquad G_{(2)12}=-\frac{\kappa \lambda }{8} 
\tag{10.5}
\end{equation}%
It is seen that $G_{(1)12}$ and $G_{(2)12}$ are of uniform (negative) signs
if $\lambda $\ is a positive parameter as previously. Hence, the disrotatory
way of closure of hexatriene meets our definition of an allowed process in
accordance with expectation and experimental facts. Hence, similarity
between the pericyclic and heterolytic reactions is now supported also in
respect of applicability of the common selection rule in terms of signs of
direct and indirect interactions.

\section{The local retransformation procedure and its applications to
define environment- determined intrafragmental effects}

In comparative studies of related chemical compounds, we often have to do
with distinct systems containing the same fragment(s). Changes in the
structures of the environment of the given fragment when passing from one
compound to another seems to give rise to certain intrafragmental effects.
Such an anticipation follows from numerous experiment- based conclusions. As
for instance, replacement of a hydrogen atom of an alkane by a heteroatom is
known to cause a definite polarization of the neighboring C-C and C-H bonds
(cf. the so-called induced dipoles [85, 98]). An additional polarization of
the heteroatom-containing bond under influence of an electron-donating
effect of an approaching nucleophile during early stages of S$_{N}$2
reactions also may be mentioned here along with a specific reorganization of
the aromatic ring into a non-aromatic system under influence of an
electrophilic attack (cf. the Wheland intermediate) [11, 15, 16, 23].
Moreover, the above-enumerated effects are always observed in similar
constitutional situations and thereby are likely to be conditioned mostly by
the structure of the given fragment and of its nearest environment.

It deserves an immediate mentioning that fragmental orbitals (FOs)
underlying the PNCMO theory of Sections 5-8 are not the optimum basis
functions in respect of describing the intrafragmental effects. Quite the
reverse, just the usual bases of AOs and/or hybrid AOs (HAOs) offer much
more convenient representations of the same effects. These standard basis
functions, however, do not meet the requirements of the PNCMO theory. To
find the way out of this situation, a certain additional procedure evidently
is required. Before turning to the latter, let us recall some principles of
the standard (canonical) MO method.

The essence of the above-mentioned popular method consists in passing from
the initial basis of AOs or HAOs $\{\chi \}$ into that of canonical
MOs(CMOs) $\{\psi \}$, wherein the Hamiltonian (or Fockian) matrix of our
system $\overline{\mathbf{H}}$ takes the diagonal form [1-3]. The respective
representation matrix of the one-electron DM $\overline{\mathbf{P}}$ also is
diagonal in the CMO basis and involves occupation numbers of these orbitals
equal to either two or zero. This implies no bond orders to arise between
CMOs. To obtain the usual CBO matrix of the same system $\mathbf{P}^{\prime
} $ containing populations of AOs(HAOs) and bond orders between the latter,
we retransform the matrix $\overline{\mathbf{P}}$ into the AO basis again.
The relation concerned is as follows [1]%
\begin{equation}
\mathbf{P}^{\prime }\mathbf{=T}\overline{\mathbf{P}}\mathbf{T}^{+}, 
\tag{11.1}
\end{equation}%
where the matrix $\mathbf{T}$\ contains the standard MO LCAO coefficients in
its columns, and the superscript + designates the transposed (or
Hermitian-conjugate) matrix as previously. This relation along with the
above-specified diagonal constitution of the matrix $\overline{\mathbf{P}}$\
yields the usual expressions for elements of the CBO matrix $\mathbf{P}%
^{\prime }$ in the form of sums of the MO LCAO coefficients over occupied
MOs.

In our context, the most important property of Eq.(11.1) consists in its
validity for any pair of basis sets, provided that the DM of the right-hand
side is known or easily constructable. The latter requirement is met in the
basis of FOs $\{\Phi \}$\ as the results of Section 6 indicate. Thus, the
retransformation procedure like that of Eq.(11.1) may be alternatively
applied [99-101]\ to derive the usual CBO matrix of our system(s) $\mathbf{P}%
^{\prime }$\ using the relevant DM\ of Section 6. Instead of Eq.(11.1) we
then obtain 
\begin{equation}
\mathbf{P}^{\prime }\mathbf{=UPU}^{+},  \tag{11.2}
\end{equation}%
where $\mathbf{U}$\ stands for the matrix of the FO LCAO coefficients. As
opposed to the above-desribed standard procedure, the DM under
transformation (i.e. $\mathbf{P}$) now contains non-zero off-diagonal
elements coinciding with bond orders between FOs defined by Eqs.(6.2) and
(6.3). Moreover, diagonal elements of this matrix are somewhat different
from either 2 or 0 as follows from Eq.(6.3). Consequently, the actual
expression for the CBO matrix being sought generally contains sums of FO
LCAO coefficients over all FOs, as well as over their pairs instead of
occupied MOs only.

Another important difference of the above-described procedure from the
standard one consists in the block-diagonal structure of the
retransformation matrix $\mathbf{U}$. To clarify this point, let us assume
our system to consist of certain fragments I,II,III, etc. The total basis of
FOs $\{\Phi \}$\ will be then accordingly composed of subsets $\{\Phi
_{I}\},\{\Phi _{II}\},\{\Phi _{III}\},etc.,$ each of basis orbitals being
localized on a single fragment only. Consequently, the total matrix $\mathbf{%
U}$\ is representable as follows%
\begin{equation}
\mathbf{U=}\left\vert 
\begin{array}{cccc}
\mathbf{U}_{I} & 0 & 0 & ... \\ 
0 & \mathbf{U}_{II} & 0 & ... \\ 
0 & 0 & U_{III} & ... \\ 
. & . & . & ...%
\end{array}%
\right\vert   \tag{11.3}
\end{equation}%
where $\mathbf{U}_{I},\mathbf{U}_{II},\mathbf{U}_{III},etc.$ are non-zero
submatrices of FO LCAO coefficients of separate fragments. The DMs $\mathbf{P%
}$\textbf{\ }and\textbf{\ }$\mathbf{P}^{\prime }$\ of Eq.(11.2) also may be
divided into intra and interfragmental blocks without any restriction.
Substituting the matrix $\mathbf{U}$\ of Eq.(11.3) into Eq.(11.2) then
allows the overall retransformation procedure to be partitioned into local
ones, each of them embracing a single fragment only, e.g.%
\begin{equation}
\mathbf{P}_{I}^{\prime }\mathbf{=U}_{I}\mathbf{P}_{I}\mathbf{U}%
_{I}^{+},\qquad \mathbf{P}_{II}^{\prime }\mathbf{=U}_{II}\mathbf{P}_{II}%
\mathbf{U}_{II}^{+},\quad etc.  \tag{11.4}
\end{equation}%
and being performable separately. Again, we should recall here that elements
of the DM $\mathbf{P}$\textbf{\ }of Section\textbf{\ } 6 contain sums of
products of interorbital interactions over the whole molecule under study
(see e.g. the definition of the element $G_{(2)il}^{(f)}$\ of Eq.(5.14)
determining the bond orders between FOs of opposite initial occupation as
Eq.(6.2) shows). This implies a particular block of the matrix $\mathbf{P}$\
(e.g. $\mathbf{P}_{I}$) to depend implicitly upon the remaining fragments of
the given system. As a result, the relevant submatrix of the CBO matrix
obtained (i.e. $\mathbf{P}_{I}^{\prime }$, respectively) reflects the
influence of the remaining part of the molecule upon the charge and bond
order distribution inside the given fragment (I) as demonstrated below.
Inasmuch as elements of the DM $\mathbf{P}$\textbf{\ }depend mostly on the
nearest environment of the FO (FOs) concerned (Section 6), the same state of
things may be expected to refer to elements of the CBO matrix $\mathbf{P}%
^{\prime }$\ too. Additivity of influences of separate neighbors is another
anticipation here. The decisive role of FOs of the given fragment and
thereby of the structure of the latter in the formation of intrafragmental
effects also may be easily predicted on the basis of Eq.(11.4). Indeed,
relations shown there indicate that both populations of AOs (HAOs) and bond
orders between the latter referring to a definite fragment (I) are
expressible as linear combinations of elements of the respective block of
the DM $\mathbf{P(\mathbf{P}}_{I}\mathbf{),}$\ the nature of these
combinations being determined by elements of the matrix $\mathbf{U}_{I},$\
i.e. by the structure of FOs of the given fragment. That is why it appears
that the relations of Eq.(11.4) may be analyzed for each type of fragments
separately without specifying either the remaining fragments of the system
or the interfragmental interaction. As a result, general algebraic
expressions have been derived for intrafragmental effects inside the most
popular fragments as overviewed below.

The above-outlined retransformation procedure may be easily extended to
embrace other characteristics of electronic structures too. For the LMO
representation matrix, we accordingly obtain [102] 
\begin{equation}
\mathbf{C}^{\prime }\mathbf{=UC}  \tag{11.5}
\end{equation}%
instead of Eq.(11.2), where $\mathbf{C}^{\prime }$\ refers to the basis of
AOs(HAOs). Employment of the block-diagonal matrix $\mathbf{U}$\ of
Eq.(11.3) followed by the respective partition of matrices $\mathbf{C}$ and $%
\mathbf{C}^{\prime }$\ into intra- and interfragmental blocks yields a
series of relations%
\begin{equation}
\mathbf{C}_{I}^{\prime }\mathbf{=U}_{I}\mathbf{C}_{I},\qquad \mathbf{C}%
_{II}^{\prime }\mathbf{=U}_{II}\mathbf{C}_{II},\quad etc.\qquad   \tag{11.6}
\end{equation}%
representing the relevant local retransformation procedures for LMOs.

To consider the total energy, let us start with the notation that general
relations between $\alpha $- and $\beta $-components of corrections $%
\mathcal{E}_{(k)}$\ shown in Eqs.(8.6) and (8.7) are invariant against
unitary transformations of the basis set including the above-described
retransformation. The same evidently refers to definitions of components $%
\mathcal{E}_{(k)}^{(\alpha )}$\ and $\mathcal{E}_{(k)}^{(\beta )}$ of
Eq.(8.5), where matrices $\mathbf{P}_{(k)},\mathbf{P}_{(k-1)},$ $\mathbf{H}%
_{(0)}$\ and $\mathbf{H}_{(1)}$ should be replaced by their stroked
counterparts referring to the basis of AOs (HAOs). If we confine ourselves
to fragmentary molecules defined as sets of weakly-interacting elementary
fragments I, II, III, etc. (Section 5), the relevant zero order Hamiltonian
matrix ($\mathbf{H}_{(0)}^{\prime }$ ) may be assumed to take a
block-diagonal form like that of Eq.(11.3), where intrafragmental blocks $%
\mathbf{H}_{(0)I}^{\prime },\mathbf{H}_{(0)II}^{\prime },etc.$ occupy the
diagonal positions. At the same time, the relevant first order member ($%
\mathbf{H}_{(1)}^{\prime }$)\ contains zero submatrices in the same
positions. We may conclude on this basis that the energy component $\mathcal{%
E}_{(k)}^{(\beta )}$\ depends on bond orders formed between AOs(HAOs) of
different fragments within the correction of the CBO matrix of the (k-1)th
order ($\mathbf{P}_{(k-1)}^{\prime }$) and thereby may be called the
interfragmental part of the kth order energy [101]. Meanwhile, the remaining
component $\mathcal{E}_{(k)}^{(\alpha )}$\ is determined only by
intrafragmental elements of the correction $\mathbf{P}_{(k)}^{\prime }$\
[including both occupation numbers of AOs(HAOs) and bond orders between the
latter] and proves to be the intrafragmental part of the same energy
correction $\mathcal{E}_{(k)}$. Consequently, the relation of Eq.(8.6)
becomes interpretable as that between energetic increments of the
interfragmental interaction and of the intrafragmental response to the
latter. Moreover, stabilization of the whole system due to formation of
interfragmental bond orders (if any) necessarily is accompanied by an
overall internal destabilization of fragments [This evidently does not imply
that any individual fragment is destabilized].

\ Let us turn now to specific fragments. Let us start with the most
widespread fragment, namely the two-center chemical bond. For the sake of
generality, let us consider a heteropolar bond between a certain heteroatom
(Z) and a carbon atom (C). [The homopolar bond corresponds to a particular
case of the heteropolar one]. Let the bond under our interest to acquire the
Ith number. Fragmental orbitals (FOs) then coincide with the BBO $\varphi
_{(+)i}$\ and the ABO $\varphi _{(-)i}$\ and are defined as simple linear
combinations of respective AOs (HAOs) $\chi _{Z}$\ and $\chi _{C}.$\
Diagonal elements of the original DM $\mathbf{P}$\ referring to BOs $\varphi
_{(+)i}$\ and $\varphi _{(-)i}$\ then follow from Eq.(6.3) and coincide with
respective occupation numbers $X_{(+)i}$\ and $X_{(-)i},$ whereas the only
off-diagonal element (bond order between the same BOs) is defined by
Eq.(6.2) and contains elements $G_{(1)ii},G_{(2)ii},etc$. After performing
the above-described local retransformation procedure, the occupation numbers
of AOs (HAOs) $\chi _{Z}$\ and $\chi _{C}$\ have been expressed as follows
[101-105]%
\begin{align}
X_{Z}\ (X_{C}) =&1\pm \cos \gamma _{I}+\frac{1}{2}\Delta X_{(2)I}\pm
p_{(2)I}\pm d_{(2)I}+  \nonumber \\
&\frac{1}{2}\Delta X_{(3)I}\pm p_{(3)I}\pm d_{(3)I}+...  \tag{11.7}
\end{align}%
where the upper signs of the right-hand side refer to $X_{Z}$, whereas the
lower ones correspond to $X_{C}.$\ Notation $\gamma _{I}$\ is used here for
the principal parameter of our bond defined as follows%
\begin{equation}
\gamma _{I}=\arctan \frac{2\beta _{I}}{\alpha _{IZ}-\alpha _{IC}}, 
\tag{11.8}
\end{equation}%
where $\alpha _{IZ}$ and $\alpha _{IC}$\ stand for Coulomb parameters of AOs
(HAOs) $\chi _{Z}$\ and $\chi _{C},$\ respectively, and $\beta _{I}$\ is\
the relevant intrabond resonance parameter. The energy reference point and
the energy unit are assumed here to be chosen so that parameters $\alpha
_{IZ},$ $\alpha _{IC}$\ and $\beta _{I}$\ take positive values and the
equality $\alpha _{IZ}\geqslant \alpha _{IC}$\ is valid (a negative energy
unit is actually accepted).

The contributions $1/2\Delta X_{(k)I},(k=2,3...)$\ of Eq.(11.7) are related
to respective members of power series for occupation numbers of BOs ($%
X_{(+)i}$\ and $X_{(-)i}$), viz.%
\begin{equation}
\Delta X_{(k)I}=X_{(+)i}^{(k)}\ +X_{(-)i}^{(k)}-2  \tag{11.9}
\end{equation}%
and represent the total kth order populations lost (acquired) by the Ith
bond due to its interaction with neighboring fragments. As for instance, the
second order member $\Delta X_{(2)I}$\ takes the form%
\begin{equation}
\Delta X_{(2)I}=2\mathop{\displaystyle \sum }%
\limits_{(+)j}[(G_{(1)ji})^{2}-(G_{(1)ij})^{2}],  \tag{11.10}
\end{equation}%
(see Eqs.(6.6)-(6.8)). It is seen that the increment $\Delta X_{(2)I}$
actually consists of difference between absolute values of the population
lost by the BBO $\varphi _{(+)i}$\ and of that acquired by the ABO $\varphi
_{(-)i}.$\ The remaining increments of Eq.(11.7) are as follows%
\begin{equation}
p_{(k)I}=-2G_{(k)ii}\sin \gamma _{I},\quad d_{(k)I}=\frac{1}{2}\Delta
R_{(k)I}\cos \gamma _{I},  \tag{11.11}
\end{equation}%
where the term \ 
\begin{equation}
\Delta R_{(k)I}=X_{(+)i}^{(k)}\ -X_{(-)i}^{(k)}-2  \tag{11.12}
\end{equation}%
describes the total kth order redistributed population referring to the Ith
bond. The respective second order member is accordingly representable as
follows%
\begin{equation}
\Delta R_{(2)I}=-2\mathop{\displaystyle \sum }%
\limits_{(+)j}[(G_{(1)ji})^{2}+(G_{(1)ij})^{2}].  \tag{11.13}
\end{equation}
This term is determined by the sum of absolute values of the above-mentioned
lost and acquired populations. It is seen that negative contributions to $%
\Delta R_{(2)I}$\ arise owing to both the additional occupation of the ABO $%
\varphi _{(-)i}$ and the partial deoccupation of the BBO $\varphi _{(+)i}$
(see Eqs.(6.6)-(6.8)) to within the second order approximation. Hence, a
negative sign of the second order redistributed population $\Delta R_{(2)I}$%
\ unambiguously follows. The analogous formula for the internal bond order ($%
B_{I}$) between AOs (HAOs) $\chi _{Z}$\ and $\chi _{C},$\ in turn, takes the
form%
\begin{equation}
B_{I}=\sin \gamma _{I}+\lambda _{(2)I}+\omega _{(2)I}+\lambda _{(3)I}+\omega
_{(3)I}+...  \tag{11.14}
\end{equation}%
where 
\begin{equation}
\lambda _{(k)I}=2G_{(k)ii}\cos \gamma _{I},\quad \omega _{(k)I}=\frac{1}{2}%
\Delta R_{(k)I}\sin \gamma _{I}  \tag{11.15}
\end{equation}%
for $k=2,3..etc.$

Let us turn now to interpretation of these expressions. The zero order
dipole-like increment $\pm \cos \gamma _{I}$\ to populations of orbitals $%
\chi _{Z}$\ and $\chi _{C}$\ does not depend on the structure of the whole
compound and has been interpreted as the primary dipole of the Ith bond.
Similarly, the zero order term of Eq.(11.14), i.e. $\sin \gamma _{I}$,
coincides with the respective primary bond order. Alterations both in
occupation numbers of orbitals $\chi _{Z}$\ and $\chi _{C}$\ and in the
internal bond order after embedding the Ith bond into our molecule are then
determined by subsequent terms of Eqs.(11.7) and (11.14) starting with $k=2$%
. Additive nature of these alterations with respect to contributions of the
remaining fragments of the given compound follows straightforwardly from
Eqs.(11.10) and (11.13) along with the relevant property of indirect
interorbital interactions $G_{(k)ij}$\ (Section 5). Additivity between the
environment- determined corrections and the primary characteristics of our
bond also deserves \ mentioning. Again, dependence of the same alterations
mostly on the structure of the nearest environment of the given bond may be
concluded on the basis of extinction of matrix elements $G_{(k)ij}$\ and $%
G_{(k)ii}$\ when the distance between the BOs concerned grows. Thus, the
environment- determined intrabond effects seem to be in line with the
classical rule of locality in addition. Transferability of these effects for
the same neighborhoods of the Ith bond also is among expectations. It
deserves emphasizing here that the above conclusions are drawn without
specifying the structure of the whole compound and thereby are of a quite
general scope of validity. More information concerning the nature of the
intrabond effects follows from analysis of separate second order terms of
Eqs. (11.7) and (11.14).

Let us start with terms originating from the interfragmental charge
redistribution (charge transfer): First, uniform increments $(1/2)\Delta
X_{(2)I}$\ to occupation numbers of both $\chi _{Z}$\ and $\chi _{C}$\ may
be mentioned that coincide with a half of the total population lost
(acquired) by the given bond. Meanwhile, it is the dipole-like term $\pm
d_{(2)I}$\ that is responsible for the non-uniform actual distribution of
the same lost (acquired) population among AOs (HAOs) $\chi _{Z}$\ and $\chi
_{C}$. Moreover, the \textit{a priori} negative sign of $d_{(2)I}$\ follows
from the above-discussed negative sign of $\Delta R_{(2)I}$. This implies
the lost (acquired) population always to give rise to lowering of the
primary dipole of our bond. This effect may be easily accounted for by
constitution of the BBO $\varphi _{(+)i}$\ and of the ABO $\varphi _{(-)i}$\
and has been accordingly called depolarization. Similarly, the contribution $%
\omega _{(2)I}$\ to the bond order $B_{I}$\ also is a negative quantity
depending on the same total redistributed population $\Delta R_{(2)I}$. This
implies reduction of the primary bond order under influence of the
interfragmental charge redistribution. The above-mentioned simultaneous
effects (i.e. depolarization of an initially-heteropolar bond and the
related reduction of the internal bond order) reflect a trend towards a
homolytic dissociation of the given bond after including it into the
molecule under study. In this connection, the term 'homolytic
predissociation' seems to describe these interdependent effects [101].

Let us turn now to the remaining second \ order terms $p_{(2)I}$\ and $%
\lambda _{(2)I}$ originating from the newly-formed bond order between BOs of
the given bond ($\varphi _{(+)i}$\ and $\varphi _{(-)i}$). The sign of the
matrix element $G_{(2)ii}$\ contained within the relevant expressions of
Eqs.(11.11) and (11.15) cannot be established \textit{a priori} (i.e.
without specifying the structure of the system). The same then refers to
signs of both $p_{(2)I}$\ and $\lambda _{(2)I}.$ Nevertheless, an
interdependence between these signs is evident, namely reduction of the bond
order (predissociation) is expected to take place for a negative element $%
G_{(2)ii}$\ and this effect is predicted to be accompanied by emergence of a
positive dipole $p_{(2)I}$. Just this fact makes the term polarization
dipole used to refer to $p_{(2)I}$\ [103] even more appropriate. Inasmuch as
the total dipole of our bond grows in this case, we have actually to do with
the 'heterolytic predissociation' of our initially-heteropolar bond [101].
Thus, two alternative ways of predissociation reveal themselves as the
principal intrabond effects in the case of a heteropolar bond. Moreover,
emergence of just these effects may be entirely traced back to the structure
of BOs $\varphi _{(+)i}$\ and $\varphi _{(-)i}$\ and thereby of the given
bond. In this connection, the effects themselves may be regarded as
manifestation of the principle of locality in the sense that their nature is
entirely determined by the relevant local structure. This state of things is
largely similar to that concerning the local constitution of MOs (Section
3). Additivity of contributions representing the homo- and heterolytic ways
of the overall predissociation of our bond also is among important
conclusions. Owing to the opposite nature of these two alternatives, we may
then expect a certain competition to take place between them in an actual
compound, the outcome of which seems to depend upon the specific
constitution of the latter. An example of such a competition may be found in
Ref.[105], where the additional dipole of a single or double
heteroatom-containing bond has been studied that arises under influence of
an electron-donating effect of an external orbital (e.g. of an approaching
nucleophile). It turned out that the direction of the additional dipole
depends decisively on the relative electronegativity of the heteroatom Z and
thereby on the initial polarity of the bond. Moreover, predominance of the
polarization dipole and thereby of the heterolytic predissociation in
general has been found for bonds of relatively low initial polarity but not
for those of high polarity. This result served to account for the well-known
experimental fact that highly electronegative heteroatoms usually are bad
nucleofuges in S$_{N}$2 processes [11,15,85].

Let us return again to expressions of Eqs.(11.7)-(11.15) and consider the
particular case of a homopolar bond characterized by equalities $\alpha
_{I1}=\alpha _{I2},\gamma _{I}=\pi /2$\ and $\cos \gamma _{I}=0.$ The
depolarization dipole $d_{(2)I}$\ is then easily seen to vanish and the
homolytic predissociation actually resolves itself into lowering of the bond
order in accordance with the expectation. Again, a secondary polarization $%
p_{(2)I}$\ also may arise in homopolar bonds. In particular, these terms
were shown to be responsible for the induced dipoles of C-C (C-H) bonds
under influence of a heteroatom-containing bond (cf. the so-called inductive
effect). Emergence of such a dipole, however, is not accompanied by
reduction of the internal bond order (as $\lambda _{(2)I}=0$\ for homopolar
bonds) and thereby does not imply its heterolytic predissociation.

Terms of Eqs.(11.7) and (11.14) of higher orders may be analyzed similarly.
As opposed to the above-discussed second order term $\Delta R_{(2)I}$,
however, the sign of the relevant third order analogue ($\Delta R_{(3)I}$)
cannot be established \textit{a priori}. \ Nevertheless, lowering of the
bond order $B_{I}$\ (predissociation) corresponds to negative signs of $%
\Delta R_{(3)I}$\ as Eq.(11.15) indicates. Consequently, the third order
homolytic predissociation also may be defined for heteropolar bonds in
analogy with its second order counterpart discussed above. An example of
systems, when the third order increments are important, may be found in
Ref.[104].

Finally, application of Eqs.(11.5) and (11.6) to our heteropolar bond
allowed us to express the so-called heads of LMOs $\Psi _{(+)i}$\ and $\Psi
_{(-)i}$\ attached to our bond [102] as linear combinations of AOs (HAOs) $%
\chi _{Z}$\ and $\chi _{C}$\ , e.g. 
\begin{equation}
\Psi _{(+)i}=r_{I}\chi _{Z}\ +s_{I}\chi _{C}+...  \tag{11.16}
\end{equation}%
where dots stand for the tail of the LMO. Thereupon, we have found twofold
squares of coefficients $r_{I}\ \ $and $s_{I},$ as well as their twofold
product. Comparison of these expressions to those of Eqs.(11.7) and (11.14)
allowed us then to establish the contribution of the own LMO of the Ith bond
and thereby of the respective "own" pair of electrons to the overall homo
and/or heterolytic predissociation. The relevant study [102] showed that the
heterolytic predissociation, in general, and the polarization dipole, in
particular, may be entirely traced back to reshaping of the "own" LMO $\Psi
_{(+)i}$ and thereby to shift of the "own" pair of electrons of the Ith
bond. This conclusion embraces also the formation of secondary dipoles of
initially homopolar bonds (e.g. of C-C and C-H bonds in substituted
alkanes). Meanwhile, the overall situation concerning the homolytic
predissociation is somewhat more complicated. Thus, the parts of the
depolarization dipole $d_{(2)I}$\ and of the related reduction of the bond
order $\omega _{(2)I}$\ originating from deoccupation of the BBO $\varphi
_{(+)i}$\ and proportional to $X_{(+)i}$\ ( see Eq.(11.12)) may be traced
back to contribution of the LMO $\Psi _{(+)i}$, whilst the remaining ($%
X_{(-)i}$-containing) parts of the same characteristics result from
contributions of occupied LMOs of other bonds extending over the ABO $%
\varphi _{(-)i}$. Consequently, the hypothesis of the classical chemistry
about bond dipoles resulting from shifts of separate pairs of electrons may
be proven provided that the dipole concerned originates mainly from the
polarization term, i.e. for bonds of relatively low initial polarity.

Similar local retransformation procedures have been developed also for other
fragments, e.g. for phenyl rings [106]. The latter evidently are more
extended systems as compared to a homopolar bond. In this connection, a few
analogues of the polarization dipole $p_{(2)I}$\ and of the term $1/2\Delta
X_{(2)I}$ arise in the\ relevant expression for occupation numbers of $%
2p_{z} $\ AOs of carbon atoms. In particular, three environment-determined
secondary (induced) dipoles have been revealed instead of the single one
(i.e. of $p_{(2)I}$), namely the so-called ipso-ortho (para-meta), para-ipso
and ortho-meta dipoles. Just the latter two moments proved to play the
principal role in the formation of the observed picture of the electron
density distribution in substituted benzenes.

Applications of analogous retransformation procedures to two weakly-
interacting molecules A and B [99,100] also deserve mentioning, where CMOs
of isolated compounds play the role of FOs. Explicit algebraic expressions
have been derived and analyzed in this case for the reorganization of
bonding inside the reactant (say A) under influence of an approaching
reagent (B). The crucial role of the initial structure of the substrate (A)
in the subsequent reorganization of bonding during the reaction process was
the principal conclusion here that is in line with the usual discussions of
chemical reactivity of a certain compound in terms of peculiarities of its
initial constitution. A detailed overview of the relevant achievements may
be found in Ref.[99]. The case of allyle cation (anion) under attack of
nucleophile (electrophile) [100] is an especially illustrative example of
these studies. Two effects have been revealed to manifest itself inside the
allyle ion if a particular terminal carbon atom is attacked by the reagent,
namely an induced lengthwise polarization and a partial switch of bond order
from one C-C bond to another. As a result, a trend is observed towards
formation of a lone electron pair and of vacancy at the carbon atom under
attack for systems anion+ electrophile and cation+ nucleophile,
respectively, along with weakening of the nearest C-C bond and strengthening
of the remaining bond. This phenomenon has been called the deconjugation
effect.

Given that the system under study consists of uniform fragments (as it is
the case with simple homogeneous fragmentary molecules defined in Section
5), retransformation procedures of Eqs.(11.4) and (11.6) also are similar.
As a result, these may be joined together to embrace total matrices $\mathbf{%
P}$ and $\mathbf{C}$ instead of their separate blocks ($\mathbf{P}_{I}$ and $%
\mathbf{C}_{I}$). To this end, a certain renumbering of basis orbitals (both
of FOs and of AOs (HAOs)) is required. As a result, separate blocks of the
retransformed DM start to yield individual intrafragmental effects
[107,108]. If we assume, for example, that our system consists of $N$
homopolar bonds, matrix analogues may be constructed for terms $1/2\Delta
X_{(k)I},$ $p_{(k)I}$\ and $\omega _{(k)I}$\ of Eqs.(11.9), (11.11) and
(11.15) that accordingly represent the interbond charge redistribution, the
intrabond polarization and the so-called rebonding effect [107] [Note that
lowering of internal bond orders due to the increment $\omega _{(2)I}$\ is
related to formation of those between BOs of different bonds as discussed
above and this fact serves to justify the term 'rebonding']. Given that our
system is supposed to be alternant in addition [this case embraces the
aliphatic conjugated hydrocarbons], the matrix analogue of the increment $%
\omega _{(k)I}$\ (denoted by $\Omega _{(k)}$) proved to be the only non-zero
submatrix of the retransformed correction $\mathbf{P}_{(k)}^{\prime }.$\
Accordingly, the rebonding effect starts to play the decisive role in the
formation of the relevant electronic structures [108].

It is seen, therefore, that universal environment-determined intrafragmental
effects may be defined and studied for systems containing a certain common
fragment. Moreover, the rules of qualitative chemical thinking (viz.
additivity, transferability and locality) are applicable in discussions of
these effects.

\section{The non-canonical MO theory of alternant hydrocarbons}

Alternant hydrocarbons (AHs) are among the first and most popular objects of
quantum chemistry. Application of the standard HMO theory (based on the H%
\"{u}ckel model in the framework of the canonical MO method) was especially
fruitful here. These studies resulted into the well-known common rules
governing the structures of CMOs of AHs and the relevant one-electron
energies that are included nowadays into almost all quantum chemistry
textbooks (see e.g. [22, 48]). It also deserves mentioning that CMOs of AHs
are delocalized over the whole system and depend on their individual
structures as usual. The less-known general form of the relevant CBO matrix $%
\mathbf{P}$ [31] may be added here as a related achievement and/or as a
corollary of the above-mentioned common constitution of CMOs of AHs. Two
alternative expressions have been obtained for the matrix $\mathbf{P,}$ viz. 
\begin{equation}
\mathbf{P=}\left\vert 
\begin{array}{cc}
\mathbf{I} & \mathbf{RB} \\ 
\mathbf{B}^{+}\mathbf{R} & \mathbf{I}%
\end{array}%
\right\vert ,\quad \mathbf{P=}\left\vert 
\begin{array}{cc}
\mathbf{I} & \mathbf{BQ} \\ 
\mathbf{QB}^{+} & \mathbf{I}%
\end{array}%
\right\vert ,  \tag{12.1}
\end{equation}%
where $\mathbf{B}$\ is the only non-zero submatrix of the common H\"{u}ckel
type Hamiltonian matrix of AHs shown in Eq.(3.1) and 
\begin{equation}
\mathbf{R}=\mathbf{(BB}^{+}\mathbf{)}^{-1/2},\qquad \mathbf{Q=(B}^{+}\mathbf{%
B)}^{-1/2}.  \tag{12.2}
\end{equation}%
[Note that the positive square root is assumed to be chosen here].
Uniqueness of the matrix $\mathbf{P}$\ is ensured here by the equality 
\begin{equation}
\mathbf{RB=BQ.}  \tag{12.3}
\end{equation}

The common CBO matrix of AHs $\mathbf{P}$\ of Eq.(12.1) indicates these
systems to resemble a single object. An analogous conclusion about AHs being
a class of chemical compounds has been drawn also in Section 3. Thus,
existance of the relevant common LMO representation matrix $\mathbf{C}$\ is
among natural expectations here.

To prove this hypothesis, we should evidently turn to the
block-diagonalization problem like that of Eq.(5.1) for the common
Hamiltonian matrix of AHs shown in Eq.(3.1). The basis of $2p_{z}$ \ AOs $%
\{\chi \}$\ may be assumed to be orthogonal as usual and thereby the
unitarity condition of Eq.(5.2) may be imposed on the matrix $\mathbf{C}$\
being sought. So far as the standard requirements of perturbation theory are
concerned, these are not met by the Hamiltonian matrix of AHs of Eq.(3.1).
Thus, we have to look for a non-perturbative solution of the relevant
block-diagonalization problem.

Solution of the above-desired type has been found in Ref.[109]. As a result,
two alternative forms of the matrix $\mathbf{C}$\ like those of Eq.(12.1)
have been derived, viz. \ 
\begin{equation}
\mathbf{C=}\frac{1}{\sqrt{2}}\left\vert 
\begin{array}{cc}
\mathbf{I} & \mathbf{RB} \\ 
\mathbf{B}^{+}\mathbf{R} & -\mathbf{I}%
\end{array}%
\right\vert ,\quad \mathbf{C=}\frac{1}{\sqrt{2}}\left\vert 
\begin{array}{cc}
\mathbf{I} & \mathbf{BQ} \\ 
\mathbf{QB}^{+} & -\mathbf{I}%
\end{array}%
\right\vert .  \tag{12.4}
\end{equation}%
A large extent of similarity between matrices $\mathbf{C}$\ and $\mathbf{P}$%
\ is evident. Moreover, the matrix $\mathbf{P}$\ of Eq.(12.1) was shown to
follow also from the commutation equation of Eq.(6.1), i.e. without invoking
CMOs. Expressions of Eqs.(12.1) and (12.4) form the basis for the
non-canonical MO theory of AHs and their derivatives developed in Refs
[109-113].

Let us turn now to the principal achievements of this alternative theory and
start with properties of NCMOs (LMOs) of AHs [109]. Expressions for these
MOs easily follow from Eq.(12.4). In particular, an occupied NCMO ($\Psi
_{(+)i}$) and a vacant one ($\Psi _{(-)m}$) take the form%
\begin{align}
\Psi _{(+)i} =&\frac{1}{\sqrt{2}}[\chi _{i}^{\ast }+\mathop{\displaystyle
\sum }\limits_{k=1}^{n}\chi _{k}^{\circ }(\mathbf{B}^{+}\mathbf{R)}_{ki}], 
\nonumber \\
\Psi _{(-)m} =&\frac{1}{\sqrt{2}}[-\chi _{m}^{\circ }+\mathop{\displaystyle
\sum }\limits_{j=1}^{n}\chi _{j}^{\ast }(\mathbf{B}^{+}\mathbf{R)}_{jm}], 
\tag{12.5}
\end{align}%
where sums over k and over j embrace AOs of subsets $\{\chi ^{\circ }\}$ and 
$\{\chi ^{\ast }\},$ respectively. It is seen that each NCMO is attached to
an individual AO. Thus, one-to-one correspondence between NCMOs and AOs
immediately follows along with the zero intrasubset delocalization of the
former. Moreover, occupied and vacant NCMOs originate from different subsets
of AOs, i.e. from $\{\chi ^{\ast }\}$ and $\{\chi ^{\circ }\},$ respectively.

Furthermore, uniform extents of delocalization are among the most important
distinctive features of NCMOs of AHs. Thus, both the relative weights of the
principal basis orbitals (AOs) and the respective total intersubset
delocalization coefficients are uniform for all NCMOs and equal to 1/2. This
result also implies that partial delocalization of NCMOs cannot exceed the
relative weight of the principal AO. Hence, NCMOs of AHs are actually of the
principal- orbital- and-tail constitution as it was the case with other
classes of molecules.

Finally, from similarity of Eqs.(12.1) and (12.4) it follows that the
vectors of coefficients of NCMOs of AHs coincide with respective columns
(rows) of the CBO matrix $\mathbf{P}$\ (up to the normalization factor $1/%
\sqrt{2}$). Consequently, the shapes of particular NCMOs may be predicted on
the basis of bond orders that are formed by the respective principal AO and
the AOs of the opposite subset. In particular, the AO of the second subset $%
\chi _{k}^{\circ }$\ contributes to the NCMO $\Psi _{(+)i}$\ if the bond
order between AOs $\chi _{k}^{\circ }$\ and $\chi _{i}^{\ast }$\ takes a
non-zero value. Substantial bond orders of conjugated hydrocarbons are known
to arise for neighboring pairs of $2p_{z}$ AOs corresponding to chemical
bonds, whilst those for other (non-neighboring) pairs are at least two times
smaller. These trends allow us to expect that the shape of a certain NCMO
(e.g. of $\Psi _{(+)i}$) depends decisively upon the number of the nearest
neighbors for the carbon atom (C$_{i}$), the respective principal AO ($\chi
_{i}^{\ast }$)\ is pertinent to and thereby on valency of this atom.
Inasmuch as mono-, di- and tri-valent carbon atoms are present in AHs, the
relevant NCMOs may be accordingly classified into two-, three- and
four-center non-canonical orbitals. As compared to LMOs of the bond-orbital-
and- tail constitution of alkanes (Section 5), the above-mentioned NCMOs are
generally less localized. This principal conclusion is in line with the
well-known hypothesis about the increasing overall extent of delocalization
of separate pairs of electrons when passing from saturated hydrocarbons
(alkanes) to their unsaturated analogues and especially to aromatic
compounds. It deserves adding in this context that the chemical
classification of molecules, in general, is based on the presumed different
extents of delocalization of these pairs in real space [10,85].

Let us turn now to comparison of NCMOs of AHs shown in Eq.(12.5) to the $%
\varepsilon -$dependent basis orbitals\ (GBOs) of the same systems [33]
discussed in Section 3 and shown in Eq.(3.5). Let us note first that the
eigenblocks ($\mathbf{E}_{1}$ and $\mathbf{E}_{2}$) of the common
Hamiltonian matrix of AHs of Eq.(3.1) (corresponding to the subsets of
occupied and vacant NCMOs, respectively) are as follows%
\begin{equation}
\mathbf{E}_{1}=\mathbf{R}^{-1}\mathbf{=(BB}^{+})^{1/2},\quad \mathbf{E}_{2}=-%
\mathbf{Q}^{-1}=-\mathbf{(B}^{+}\mathbf{B})^{1/2}.  \tag{12.6}
\end{equation}%
This implies that the occupied CMOs of AHs are expressible as linear
combinations of only occupied NCMOs (LMOs) $\Psi _{(+)i},i=1,2...N,$\ the
coefficients of these combinations coinciding with eigenvectors of the
matrix $\mathbf{(BB}^{+})^{1/2}$. [In the case of vacant CMOs(NCMOs), the
matrix $\mathbf{(B}^{+}\mathbf{B})^{1/2}$\ playes the same role]. Inasmuch
as matrices $\mathbf{BB}^{+}$ and $\mathbf{(BB}^{+})^{1/2}$\ are
characterized by a common set of eigenvectors, we may then conclude both the
above-mentioned linear combination and its counterpart in terms of GBOs (see
Eq.(3.4) of Section 3) to contain the same set of coefficients. This fact
makes the comparison of NCMOs to GBOs even more intriguing.

Similarity between NCMOs of AHs and their GBOs is beyond any doubt. Indeed,
both (occupied) NCMOs \ $\Psi _{(+)i}$\ and the relevant GBOs $\varphi
_{i}(\varepsilon )$\ are attached to individual AOs $\chi _{i}^{\ast }$\ as
Eqs.(3.5) and (12.5) indicate. Moreover, the overall shapes of both orbitals
are determined by the valency of the respective carbon atom and thereby by
the local structure. Thus, the conclusion of Section 3 about the local
structures of CMOs of AHs to be determined by local spatial constitutions of
these hydrocarbons proves to be additionally supported. Again, significant
differences between these orbitals also deserve mentioning. The most
important one concernes the way, the influence of the global structure upon
the particular orbital is represented by. Indeed, this influence manifests
itself via the eigenvalue $\varepsilon _{i}$\ and via the bond orders $%
(B^{+}R)_{ki}$\ in the GBO $\varphi _{i}(\varepsilon _{i})$\ of Eq.(3.5) and
in the NCMO $\Psi _{(+)i}$\ of Eq.(12.5), respectively. Accordingly, the
NCMOs of Eq.(12.5) possess tails embracing the remaining AOs of the opposite
subset along with their principal parts localized on the nearest
neighborhood of the given carbon atom. Meanwhile, GBOs possess no such tails
and thereby are more localized. Finally, the NCMOs of the present section
are orthogonal and eigenvalue-independent, whereas GBOs represent just the
opposite case.

Transformability of the common H\"{u}ckel type Hamiltonian matrix of
Eq.(3.1) into a block-diagonal form by means of the matrix $\mathbf{C}$\ \
of \ Eq. (12.4) allowed a new and efficient methodology to be developed for
solution of block-diagonalization problems and/or commutation equations for
more involved Hamiltonian matrices representing various derivatives of AHs
[110-113]. In these cases, the total Hamiltonian matrix contains a zero
order term $\mathbf{H}_{(0)}$\ referring to the respective\ parent AHs and a
first order term $\mathbf{H}_{(1)}$. The overall solution procedure then
consists of three principal steps: i) An initial passing to the basis of
NCMOs of the parent AHs by applying the transformation of Eq.(12.4) to both
terms of the total Hamiltonian matrix, ii) Application of the perturbation
theory like that of Sections 5 and 6 to solve the relevant non-canonical
problems for the transformed Hamiltonian matrix and iii) The subsequent
retransformation of the results of the second step into the basis of AOs $%
\{\chi \}$\ again.

The above-described methodology has been successfully applied to study the
so-called perturbed AHs (PAHs) in Ref.[110], including systems characterized
by a certain alteration of an individual Coulomb parameter originating from
replacement of the respective carbon atom by a more electronegative
heteroatom. The first order member $\mathbf{P}_{(1)}$\ of the power series
for the CBO matrix $\mathbf{P}$\ of AHs due to perturbation has been
analyzed. The well-known results concerning the consequent charge and bond
order redistributions in AHs have been rederived in this study without
invoking CMOs (including the famous rule of the alternating polarity). On
this basis, the above-mentioned results have been considered as a part of
the NCMO theory of PAHs.

Relations between reshapings of NCMOs due to perturbations and the relevant
charge redistributions rank among the principal new achievements of the same
theory. In particular, reshaping of a single NCMO was shown to reflect the
rule of the alternating polarity, namely of that NCMO the principal AO of
which coincides with the site of perturbation. Moreover, the overall
reshaping pattern of NCMOs proved to be in line with predictions of the
simple resonance theory about increased contributions of certain quinoidal
structures to the electronic strctures of PAHs due to perturbations.

Obtaining of general algebraic expressions for the second order corrections (%
$\mathbf{P}_{(2)}$) to the CBO matrix of AHs ($\mathbf{P}$) also is among
new achievements [111]. These corrections were shown to determine
alterations in bond orders between chemically bound pairs of atoms of AHs
under influence of the most popular types of perturbation, as well as to
play an important role in the formation of the stabilization energy of PAHs
[Invariance of Eqs.(8.4)-(8.7) towards an unitary transformation including
the matrix $\mathbf{C}$\ of Eq.(12.4) allows the component $\mathcal{E}%
_{(2)}^{(\alpha )}$\ of AHs to be related to alterations in the neighboring
bond orders due to perturbation]. On this basis, an additional insight was
given into the content of the well-known formulae for total energies of PAHs
in terms of self-polarizabilities of atoms and bonds, viz. an energy
correction was shown to be made up of a difference between the primary
stabilizing contribution of perturbation (which is twice as large as the
final stabilization energy) and the destabilizing increment related to
weakening of remaining chemical bonds. Again, the same stabilization energy
has been established to be additive with respect to contributions of
separate chemical bonds. Finally, a non-trivial and intriguing similarity
has been concluded between electronic structures of compounds originating
from the same parent hydrocarbon (R) after perturbation of the Coulomb
parameter of a certain AO $\chi _{r}$\ and after building up a composite AH
(R-R') by formation of a new bond between AOs $\chi _{r}$\ and $\chi
_{r^{\prime }}$\ of two identical AHs R and R' (e.g. pyridine and biphenyl).
Besides, this similarity is in line with conclusions of the simple resonance
theory too.

Another application of the same methodology concerns the substituted AHs
(SAHs) [112]. New rules have been established that govern the effects of
substituents upon charge and bond order redistributions in AHs. In
particular, two additive components have been revealed within these effects,
namely the charge transfer between the AH and the substituent and the
intersubset polarization inside the AH, the latter being governed by a
second order analogue of the rule of the alternating polarity. As a result,
a certain analogy has been concluded between electron density distributions
in PAHs and SAHs containing an electron-accepting substituent.

Finally, the latest achievement in the same field [113] concerns relative
reactivities of carbon atoms of PAHs vs. those of the parent AHs in the
electrophilic substitution (S$_{E}$2) reactions. Indeed, replacement of a
carbon atom of an AH by a more electronegative heteroatom (e.g. when passing
from benzene to pyridine) is known to give rise to suppression of
reactivities of the remaining carbon atoms. To prove this experiment-based
rule, the reactivities were related to the relevant extents of the
intermolecular charge transfer. General algebraic expressions have been
derived and analyzed for these extents for various directions of the attack
of electrophile, the latter being represented by a single initially-vacant
orbital ($\varphi _{(-)E}$) (Section 9). Changes in the intermolecular
charge transfer when passing from an AH to a PAH were shown to consist of
two components, viz. of an initial-population- dependent component
contributing increments of alternating signs for shifting positions of
electrophile along the hydrocarbon chain and of an additional negative
contribution originating from the increased interval between one-electron
energies of orbitals participating in the charge transfer. For soft
electrophiles, the negative contribution was shown to predominate over the
alternating one. This implies the charge transfer between heteroaromatic
reactants (PAHs) and electrophiles to be suppressed vs. the relevant values
for hydrocarbon-containing systems. Hence, the above-discussed result of
Ref. [81] concerning the suppressed reactivity of pyridine vs. benzene
(Section 9) is now generalized to any pair of a PAH and of its parent AH.

\section{Applications of the block-diagonalization transformation beyond
the limits of the Brillouin theorem}

In Section 5, the block-diagonalization problem arose as a matrix
representation of the Brillouin theorem (see Eqs.(5.1)-(5.3) and discussion
nearby). The initial Hamiltonian matrix ($\mathbf{H}$) has been transformed
in this case into a specific block-diagonal matrix ($\mathbf{E}$), namely
into that consisting of two eigenblocks $\mathbf{E}_{1}^{(n\times n)}$\ and $%
\mathbf{E}_{2}^{(s\times s)}$, the dimension of the former ($n$) being
unambiguosly determined by the total number of electrons in the given system
($2n$). To formulate and to solve this non-canonical one-electron problem
perturbatively, the system under study was additionally assumed to be
representable by two weakly-interacting subsets of basis functions $\mathbf{%
\{\Phi }_{1}\mathbf{\}}$\ and $\{\mathbf{\Phi }_{2}\}$. After partitioning
the initial matrix $\mathbf{H}$ into respective four submatrices (blocks),
the intersubset blocks ($\mathbf{R}$) have been accordingly supposed to be
first order terms vs. the intrasubset ones (see Eq.(5.4)). Solution of the
problem has been consequently expressed in terms of entire blocks of the
matrix $\mathbf{H}$ without specifying their internal structures.

It is evident that neither the block-diagonalization problem itself nor its
solution of the above-discussed rather general nature are necessarily
related to the Brillouin theorem. Indeed, we may look for transformation of
our initial matrix ($\mathbf{H}$) into a block-diagonal form of any
constitution, e.g. into that consisting of $N$ eigenblocks $\mathbf{E}%
_{1}^{(m\times m)},$ $\mathbf{E}_{2}^{(s\times s)},$ $\mathbf{E}%
_{3}^{(p\times p)}$ $...\mathbf{E}_{N}^{(t\times t)}$ of arbitrary
dimensions $m,s,p,..t,$ whatever the total number of electrons. Given that
our system is additionally representable by $N$ weakly interacting subsets
of corresponding dimensions $\mathbf{\{\Phi }_{1}\mathbf{\},}$\ $\{\mathbf{%
\Phi }_{2}\},$ $\mathbf{\{\Phi }_{3}\mathbf{\}}$... $\{\mathbf{\Phi }_{N}\},$
the above-described perturbative way of solution also seems to be
accordingly generalizable. It is also obvious that transformations of this
type may serve as intermediate steps in the way of diagonalization of the
matrix $\mathbf{H}$, i.e. as additional procedures inside the canonical
one-electron problem.

To clarify what is the good of this extra expenditure, let us dwell on
molecules and/or molecular systems consisting of $N$ weakly interacting
subsystems. Let the basis functions of individual subsystems to be included
into separate subsets $\mathbf{\{\Phi }_{1}\mathbf{\},}$\ $\{\mathbf{\Phi }%
_{2}\},$ $\mathbf{\{\Phi }_{3}\mathbf{\}}$... $\{\mathbf{\Phi }_{N}\}.$
Suppose that the relevant total Hamiltonian matrix is transformed into a
block-diagonal form consisting of $N$ eigenblocks $\mathbf{E}_{1}^{(m\times
m)},$ $\mathbf{E}_{2}^{(s\times s)},$ $\mathbf{E}_{3}^{(p\times p)}$ $...%
\mathbf{E}_{N}^{(t\times t)}$ of the appropriate dimensions. These
eigenblocks then seem to coincide with effective Hamiltonian matrices for
separate subsystems influenced by the intersubsystem interaction. Moreover,
elements of these matrices are likely to comply with the rule (principle) of
locality just owing to the perturbative nature of expressions for
eigenblocks. In this respect, a certain analogy may be expected between the
approach under present discussion and the alternative way of dealing with
eigenvalue equations outlined in Section 3.

Let us now turn to illustrations of these rather abstract assertions. Let us
confine ourselves first to the simplest case of systems containing two
weakly interacting subsystems ($N=2$). The initial Hamiltonian matrix of our
system is then representable as shown in Eq.(5.4), except for the dimensions
of blocks $m$ and $s$ being now determined by sizes of subsystems instead of
the total number of electrons. The same evidently refers to dimensions of
eigenblocks $\mathbf{E}_{1}^{(m\times m)}$\ and $\mathbf{E}_{2}^{(s\times s)}
$ being sought. The block-diagonalization problem to be considered also
coincides with that shown in Eqs. (5.1) and (5.2). Thus, no more is now
required as to supplement the solution of Section 5 by expressions for
eigenblocks. These are as follows [114]%
\begin{align}
\mathbf{E}_{1}^{(m\times m)}\  =&\mathbf{E}_{(+)}+\mathbf{T-}\frac{1}{2}(%
\mathbf{RG}_{(1)}^{+}+\mathbf{G}_{(1)}\mathbf{R}^{+})\mathbf{-}\frac{1}{2}(%
\mathbf{RG}_{(2)}^{+}+\mathbf{G}_{(2)}\mathbf{R}^{+})-...,  \nonumber \\
\mathbf{E}_{2}^{(s\times s)} =&-\mathbf{E}_{(-)}+\mathbf{Q+}\frac{1}{2}(%
\mathbf{G}_{(1)}^{+}\mathbf{R}+\mathbf{R}^{+}\mathbf{G}_{(1)})\mathbf{+}%
\frac{1}{2}(\mathbf{G}_{(2)}^{+}\mathbf{R}+\mathbf{R}^{+}\mathbf{G}%
_{(2)})+...,  \tag{13.1}
\end{align}%
where terms of power series to within the third order inclusive are
explicitly shown. Matrices $\mathbf{G}_{(1)}$ and $\mathbf{G}_{(2)}$\
coincide with those of Sections 5 and are conditioned by Eq.(5.7) as
previously.

As is seen from Eq.(13.1), the zero and first order members of the power
series for eigenblocks coincide with respective blocks of the matrix $%
\mathbf{H}$\ of Eq.(5.4) referring to subsets $\mathbf{\{\Phi }_{1}\mathbf{\}%
}$ and\ $\{\mathbf{\Phi }_{2}\}.$\ Thus, correspondence between the
eigenblocks $\mathbf{E}_{1}^{(m\times m)}$\ and $\mathbf{E}_{2}^{(s\times
s)},$ on the one hand, and subsystems of our system, on the other hand, is
beyond any doubt. At the same time, the eigenblocks imbibe the
intersubsystem interaction as the second and third order terms of the same
expressions indicate. Hence, matrices $\mathbf{E}_{1}^{(m\times m)}$\ and $%
\mathbf{E}_{2}^{(s\times s)}$ are nothing more than those of effective
Hamiltonians of subsystems in accordance with our above statement.
Furthermore, expressions of Eq.(13.1) resemble those for diagonal blocks of
the CBO matrix $\mathbf{P}$ (see Eq. (6.3)) in the sense that products of
matrices of intersubset interaction of lower orders, i.e. $\mathbf{R,G}%
_{(1)},\mathbf{G}_{(2)},etc.$ are present there. Thus, elements of the
eigenblocks $\mathbf{E}_{1}^{(m\times m)}$\ and $\mathbf{E}_{2}^{(s\times s)}
$ seem to comply with the rules of additivity and transferability along with
that of locality as it was the case with occupation numbers of basis
orbitals and bond orders between the latter (Section 6). To demonstrate
these peculiarities of eigenblocks in a more detail, let us turn to the case
of diagonal zero order Hamiltonian matrices as shown in Eq.(5.12) and
referring to fragmentary systems. Let us also assume that one-electron
energies are uniform inside individual subsystems and coincide with certain
constants $\varepsilon _{(0)1}$\ and $\varepsilon _{(0)2}$ as it was the
case with simple homogeneous systems (Section 5). Furher, the energy
reference point will be chosen in the middle of the intersubsystem energy
gap. Consequently, matrices $\mathbf{G}_{(1)}$ and $\mathbf{G}_{(2)}$\ are
expressible in terms of entire blocks of the first order Hamiltonian matrix
as shown in Eq.(5.15). The same then accordingly refers to eigenblocks $%
\mathbf{E}_{1}^{(m\times m)}$\ and $\mathbf{E}_{2}^{(s\times s)}.$ As for
instance, we obtain%
\begin{align}
\mathbf{E}_{1}^{(m\times m)}  =&\varepsilon _{(0)1}\mathbf{I}+\mathbf{T+}%
\frac{1}{\varepsilon _{(0)1}+\varepsilon _{(0)2}}\mathbf{RR}^{+}+\frac{1}{%
(\varepsilon _{(0)1}+\varepsilon _{(0)2})^{2}}[\mathbf{RQR}^{+}-  \nonumber
\\
&-\frac{1}{2}(\mathbf{RR}^{+}\mathbf{T+TRR}^{+})]+...  \tag{13.2}
\end{align}%
instead of the first relation of Eq.(13.1). As a result, diagonal and
off-diagonal elements of the first eigenblock are as follows 
\begin{align}
E_{1,ii}^{(m\times m)}  =&\varepsilon _{(0)1}+T_{ii}+\frac{1}{\varepsilon
_{(0)1}+\varepsilon _{(0)2}}\mathop{\displaystyle \sum }\limits%
_{k}^{(2)}(R_{ik})^{2}+...  \nonumber \\
E_{1,ij}^{(m\times m)}  =&T_{ij}+\frac{1}{\varepsilon _{(0)1}+\varepsilon
_{(0)2}}\mathop{\displaystyle \sum }\limits_{k}^{(2)}R_{ik}R_{jk}+... 
\tag{13.3}
\end{align}%
and contain sums of various types of direct and of indirect interactions of
orbitals concerned, i.e. of $\varphi _{1,i}$\ \ and $\varphi _{1,j}$. In
particular, the first order increment to the element $E_{1,ij}^{(m\times m)}$%
\ (i.e. $T_{ij}$) represents the direct (through-space) interaction of the
above-mentioned orbitals, whilst the respective second order correction
describes the relevant indirect interaction, wherein orbitals of the second
subsystem play the role of mediators (sums over $k$ embrace here the basis
functions of the second subsystem $\varphi _{2,k}$). It is also seen that
second order contributions to elements $E_{1,ii}^{(m\times m)}$\ and $%
E_{1,ij}^{(m\times m)}$\ depend on increments of individual orbitals of the
opposite subsystem in an additive way [Accordingly, the third order
increments contain sums over pairs of basis functions, at least one of them
belonging to the opposite subsystem]. If we recall here the known extinction
of resonance parameters when the distance between the relevant basis
orbitals grows (Section 5), elements of the effective Hamiltonian matrix may
be concluded to be determined mainly by that part of the second subsystem
which is attached to orbitals underlying the elements concerned (i.e. $%
\varphi _{1,i}$\ \ and $\varphi _{1,j}$). Given that the above-specified
local environments are similar, the relevant elements of the eigenblock $%
\mathbf{E}_{1}^{(m\times m)}$ are transferable. Besides, analogous
conclusions may be drawn also in the case of non-uniform one-electron
energies of basis functions, when elements of matrices $\mathbf{G}_{(1)}$
and $\mathbf{G}_{(2)}$\ are expressible as shown in Eqs.(5.13) and (5.14).
Therefore, the effective Hamiltonian matrices of separate subsystems are in
line with the principles of qualitative chemical thinking.

As the first specific example, let us consider the AMs $\mathbf{B}%
(G_{Ch}^{a})$\ of chemical graphs of alkanes in terms of atoms (Section 4).
Diagonal elements of these matrices were shown to coincide with 3 and 0 for
carbon and hydrogen atoms, respectively. These atoms will be now
correspondingly regarded as the first and the second subsystem. Accordingly,
parameters $\varepsilon _{(0)1}$\ and $\varepsilon _{(0)2}$ of Eq.(13.2)
coincide with 3 and 0, respectively. Further, the off-diagonal elements of
the AMs $\mathbf{B}(G_{Ch}^{a})$\ referring to both C-C and C-H bonds were
shown to take unit values. This implies the relevant elements of matrices $%
\mathbf{T}$\ and $\mathbf{R}$\ to coincide with 1, whilst those of the
matrix $\mathbf{Q}$\ vanish owing to absence of H-H bonds. Thus, the
intersubsystem energy gap (vs. the relevant interaction) coincides with 3
and seems to be sufficient for our illustrative purposes. Equation (13.2)
then yields the following formula for the eigenblock $\mathbf{E}%
_{1}^{(m\times m)}$\ corresponding to the C-skeleton of our alkane, viz. 
\begin{equation}
\mathbf{E}_{1}^{(m\times m)}\ =3\mathbf{I}+\mathbf{T+}\frac{1}{3}\mathbf{RR}%
^{+}-\frac{1}{18}[\mathbf{RR}^{+}\mathbf{T+TRR}^{+}]+...  \tag{13.4}
\end{equation}%
It is seen that the influence of hydrogen atoms upon the AM of the
C-skeleton ($3\mathbf{I}+\mathbf{T}$) is described mainly by the second
order term $1/3\mathbf{RR}^{+}.$ Off-diagonal elements of this term vanish
because two carbon atoms usually possess no common hydrogen atoms in
alkanes. Meanwhile, diagonal elements of the same correction are additive
with respect to increments of the attached hydrogen atoms and thereby are
proportional to the relevant number of the latter. For example, 2/3 and 1
follow from this term for internal and terminal carbon atoms of normal
alkanes, respectively. Thus, the terminal atoms prove to be characterized by
elements of effective Hamiltonian matrices of higher absolute values as
compared to internal atoms. This result seems to be in an excellent
aggreement with that following from Eq.(3.8). To make sure this is the case,
we should recall here that the notation $v$ of Eq.(3.8) stands for the
valency of the given carbon atom in the reduced graph $G_{ch}^{a\ast }.$\
For internal and terminal carbon atoms of normal alkanes, these valencies
equal to 2 and 1, respectively. Further, the energy variable $\varepsilon $\
may be approximately replaced by 3. The second term of Eq.(3.8) then also
correspondingly yields corrections 2/3 and 1. Thus, parallelism between the
results of Section 3 and those of the present approach is beyond any doubt.
It should be emphasized here, however, that a direct application of
Eq.(13.2) to the AMs $\mathbf{A}(G_{Ch}^{b})$\ of graphs of alkanes in terms
of bonds $(G_{Ch}^{b})$\ is hardly possible because the requirements of the
perturbation theory are not met in this case. Hence, the overall scope of
applicability of the present procedure seems to be narrower as compared to
that of the alternative approach of Section 3. Again, Eq.(13.1) provides us
with energy-variable- independent expressions for characteristics under
interest in contrast to formulae of Section 3.

Our second example belongs to regular quasi-one- dimensional systems defined
in Section 3, i.e. to chains of cyclic constitution characterized by
translational symmetry. Several chains of this type have been considered in
Ref.[115]. Let us confine ourselves here to a regular chain of $N$ uniform
atoms each of them contributing two AOs of the $s$ type (e.g. $ns=1s$ and $%
n\prime s=2s$) separated by a substantial energy gap. The overall chain may
be then considered as consisting of two weakly- interacting translationally-
symmetric subchains embracing the $ns$ ans $n\prime s$ AOs, respectively.
The first step of our analysis coincides with transforming the relevant
initial Hamiltonian matrix into a block-diagonal form containing two
eigenblocks $\mathbf{E}_{1}^{(N\times N)}$\ and $\mathbf{E}_{2}^{(N\times N)}
$\ that is equivalent to a direct application of Eq.(13.2) in practice.
Thus, we actually start with taking into account the intersubchain
interactions. The translational symmetry is automatically preserved during
this step and is dealt with later as described below. Hence, the present
approach also proves to be an alternative to the standard solid state theory
as it was the case with that of Section 3.

Given that the energy reference point is chosen in the middle of the
above-mentioned gap and the latter coincides with the double negative energy
unit, elements of the resulting effective Hamiltonian matrices $\mathbf{E}%
_{1}^{(N\times N)}$\ and $\mathbf{E}_{2}^{(N\times N)}$\ are uniform over
the subchains and take the form [115]%
\begin{align}
E_{1,ij}^{(N\times N)} =&(1+\beta ^{2}+\frac{1}{2}\gamma ^{2})\delta
_{ij}+(\sigma +\beta \gamma )(\delta _{i,j+1}+\delta _{i,j-1})+\frac{1}{2}%
\beta ^{2}(\delta _{i,j+2}+\delta _{i,j-2})+...  \nonumber \\
E_{2,ij}^{(N\times N)} =&-(1+\beta ^{2}+\frac{1}{2}\gamma ^{2})\delta
_{ij}+(\omega -\beta \gamma )(\delta _{i,j+1}+\delta _{i,j-1})-\frac{1}{2}%
\beta ^{2}(\delta _{i,j+2}+\delta _{i,j-2})+...  \nonumber \\
&  \tag{13.5}
\end{align}%
where $\omega ,\sigma ,\beta $ and $\gamma $ are resonance parameters of
different types [$\omega $ and $\sigma $\ correspondingly stand for
resonance parameters between the neighboring pairs of AOs inside the first
and the second subchain, whilst $\gamma $\ and $\beta $ are intersubchain
parameters inside a single atom and for a neighboring pair of atoms,
respectively]. The influence of the intersubchain interaction upon the
intrasubchain Hamiltonian matrix elements may be easily seen from Eq.(13.5).
Thus, one-electron energies of the $ns$ AOs become increased by%
\begin{equation}
\Delta =\beta ^{2}+\frac{1}{2}\gamma ^{2}  \tag{13.6}
\end{equation}%
under influence of the $n\prime s$ AOs, whereas those of the $n\prime s$ AOs
are decreased accordingly. This correction originates from the second order
term of Eqs.(13.2) and/or (13.3) and describes the indirect self-interaction
of the $ns(n\prime s)$ AOs. Furthermore, the interactions between the
first-neighboring AOs of the first subchain becomes increased by $\beta
\gamma $\ owing to the indirect interaction of these AOs by means of the
nearest AOs of the second subchain. The most important peculiarity of
elements $E_{1,ij}^{(N\times N)}$\ and \ $E_{2,ij}^{(N\times N)},$ however,
consists in the emergence of new effective interactions between the
second-neighboring pairs of AOs of the first (second) subchain under
influence of the second (first) one. Basis orbitals of the opposite subchain
situated in between the interacting AOs play the role of mediators in this
case. Effective parameters of these new interactions coincide with $\pm
1/2\beta ^{2}$. [It deserves mentioning here that third and subsequent
members of the same power series give birth to new interactions between
pairs of more remote AOs that are not exhibited explicitly in Eq.(13.5). The
main reason for that consists in their relatively small absolute values (the
effective parameters accordingly depend upon products of three and more
parent resonance parameters)].

If we recall now the overall block-diagonal constitution of the transformed
Hamiltonian matrix (that contains the eigenblocks $\mathbf{E}_{1}^{(N\times
N)}$\ and $\mathbf{E}_{2}^{(N\times N)}$ in its diagonal positions), we may
conclude our initial chain to be partitioned into two non-interacting chains
of regular and rather simple constitutions that may be subsequently studied
separately and independently.

To this end, the well-known methods of obtaining the dispersion relations
for simple chains may be applied. After employment of terms of the solid
state theory [39-41] (i.e. of $\mathbf{k}$\ and $\mathbf{a}$ defined in
Section 3), the final dispersion relations for the relevant two energy bands
are as follows%
\begin{align}
\varepsilon _{1}(\mathbf{k}) =&(1+\Delta )\delta _{ij}+2(\sigma +\beta
\gamma )\cos (\mathbf{ka})+\beta ^{2}\cos (2\mathbf{ka})+...  \nonumber \\
\varepsilon _{2}(\mathbf{k}) =&-(1+\Delta )\delta _{ij}+2(\omega -\beta
\gamma )\cos (\mathbf{ka})-\beta ^{2}\cos (2\mathbf{ka})+...  \tag{13.7}
\end{align}%
Similarity of these relations to those of Eqs.(3.10) and (3.11) is obvious.
Moreover, an evident interelation between separate terms of Eqs.(13.5) and
(13.7) along with the above-discussed local nature of the former allows an
interpretation of individual additive components of dispersion relations of
Eq.(13.7) in terms of local structure. Thus, the first $\mathbf{k-}$\
independent terms of these relations originate from effective energies of
AOs determined by parameter $\Delta $ of Eq.(13.6). Similarly, coefficients
of the $\cos (\mathbf{ka})-$\ like terms are related to effective
interactions between the first- neighboring AOs. Finally, the "weights" of
the $\cos (2\mathbf{ka})-$ like increments depend on the relative value of
the new effective interaction between the second- neighboring AOs. These
terms were shown to determine significant changes of the final dispersion
curves as compared to the simple \ $\cos (\mathbf{ka})-$like shape. Hence,
the above-concluded parallelism between the results of Section 3 and those
of the present procedure is now supported by another example. Besides, some
simple chains have been found in Ref.[115], when both approaches yield
coinciding results.

The third important specific example may be found in Ref.[114]. The model
studied there embraces molecules and molecular systems described by a
general formula A-(X)$_{n}$-B, where A and B stand for some terminal
functional groups and -(X)$_{n}$- is a bridge usually consisting of a
certain number of similar elementary units X. The effective interaction
between groups A and B was the principal characteristic under interest. In
this connection, the terminal groups and the bridge have been considered as
separate subsystems. Moreover, each terminal fragment has been represented
by a single orbital for simplicity. As a result, the only off-diagonal
element of the eigenblock $\mathbf{E}_{1}^{(2\times 2)}$\ has been expected
to represent the effective interaction concerned. Besides, the functional
groups A and B were assumed to be generally dissimilar so that the overall
model served to illustrate the case of non-uniform zero order one-electron
energies inside subsets. Explicit algebraic expressions have been derived
and analyzed for the above-described decisive matrix element to within the
fourth order members of power series inclusive. Additive nature of the
effective interaction between groups A and B has been demonstrated. The
direct (through-space) interaction between the orbitals concerned and
various types of the relevant indirect (bridge-assisted) interactions proved
to be among additive components here. In other words, the effective
interaction energy has been established to contain sums of increments, each
of them corresponding to a certain pathway through the bridge from one
terminal group to another. In this respect, a certain analogy reveals itself
between the present approach and the so-called partitioning technique [1]. A
more detailed discussion of this point may be found in Ref.[114].

Let us return again to the case of systems consisting of an arbitrary number
of weakly-interacting subsystems and/or of subsets. A general form of the
perturbation theory (PT) suitable for solution of the relevant
block-diagonalization problem has been developed in Ref.[53-55]. Comparison
of this new PT to the standard Rayleigh-Schr\"{o}dinger PT (RSPT) \ [2, 116,
117] showed that we actually have to do with passing from the usual
one-dimensional Hamiltonian matrix elements to multidimensional parameters
(i.e. submatrices) . Since commutative quantities become replaced by
non-commutative ones in this case, the new PT has been called the
non-commutative Rayleigh-Schr\"{o}dinger PT (NCRSPT). Peculiarities of this
alternative PT vs. the standard RSPT has been discussed in Ref.[55] in a
detail. Let us confine ourselves here to properties of eigenblocks $\mathbf{E%
}_{i}(i=1,2...N).$

Let us assume the initial Hamiltonian matrix $\mathbf{H}$ to consist of a
certain zero order member $\mathbf{H}_{(0)}$\ [being a direct sum of $N$
non-zero blocks $\mathbf{E}_{(0)i}(i=1,2...N)$\ ] and of a first order
(perturbation) term denoted by $\mathbf{V}$ as usual.\ The ith eigenblock $%
\mathbf{E}_{i}$ takes then the following form [115]%
\begin{equation}
\mathbf{E}_{i}=\mathbf{E}_{(0)i}+\mathbf{V}_{ii}+\frac{1}{2}%
\mathop{\displaystyle \sum }\limits_{j}(1-\delta _{ij})(\mathbf{V}_{ji}^{+}%
\mathbf{C}_{ji}^{(1)}+\mathbf{C}_{ji}^{(1)+}\mathbf{V}_{ji})+...  \tag{13.8}
\end{equation}%
where the sum over $j$ embraces here all the subsets of basis functions.
Notations $\mathbf{C}_{ji}^{(1)}$, in turn, are used for blocks of the first
order member \ $\mathbf{C}^{(1)}$ of the power series for the transformation
matrix $\mathbf{C}$\ as it was the case in Eq.(5.5). These blocks are
conditioned by matrix equations like that of Eq.(5.7) as previously. If we
recall that $\mathbf{G}_{(1)}$ of Eq.(13.1) coincides with $\mathbf{C}%
_{12}^{(1)}$\ (see Eq.(5.6)), the new formula (13.8) proves to be a direct
generalization of Eq.(13.1) to the case of an arbitrary number of
subsystems. \ 

In this context, comparison of the present approach to the so-called PMO
theory [23] also deserves attention. Let us start with a brief specifying of
the latter.

Traditional ways of investigation of systems consisting of weakly
interacting subsystems are based on application of the quasi-degenerate RSPT
[2, 116, 117]. The overall treatment then starts with diagonalization of the
intrasubsystem blocks of the initial Hamiltonian matrix, whilst taking into
account the intersubsystem (intersubset) interaction makes up the second
step. This procedure actually implies passing to the basis of delocalized
(canonical) MOs of isolated subsystems from the very outset of solving the
problem and a subsequent regard for interaction between these MOs. Just this
approach is usually referred to as the PMO theory [23]. It is evident that
application of the NCRSPT offers us a possibility of an opposite order of
operations versus that of the PMO theory. Indeed, the intersubsystem
interactions are taken into account by means of an initial
block-diagonalization procedure, i.e. before regard fot the intrasubsystem
ones. Thereupon, the eigenblocks obtained may be diagonalized as usual.

Before finishing this Section, let us return to comparison of the present
alternative approach to that of Section 3 and summarize their similarities
and differencies. First, both alternative ways of dealing with secular
equations are based on an inverted order of the principal operations vs. the
usual one. Second, both approaches comply with the classical principles of
the qualitative chemical thinking, i.e. with those of locality, additivity
and trasferability. Significant differences between these approches also
deserve mentioning. First, the eigenvalue- dependent effective Hamiltonian
matrices of Section 3 describe the whole system under study and not its
separate subsystems. Second, quite extended fragments of the system and not
separate AOs correspond to diagonal elements of matrices of Section 3.
Finally, no conditions of perturbative nature have been imposed when using
the approach of Section 3 in contrast to the present one.

\section{Conclusions}

The principal achievements of above-overviewed studies are as follows:

1. Common quantum-chemical (algebraic) problems are formulated for extended
classes of molecules in analogy with the classical chemistry, where a class
of compounds is considered as a single object. The new problems coincide
either with eigenblock equations for the relevant common Hamiltonian
matrices or with commutation equations determining the one-electron density
matrices and thereby find themselves beyond the usual secular (eigenvalue)
problems for Hamiltonian matrices of individual compounds. Again, an
eigenblock equation is shown to be a generalization of an eigenvalue
equation for a definite two-dimensional matrix, wherein multidimensional
(non-commutative) quantities (submatrices) play the role of the usual matrix
elements. Accordingly, general solutions of these newly-formulated problems
are obtained in terms of entire submatrices of the common Hamiltonian matrix
of the given class of molecules without specifying either the structures or
dimensions of these submatrices.

2. Formations of numerous fundamental quantum-chemical characteristics of
molecules are shown to be governed by rules of qualitative chemical
thinking, viz. additivity, transferability and locality. This refers not
only to traditionally "semi-classical" characteristics (e.g. total energies,
dipole moments, etc.), but also to such "purely quantum-mechanical"
quantities as one-electron spectra and canonical MOs. In particular,
dispersion relations representing one-electron energy bands of regular
quasi-one- dimensional systems are shown to consist of a few additive
components, each of them being interpretable in terms of specific
peculiarities of local constitutions.

3. Quantum-chemical and thereby algebraic analogues are constructed for
numerous qualitative chemical concepts including different roles of the
reaction center and of its neighbourhoods in chemical processes, the Lewis
perspective on chemical reactions (cf. the so-called "curly arrow
chemistry") and others. Accordingly, some known intuition-based verbal
relations now acquire rigorous algebraic forms, e.g. the relation between
charge redistribution, delocalization and stabilization, the relation
between peculiarities of the inductive effect and the extent of
delocalization of electrons in the parent hydrocarbon, etc.

4. Scopes of applicability of some classical concepts and/or models are
explored and conditions of their validity are formulated explicitly. This
primarily refers to various local and semilocal models of chemical
reactions, as well as to models based on the "curly arrow chemistry".
Moreover, the above-mentioned intuition-based relations also are shown to be
of a limited validity.

5. A great cognitive potential of quasi-classical alternatives under
developement is demonstrated by defining new generalities and by introducing
new concepts and/or rules. Universal environment- determined intrafragmental
effects (such as homo- and heterolytic predissociation of bonds) and
recently-defined classes of fragmentary and homogeneous molecules may be
mentioned as examples of new generalities. Again, the newly- introduced
rules may be exemplified by the common selection rule for organic reactions
in terms of signs of direct and indirect interorbital interactions. Novel
applications of the classical concepts in the "purely quantum-mechanical"
fields also may be added here, e.g. interpretation of one-electron spectra
in terms of peculiarities of the relevant chemical formulae (molecular
graphs).

\section*{REFERENCES}

1. R. McWeeny, \textit{Methods in Molecular Quantum Mechanics,} 2nd. ed.
(Academic Press, London, 1992).

2. L. Z\"{u}licke, \textit{Quantenchemie, Bd.1, Grundlagen und Algemeine
Methoden} (Deut- scher Verlag der Wissenschaften, Berlin, 1973).

3. J. Simons, \textit{An Introduction to Theoretical Chemistry} (Cambridge
Univ. Press, Cambridge, 2003).

4. C.J. Cramer, \textit{Essentials of Computational Chemistry,} \textit{%
Theories and Models,} 2nd. ed. (John Wiley \& Sons, 2004).

5. F. Jensen, \textit{Introduction to Computational Chemistry}, 2nd. ed.
(John Wiley \& Sons, 2007).

6. E.G. Lewars, \textit{Computational Chemistry: Introduction to the Theory
and Applications of Molecular and Quantum Mechanics, }2nd. ed. (Springer,
Heidelberg, 2011).

7. A. Szabo, N.S. Ostlund, \textit{Modern Quantum Chemistry } (McGraw-Hill,
1982).

8. O. Chalvet (Ed.), \textit{Localization and Delocalization in Quantum
Chemistry, Atoms and Molecules in the Ground State}, Vol.1 (Reidel,
Dordrecht, 1975).

9. V. M. Tatevskii, \textit{Constitution of Molecules} (Khimia, Moscow,
1977) [in Russian].

10. J. March, \textit{Advanced Organic Chemistry, Reactions, Mechanisms and
Structure} (Wiley Interscience, New York, 1985).

11. F.A. Carroll, \textit{Perspectives on Structure and Mechanism in Organic
Chemistry} (Brooks/Cole, Pacific Grove, CA, 1998).

12. V. Gineityte, J. Mol. Struct. (Theochem) \textbf{491}, 205 (1999).

13. C. Yang, J. Mol. Struct. (Theochem) \textbf{169}, 1 (1988).

14. G. Del Re, F. Zuccaello, Croat. Chem. Acta, \textbf{64}, 449 (1991).

15. M. Edenborough, \textit{Organic Reaction Mechanisms. A Step by Step
Approach} (Taylor and Francies, London, 1999).

16. L. Pauling, \textit{The Nature of the Chemical Bond,} 3rd. ed. (Cornell
Univ. Press, Ithaca, New York, 1960).

17. A. Streitwieser, Jr., \textit{Molecular Orbital Theory} (John Wiley
\&Sons, New York, 1961).

18. W. Kutzelnigg, J. Comput. Chem. \textbf{28}, 25 (2007).

19. E. H\"{u}ckel, Z. Phys. \textbf{70}, 204 (1931).

20. E. H\"{u}ckel, Z. Phys. \textbf{76}, 628 (1932).

21. \ R. Zahradnik, R. Polak, \textit{Elements of Quantum Chemistry} (Plenum
Press, New York, 1980).

22. N. Trinajstic, \textit{The H\"{u}ckel Theory and Topology}, in: G.A.
Segal (Ed.), Semiempirical Methods in Electronic Structure Calculations,
Part A, Techniques (Plenum Press, New York, London, 1977).

23. M.J.S. Dewar, and R.C. Dougherty, \textit{The PMO Theory of Organic
Chemistry} (Plenum Press, New York, 1975).

24. J.-P. Malrieu, J. Mol. Struct. (Theochem) \textbf{424}, 83 (1998).

25. T.A. Carlson, Ann. Rev. Phys. Chem., \textbf{26}, 211 (1975).

26. T.C. Koopmans, Physica, \textbf{1}, 104 (1934).

27. G. Bieri, F. Burger, E. Heilbronner, J.P. Maier, Helv. Chim. Acta, 
\textbf{60}, 2213 (1977).

28. G. Bieri, J.D. Dill, E. Heilbronner, A. Schmelzer, Helv. Chim. Acta, 
\textbf{60}, 2234 (1977).

29. E. Heilbronner, Helv. Chim. Acta, \textbf{60}, 2248 (1977).

30. E. Heilbronner, \textit{The Photoelectron Spectra of Saturated
Hydrocarbons,} In: The Chemistry of Alkanes and Cycloalkanes\textit{, }Eds.
S.Patai, Z.Rappoport (Wiley, New York, 1992), pp. 455-529.

31. G.G. Hall, Proc. Roy. Soc. (London), \textbf{A229}, 251 (1955).

32. V. Gineityte, Int. J. Quant. Chem. \textbf{60}, 717 (1996).

33. V. Gineityte, MATCH Commun. Math. Comput. Chem. \textbf{70}, 119 (2013).

34. V. Gineityte, J. Mol. Struct. (Theochem) \textbf{487}, 231 (1999).

35. R.B. Potts, J. Chem. Phys. \textbf{21}, 758 (1953).

36. V. Gineityte, Int. J. Quant. Chem. \textbf{53}, 245 (1995).

37. V. Gineityte, Croat. Chem. Acta, \textbf{71}, 673 (1998).

38. V. Gineityte, Int. J. Quant. Chem. \textbf{64}, 481 (1997).

39. R. Hoffmann, \textit{Solids and Surfaces: A Chemist's View of Bonding in
Extended Structures} (VCH Publishers, 1988).

40. A.A. Levin, \textit{Introduction to Quantum Chemistry of Solids}
(Khimia, Moscow, 1974) [in Russian].

41. Ch. Kittel, \textit{Introduction to Solid State Physics} (Wiley, New
York, 1977).

42. D. M. Cvetkovic, M. Doob, H. Sachs,\ \textit{Spectra of Graphs.} \textit{%
Theory and Application} (VEB Deutscher Verlag der Wissenschaften, Berlin,
1980).

43. V. Gineityte, Int. J. Quant. Chem. \textbf{60}, 743 (1996).

44. I. Mayer, Chem. Phys. Lett. \textbf{89}, 390 (1982).

45. P.R. Surjan, I. Mayer, and M. Ketesz, J. Chem. Phys.\textbf{\ 77}, 2454
(1982).

46. I. Mayer, and P.R Surjan, J. Chem. Phys. \textbf{80}, 5649 (1984).

47. J.P. Daudey, Chem. Phys. Lett. \textbf{24}, 574 (1974).

48. S. Huzinaga, \textit{The MO Method} (Mir, Moscow, 1983) [in Russian].

49. V Gineityte, J. Mol. Struct. (Theochem), \textbf{288}, 111 (1993).

50. V. Gineityte, J. Mol. Struct. (Theochem), \textbf{333}, 297 (1995).

51. V. Gineityte, J. Mol. Struct. (Theochem) \textbf{343}, 183 (1995).

52. V. Gineityte, Int. J. Quant. Chem. \textbf{72}, 559 (1999).

53. V. Gineityte, Int. J. Quant. Chem. \textbf{68}, 119 (1998).

54. V. Gineityte, Lith. J. Phys. \textbf{51}, 107 (2011).

55. V. Gineityte, Lith. J. Phys. \textbf{44}, 219 (2004).

56. V. Gineityte, J. Mol. Struct. (Theochem) \textbf{342}, 219 (1995).

57. P. Lankaster, \textit{Theory of Matrices }(Academic Press, New York,
1969).

58. V. Gineityte, J. Mol. Struct. (Theochem) \textbf{680}, 199 (2004).

59. V. Gineityte, D.B. Shatkovskaya, Zh. Strukt. Khim. \textbf{25}, 152
(1984).

60. M.M. Mestetchkin, \textit{The Density Matrix Method in Quantum Chemistry}
(Naukova Dumka, Kiev, 1977) [in Russian].

61. V. Gineityte, J. Mol. Struct. (Theochem) \textbf{585}, 15 (2002).

62. G.N. Lewis, J. Amer. Chem. Soc. \textbf{38}, 762 (1916).

63. V. Gineityte, J. Mol. Struct. (Theochem) \textbf{364}, 85 (1996).

64. V. Gineityte, Int. J. Chem. Model. \textbf{5}, 99 (2013).

65. V. Gineityte, MATCH Commun. Math. Comput. Chem. \textbf{72}, 39 (2014).

66. V. Gineityte, Int. J. Quant. Chem. \textbf{77}, 534 (2000).

67. V.F.Traven, \textit{Electronic Structure and Properties of Organic
Molecules} (Khimia, Moscow, 1989) [in Russian].

68. M. Randic, Chem. Revs. \textbf{103}, 3449 (2003).

69. M. Randic, N. Trinajstic, \textit{On Conjugated Chains,} in: R.C. Lacher
(Ed.), MATCH/CHEM/COMP, 1987, Elsevier, Amsterdam, 1988, p.p. 109-123.

70. G. Klopman (Ed.), \textit{Chemical Reactivity and Reaction Paths} (John
Wiley and Sons Inc., New York, London, Sydney, Toronto, 1974).

71. N.V. Vasiljeva, \textit{Theoretical Introduction into Organic Synthesis }%
(Prosveshtchenie, Moscow, 1976) [in Russian].

72. K. Fukui, H. Fujimoto, Bull. Chem. Soc. Japan \textbf{41}, 1989 (1968).

73. K. Fukui, H. Fujimoto, Bull. Chem. Soc. Japan \textbf{42}, 3399 (1969).

74. J.M. Tedder and A. Nechvatal, \textit{Pictorial Orbital Theory} (Pitman,
London 1985).

75. V. Gineityte, J. Mol. Struct. (Theochem) \textbf{465}, 183 (1999).

76. V. Gineityte, Int. J. Quant. Chem. \textbf{94}, 302 (2003).

77. V. Gineityte, J. Mol. Struct. (Theochem) \textbf{541}, 1 (2001).

78. V. Gineityte, J. Mol. Struct. (Theochem) \textbf{588}, 99 (2002).

79. V. Gineityte, J. Mol. Struct. (Theochem) \textbf{663}, 47 (2003).

80. V. Gineityte, J. Mol. Struct. (Theochem) \textbf{726}, 205 (2005).

81. V. Gineityte, J. Mol. Struct. (Theochem) \textbf{760}, 229 (2006).

82. V. Gineityte, Lith. J. Phys. \textbf{49}, 389 (2009).

83. V.Gineityte, Z. Naturforsch. \textbf{64A}, 132 (2009).

84. C.K. Ingold, \textit{Structure and Mechanism in Organic Chemistry}
(Cornell University Press, Ithaca, 1953).

85. H.G.O. Becker, \textit{Einf\"{u}rung in die Elektronentheorie Organisch
Chemischen Reactionen} (Deutscher Verlag der Wissenschaften, Berlin, 1974).

86. V. Gineityte, J. Mol. Struct. (Theochem) \textbf{546}, 107 (2001).

87. R.B. Woodward, R. Hoffmann, J. Amer. Chem. Soc. \textbf{87} 395 (1965).

88. R.B. Woodward, R. Hoffmann, \textit{The Conservation of Orbital Symmetry}
(Verlag Chemie/ Academic Press, Weinheim, 1971).

89. R. Hoffmann, R.B. Woodward, Acc. Chem. Res. \textbf{1,} 17 (1968).

90. M.J.S. Dewar, Tetrahedron Suppl. \textbf{8,} 75 (1966).

91. M.J.S. Dewar, Angew. Chem. Int. Ed. \textbf{10,} 761 (1971).

92. M.J.S. Dewar, S. Kirschner, H.W. Kollmar, J. Amer. Chem. Soc. \textbf{96,%
} 5240 (1974).

93. E. Zimmerman, Acc. Chem. Res. \textbf{4,} 272 (1971).

94. V. Gineityte, J. Mol. Struct. (Theochem) \textbf{714}, 157 (2005).

95. V. Gineityte, Int. J. Quant. Chem. \textbf{108}, 1141 (2008).

96. V. Gineityte, Int. J. Chem. Model. \textbf{4}, 189 (2012).

97. E. Heilbronner, Tetrahedron Lett. \textbf{29,} 1923 (1964).

98. A.N. Vereshtchagin, \textit{The Inductive Effect} (Nauka, Moscow, 1987)
[in Russian].

99. V. Gineityte, Int. J. Quant. Chem. \textbf{110}, 1327 (2010).

100. V. Gineityte, J. Mol. Struct. (Theochem) \textbf{766}, 19 (2006).

101. V. Gineityte, Lith. J. Phys. \textbf{45}, 7 (2005).

102. V. Gineityte, J. Mol. Struct. (Theochem) \textbf{810}, 91 (2007).

103. V. Gineityte, J. Mol. Struct. (Theochem) \textbf{434}, 43 (1998).

104. V. Gineityte, J. Mol. Struct. (Theochem) \textbf{532}, 257 (2000).

105. V. Gineityte, J. Mol. Struct. (Theochem) \textbf{713}, 93 (2005).

106. V. Gineityte, J. Mol. Struct. (Theochem) \textbf{507}, 253 (2000).

107. V. Gineityte, J. Mol. Struct. (Theochem) \textbf{430}, 97 (1998).

108. V. Gineityte, J. Mol. Struct. (Theochem) \textbf{497}, 83 (2000).

109. V. Gineityte, Int. J. Quant. Chem. \textbf{101}, 274 (2005).

110. V. Gineityte, Int. J. Quant. Chem. \textbf{105}, 232 (2005).

111. V. Gineityte, Int. J. Quant. Chem. \textbf{106}, 2145 (2006).

112. V. Gineityte, Croat. Chem. Acta \textbf{81}, 487 (2008).

113. V. Gineityte, Croat. Chem. Acta \textbf{86}, 171 (2013).

114. V. Gineityte, Lith. J. Phys. \textbf{42}, 397 (2002).

115. V. Gineityte, Int. J. Quant. Chem. \textbf{81}, 321 (2001); Erratum, 
\textbf{82}, 262 (2001).

116. L.D. Landau, E.M. Lifshits, \textit{Quantum Mechanics. The
Non-relativistic Theory }(Nauka, Moscow, 1974).

117. W.H. Flygare, \textit{Molecular Structure and Dynamics} (Prentice-Hall,
Englewood Cliffs/ New York, 1978).

\end{document}